\documentclass[aps,twocolumn,preprintnumbers,nofootinbib,showpacs,prd]{revtex4-1}
\usepackage{graphicx,color}
\usepackage{amsmath,amssymb}
\usepackage{hyperref}
\usepackage{enumerate}

\newcommand{\minus}[0]{\scalebox{0.75}[1]{-}}

\begin{document}

\title{On Tree-Level Unitarity in Theories of Massive Spin-2 Bosons}
\author{Neil D. Christensen and Stefanus\\
PITTsburgh Particle physics, Astrophysics and Cosmology Center (PITT PACC), Department of Physics and Astronomy, University of Pittsburgh, Pittsburgh, PA 15260 USA}
\date{\today}
\begin{abstract}
We analyze the tree-level $2\to2$ scattering of massive spin-2 bosons in a theory with only relevant and marginal operators and extract the sum rules on the coupling constants and masses required to achieve tree-level unitarity to very high energy.  We do this for four illustrative cases.  In the first, we include massive spin-1 and spin-0 bosons in our theory, but do not require gauge invariance.  For this case, we find that it is, in fact, possible to construct a theory where all $2\to2$ scattering processes are tree-level unitary to very high energy.  In the second case, we consider a theory that only includes massive spin-2 and spin-0 bosons.  In the absence of spin-1 bosons, we find that it is impossible to unitarize the spin-2 scattering.  In the third and fourth cases, we reintroduce the spin-1 bosons as gauge bosons and demand that all interactions are gauge invariant.  We take the spin-2 bosons to transform under the adjoint representation of one or two gauge groups.  For both of these cases, we find that unitarization is unachievable.
\end{abstract}
\maketitle

It is generally believed that all theories with a finite number of fields of spin greater than one are nonrenormalizable and perturbatively nonunitary \cite{Bekaert:2009ud}.  In fact, there have been several no-go theorems for a finite number of massless fields with higher spin \cite{Weinberg:1964ew}.
In the case of gravitational theories, the nonrenormalizability has been shown to occur at one-loop for scalar fields, fermions and gauge fields coupled to gravity \cite{'tHooft:1974bx,deser}, while pure gravitational theories have been shown to be nonrenormalizable at two-loops \cite{'tHooft:1974bx,sagnotti}.  
With these high-profile failures, the renormalizability of theories with spin-2 fields might seem hopeless.
However, on the other hand, $N=8$ supergravity theories have been shown to be divergence free up to at least four-loop order \cite{Bern:1998ug} with work currently under way at five loops \cite{Bern:2012uc}.  
Also, new directions in scattering theory that bypass fields altogether look promising for consistently describing spin-2 particles \cite{Arkani-Hamed:2013jha}.
Furthermore, to the best of our knowledge, it has never been formally proved that \textit{all} theories of spin-2 bosons are nonrenormalizable.

Finding a fully renormalizable and perturbatively unitary theory of interacting spin-2 bosons is a great challenge and we certainly do not claim to have achieved this in the present article.  Instead, we take on a much more modest challenge.  We attempt to find a theory of massive spin-2 bosons that only contains relevant (dimension-3) operator and marginal (dimension-4) operator interactions with fields of spin-0, spin-1, and spin-2 that is tree-level unitary at high energy.  Since in the present article, we are interested in the general existence of such a theory, for the moment, we completely ignore any higher symmetry and focus instead on freely choosing the Lorentz invariant operators that will give tree-level unitarity.  With this in mind, we admit right away that our results may have no relevance to gravitational type theories.  On the other hand, we think this is an interesting quantum field theory question and we leave our minds open to what, if any, possible future applications may come from this.  This allows us to freely choose our couplings in a way that achieves our goal of a tree-level unitary theory.

This article is organized as follows:  In App.~\ref{app:no energy growth}, we review the fact that unitarity implies a lack of polynomial energy growth at high energies in $2\to2$ scattering processes.  When we say that we have achieved tree-level unitarity in our theory, we mean that we have cancelled this polynomial high energy growth in all tree-level $2\to2$ scattering processes and the theory is, with appropriate choice of couplings and masses, tree-level unitary up to some high energy scale, such as the Planck scale.  In App.~\ref{sec:Feynman Rules}, we list all possible Lorentz invariant relevant and marginal operators that give a unique non-zero contribution to the tree-level $2\to2$ scattering amplitudes of our massive spin-2 particle.  We also describe the propagators that we use for each spin.  For convenience, we take all our fields to be real.  In App.~\ref{app:polarization vectors}, we describe our conventions for the momenta and polarization vectors.  In App.~\ref{sec:equivalencies}, we describe the relationships between the helicity amplitudes due to rotational, P and T invariance and show that only 97 of the 625 helicity amplitudes are linearly independent.  In App.~\ref{app:M0000}, we give the high polynomial energy contributions to the $2\to2$ scattering amplitude from each diagram for the external helicities $0,0,0,0$ while in App.~\ref{app:M20-10}, we do the same for helicities $2,0,-1,0$.  This appendix can be used to check the readers calculation against ours.  

With these details out of the way, we focus, in the main body of this article, on the sum rules between the couplings and the masses that result from demanding tree-level unitarity of the $2\to2$ scattering amplitudes.  We do this by series expanding the amplitudes in the small ratio of masses divided by the energy of the collision.  With this expansion, the amplitudes are polynomials in the energy and $\cos\theta$ where $\theta$ is the scattering angle.  Since the high energy growth must cancel for all high energies and all angles, and since each power of energy and $\cos\theta$ are linearly independent, the only way for tree-level unitarity to be satisfied is for the coefficient of each positive power of energy and $\cos\theta$ to vanish.  This gives a system of equations relating the coupling constants and the masses that must be satisfied.  Since this must be done for all helicity combinations of the external particles, it might at first appear that this system of equations will be very badly over-constrained.  However, most of these equations are linearly dependent and the final system of equations can be satisfied in some but not all cases as we will describe for four illustrative cases.

In Sec.~\ref{sec:non gauge-invariant}, we consider theories with only one spin-2 boson and any number of spin-1 and spin-0 bosons.  In this section, we do \textit{not} enforce gauge invariance but allow the spin-1 couplings to take on any Lorentz invariant form.  We present the sum-rules between the couplings and masses that must be satisfied for tree-level unitarity to be achieved and discuss scenarios where these sum-rules are fulfilled.
In Sec.~\ref{sec:multiple spin-2}, we consider theories with no spin-1 bosons which are, therefore, trivially gauge invariant.  We do, however, allow as many spin-2 and spin-0 bosons as required to achieve unitarity.  In contrast to the previous case, we find that only the trivial solution where all the couplings are zero satisfies the system of equations required for unitarity.  This is in interesting distinction compared to what occurs in Higgsless theories where it is found that massive spin-$\frac{1}{2}$ and -1 scattering can be made perturbatively unitary in the absence of fields of lower spin \cite{SekharChivukula:2001hz,Sekhar Chivukula:2007mw}. 
In Sec.~\ref{sec:adjoint rep}, we bring back the spin-1 field as a gauge boson and require that all interaction operators are gauge invariant.  In addition to the gauge boson, we allow as many spin-2 and spin-0 fields as required for unitarity and take the spin-2 fields to transform under the adjoint representation of the group.  Interestingly, we find in this case that, due to the antisymmetry of the structure constant, the diagrams with intermediate spin-2 fields all identically vanish and can not contribute to the unitarization of the scattering process.  We further find that only the trivial theory with all potentially contributing couplings vanishing is allowed.
In Sec.~\ref{sec:Symmetric cubic term}, we again consider a theory with only gauge invariant interactions.  However, this time we include two gauge groups and take the spin-2 fields to transform under the adjoint representation of both gauge groups.  In contrast to the previous case, we find that diagrams with intermediate spin-2 fields do contribute to the scattering amplitude.  We again allow as many spin-2 and spin-0 fields as necessary.  Nevertheless, we again find that only the trivial solution with all couplings set to zero is allowed by unitarity.  In Sec.~\ref{sec:conclusion}, we conclude.

\tableofcontents

\section{\label{sec:non gauge-invariant}Non-Gauge Invariant Theory}
In this section we consider a theory with a single spin-2 boson and as many spin-1 and spin-0 bosons as required.  In this first attempt, we do not concern ourselves with gauge invariance but allow the couplings with spin-1 fields to take on whatever Lorentz invariant form necessary to achieve a tree-level unitary theory.  We will return to theories that are gauge invariant in Sections~\ref{sec:multiple spin-2}, \ref{sec:adjoint rep} and \ref{sec:Symmetric cubic term}.  

We consider the most general set of dimension-3 and -4 operators consistent with Lorentz invariance that make unique nonvanishing contributions to the tree-level scattering process $h_{\mu_1\nu_1}(p_1)h_{\mu_2\nu_2}(p_2)\to h_{\mu_3\nu_3}(p_3)h_{\mu_4\nu_4}(p_4)$, where $h$ is a spin-2 field.  This means that operators that are antisymmetric in the Lorentz indices of a single external $h$ or trace the Lorentz indices of a single external $h$ are not included.  It also means that operators that differ only in the interchange of the Lorentz indices of a single external $h$ are all represented by a single operator from the set.  However, on the other hand, an internal $h$ is allowed to have trace and antisymmetric contributions.  Our full set of operators along with their resulting Feynman rules as well as the propagators for our fields are described in detail in Appendix~\ref{sec:Feynman Rules}.  

These Feynman rules are then used to construct the full set of $2\to2$ scattering diagrams which includes the 4-point diagram as well as the S-, T- and U-channel diagrams which contain an intermediate spin-2 $h$, spin-1 $v_j$ or spin-0 $s_k$, where the subscripts $j$ and $k$ enumerate the spin-1 and spin-0 bosons, respectively.  Each of the external spin-2 bosons $h$ contains 5 polarizations or helicities.  Since there are four external $h$'s, there are $5^4=625$ helicity combinations for this scattering amplitude.  However, as described in detail in Appendix~\ref{sec:equivalencies}, not all of these helicity combinations are linearly independent.  Many of them are related by spatial rotations as well as T and P invariance.  After taking these symmetries into account, only 97 helicity combinations are linearly independent.  We list these 97 linearly independent helicity amplitudes in Table~\ref{tab:97 amps}.  Moreover, we have explicitly calculated all 625 helicity amplitudes and have verified that our calculation satisfies these linear dependencies.

For tree-level unitarity to be valid, the energy growth in these helicity amplitudes must vanish for all large energies which are much greater than the masses of the particles and all scattering angles $\theta$ at those energies (see Appendix~\ref{app:no energy growth}).  At these high energies, the scattering amplitude can be series expanded in the ratio of the masses divided by the scattering energy, where these ratios are small parameters.  After this expansion, the scattering amplitude is a polynomial in energy and the cosine of the scattering angle $\theta$ (sometimes with overall factors that are functions of $\theta$ for each power of the energy.)  Since each of these polynomial terms are linearly independent, in order for the large energy growth to vanish for all energies and angles $\theta$, each coefficient must vanish simultaneously.  This gives a system of equations that must be satisfied in order to achieve tree-level unitarity.  For example, the highest potential energy growth is in the amplitude $\mathcal{M}_{0000}$, where the subscript gives the helicities of the external states, and is given by
\begin{eqnarray}
&&\mathcal{M}_{0000} = \frac{80}{27}\frac{E^{10}}{M_2^{12}}
\left(\cos^4\theta+2\cos^2\theta-1\right)\times\nonumber\\
&&\left[2\left(2g^{(1)}_{222}+3g^{(2)}_{222}\right)^2-3\sum_j\frac{M_2^4}{M_{1j}^2}\left(g^{(1)}_{221j}+2g^{(2)}_{221j}\right)^2\right]\nonumber\\
&&+\mathcal{O}\left(E^8\right)\ ,
\label{eq:non-gauge 1st eq}
\end{eqnarray}
where $M_2$ is the mass of the scattering $h$, $M_{1j}$ is the mass of $v_j$, $g_{222}^{(1)}$ and $g_{222}^{(2)}$ are coupling constants for $hhh$ vertices (see App.~\ref{app:h^3 interactions}), $g_{221j}^{(1)}$ and $g_{221j}^{(2)}$ are coupling constants for $hhv_j$ vertices (see App.~\ref{sec:h^2v_j interactions}), and we have taken the standard propagators (see App.~\ref{app:propagators}) for compactness and clarity in these expressions.  
Although there are three linearly independent powers of $\cos\theta$ in the $E^{10}$ term, their coefficients are linearly dependent, giving us our first equation (or sum rule) that must be satisfied for tree-level unitarity
\begin{equation}
2\left(2g^{(1)}_{222}+3g^{(2)}_{222}\right)^2=3\sum_j\frac{M_2^4}{M_{1j}^2}\left(g^{(1)}_{221j}+2g^{(2)}_{221j}\right)^2\ .
\label{eq:first condition}
\end{equation}

The next highest power of energy growth is an $E^8$ term which has $\cos^4\theta$, $\cos^2\theta$ and $\cos^0\theta$ terms.  Only one is linearly independent with the coefficient in square brackets in Eq.~\eqref{eq:non-gauge 1st eq}.  We choose the $\cos^0\theta$ term which gives the additional constraint
\begin{eqnarray}
&&54\left(2g^{(1)}_{2222}+g^{(2)}_{2222}\right)M_2^2 + \nonumber\\
&&2\left(2g^{(1)}_{222}+3g^{(2)}_{222}\right)\left(32g^{(1)}_{222}+129g^{(2)}_{222}\right) = \nonumber\\
&&3\sum_j\frac{M_2^4}{M_{1j}^2}\left(g^{(1)}_{221j}+2g^{(2)}_{221j}\right)\left(43g^{(1)}_{221j}+50g^{(2)}_{221j}\right)\ ,
\label{eq:second condition}
\end{eqnarray}
where $g_{2222}^{(1)}$ and $g_{2222}^{(2)}$ are coupling constants for $hhhh$ vertices (see App.~\ref{app:h^4 interactions}) and we have again used the standard propagators.
This procedure must be followed for the $E^6$ term, the $E^4$ term and the $E^2$ term, extracting all linearly independent constraints on the couplings and masses.  Furthermore, this procedure must be followed for all 97 linearly independent helicity amplitudes.  Although it might appear that there would be a very large number of constraints, we find that there are only 7 that are linearly independent.  We also find that all 7 linearly independent constraints can be obtained from the two amplitudes $\mathcal{M}_{0000}$ (see App.~\ref{app:M0000}) and $\mathcal{M}_{20\minus10}$ (see App.~\ref{app:M20-10}).  All other constraints are linearly dependent with these.  Slightly rearranged for convenience, the first three are
\begin{equation}
g^{(2)}_{2222} = g^{(2)}_{222} = g^{(1)}_{221j} = 0\ ,
\label{eq:sec1:g=0}
\end{equation}
while the final 4 equations are 
\begin{eqnarray}
2g^{(1)}_{2222}M_2^2 &=& \left(4C_{47}+8C_{89}+C_{10}\right)g^{(1)\ 2}_{222}\nonumber\\
&&+\left(1-F\right)\sum_jg^{(2)\ 2}_{221j}M_2^2\ ,\nonumber\\
C_{10}g^{(1)\ 2}_{222} &=& F\sum_j  g^{(2)\ 2}_{221j}\frac{M_2^4}{M_{1j}^2}\ ,\nonumber\\
\sum_k g_{220k}^2 \frac{M_{0k}^2}{M_2^2} &=& \left(F-1\right)\sum_j g^{(2)\ 2}_{221j}\frac{M_{1j}^4}{M_2^2}\nonumber\\
&&\hspace{-0.3in}-\left(8C_{12}+16C_3+4C_{47}+8C_{89}+C_{10}\right)g^{(1)\ 2}_{222}\ ,\nonumber\\
\sum_k g_{220k}^2 &=& \left(F-1\right)\sum_j g^{(2)\ 2}_{221j} M_{1j}^2\nonumber\\
&&\hspace{-0.3in}-\left(8C_{12}+16C_3+4C_{47}+8C_{89}+C_{10}\right)g^{(1)\ 2}_{222}\ ,\nonumber\\
\label{eq:sec1:sum rules}
\end{eqnarray}
where $M_{0k}$ is the mass of $s_k$, $g_{220k}$ is the coupling constant for the $hhs_k$ vertex (see App.~\ref{app:hhsk interaction}), $F$ is a spin-1 propagator numerator coefficient and $C_{12}, C_{3}, C_{47}, C_{89}$ and $C_{10}$ are spin-2 propagator numerator coefficients (see App.~\ref{app:propagators}).
We have explicitly checked that inserting these sum rules into all 625 tree-level helicity amplitudes removes all high energy growth.  We have also checked that the helicity amplitudes do not identically vanish when using this solution and arbitrary propagator coefficients.  For example, after inserting these relations and using the standard propagator for spin-2 but setting $F=2$, we have
\begin{eqnarray}
\mathcal{M}_{0000} &=& \frac{4}{9}\left(\sum_j g^{(2)\ 2}_{221j}\frac{M_{1j}^6}{M_2^6}-\sum_k g_{220k}^2\frac{M_{0k}^4}{M_2^6}\right)\nonumber\\
&&+\frac{1}{3}\left(\sum_j g^{(2)\ 2}_{221j}\frac{M_{1j}^8}{M_2^8}-\sum_k g_{220k}^2\frac{M_{0k}^6}{M_2^8}\right)\nonumber\\
&&+\mathcal{O}\left(E^{-2}\right)\ .
\label{eq:nonzero E0 term}
\end{eqnarray}

On the other hand, if we use the standard propagators for spin-2 and spin-1, their numerator coefficients satisfy
\begin{eqnarray}
F-1 &=& 0\ ,\nonumber\\
8C_{12}+16C_3+4C_{47}+8C_{89}+C_{10} &=& 0\ .
\end{eqnarray}
Consequently, our sum rules reduce further to
\begin{equation}
g_{220k} = 0
\label{eq:g220k=0}
\end{equation}
and
\begin{equation}
3g^{(1)}_{2222}M_2^2 = 
2g^{(1)\ 2}_{222} =
3\sum_j g^{(2)\ 2}_{221j}\frac{M_2^4}{M_{1j}^2}\ ,
\label{eq:secI simplified constraints}
\end{equation}
which clearly satisfies Eqs.~\eqref{eq:first condition} and \eqref{eq:second condition}.  The interaction Lagrangian for this theory is nontrivial
\begin{eqnarray}
\mathcal{L}_{\mbox{int}} &=& 
g^{(1)}_{2222} h_{\mu\nu}(x)h^{\mu\nu}(x)h_{\alpha\beta}(x)h^{\alpha\beta}(x)\nonumber\\
&&+g^{(1)}_{222} h_{\mu\nu}(x)h^{\mu\nu}(x)h_{\alpha}^{\ \ \alpha}(x)\nonumber\\
&&+i \sum_jg_{221j}^{(2)} h^{\mu\nu}(x)h_{\mu\nu}(x)\partial_\alpha v_{j}^{\alpha}(x)\ .
\label{eq:secI:Lint}
\end{eqnarray}
However, interestingly, with this solution and using the standard propagators, all tree-level $h_{\mu_1\nu_1}(p_1)h_{\mu_2\nu_2}(p_2)\to h_{\mu_3\nu_3}(p_3)h_{\mu_4\nu_4}(p_4)$ amplitudes identically vanish independent of the values of the external momenta, Lorentz indices and helicities.  Nevertheless, it should be noted that the individual diagrams do not vanish.  Only the combination of diagrams vanishes after relating the couplings and masses in this way.  Other scattering processes should still be nonzero.

Because the sum rules for the standard propagator case are so simple, they can be achieved in a theory with only one spin-1 boson $v$, where we have dropped the subscript $j$.  In this case, the Lagrangian is given by
\begin{eqnarray}
\mathcal{L}_{\mbox{int}} &=& 
g^2 h_{\mu\nu}(x)h^{\mu\nu}(x)h_{\alpha\beta}(x)h^{\alpha\beta}(x)\nonumber\\
&&+\sqrt{\frac{3}{2}}g M_2 h_{\mu\nu}(x)h^{\mu\nu}(x)h_{\alpha}^{\ \ \alpha}(x)\nonumber\\
&&+i g\frac{M_1}{M_2} h^{\mu\nu}(x)h_{\mu\nu}(x)\partial_\alpha v^{\alpha}(x)\ ,
\label{eq:secI:Lint: one v}
\end{eqnarray}
where $g$ is some small coupling that defines this theory and $M_1/M_2$ is not too large.

Considering again the general sum rules in Eqs.~\eqref{eq:sec1:g=0} and \eqref{eq:sec1:sum rules}, it remains to show that the tree-level $2\to2$ scattering amplitudes with other particles in the external states do not grow at high energy.  We first consider amplitudes with one or more external spin-1 $v_j$'s.  We note that our only operator for the spin-1 $v_j$ fields (see Eqs.~\eqref{eq:op:g_221j} and \eqref{eq:sec1:g=0}) contains $\partial_\mu v^\mu_j(x)$ which identically vanishes on-shell.  That is, the vertex with the external spin-1 $v_j$ will contribute $p_\mu \epsilon^\mu_\sigma(p)$, where $\epsilon$ is the polarization vector for the external $v_j$ (see App.~\ref{app:polarization vectors}), and this will vanish because the polarization vector is orthogonal to the momentum of the external particle.  Therefore, all tree-level $2\to2$ amplitudes with external $v_j$'s trivially satisfy unitarity.  

We next consider amplitudes with external spin-0 $s_k$ bosons.  The operator in Eq.~\eqref{eq:g_220k} will result in amplitudes with one or two $s_k$'s in the external states along with $h$'s.  We have checked these amplitudes and find that the conditions in Eqs.~\eqref{eq:sec1:g=0} and \eqref{eq:sec1:sum rules} are not sufficient to nontrivially remove the high energy growth in these amplitudes.  However, we are free to add other three- and four-point operators that could contribute to these processes but not to $hh\to hh$.  We expect that it would be possible to choose the couplings of these new operators such that tree-level unitarity would also be satisfied in amplitudes with external $s_k$'s.  However, since our goal was to achieve a non-trivial theory with massive spin-2 bosons that is tree-level unitary, we note that we can accomplish this with only the spin-1 bosons by taking the further condition
\begin{equation}
g_{220k}=0\ ,
\end{equation}
along with the reduced set of constraints
\begin{eqnarray}
2g^{(1)}_{2222}M_2^2 &=& \left(4C_{47}+8C_{89}+C_{10}\right)g^{(1)\ 2}_{222}\nonumber\\
&&+\left(1-F\right)\sum_jg^{(2)\ 2}_{221j}M_2^2\ ,\nonumber\\
C_{10}g^{(1)\ 2}_{222} &=& F\sum_j  g^{(2)\ 2}_{221j}\frac{M_2^4}{M_{1j}^2}\ ,\nonumber\\
\left(F-1\right)\sum_j g^{(2)\ 2}_{221j}\frac{M_{1j}^4}{M_2^2} &=& \left(F-1\right)\sum_j g^{(2)\ 2}_{221j} M_{1j}^2 = \nonumber\\
&&\hspace{-0.7in}\left(8C_{12}+16C_3+4C_{47}+8C_{89}+C_{10}\right)g^{(1)\ 2}_{222}\ .\nonumber\\
\end{eqnarray}
The theory that satisfies these constraints removes the polynomial high energy growth in all tree-level $2\to2$ scattering amplitudes.  Its interaction Lagrangian is the same as in Eq.~\eqref{eq:secI:Lint}.

Although we have achieved tree-level unitarity in this theory, we should comment that we do not expect our sum rules to be sufficient to unitarize the theory at higher loop-level.  In fact, at higher perturbative order, new diagrams contribute which alter the set of sum rules outlined here, likely modifying these simple relations.  On the other hand, the new diagrams also include contributions from new Lagrangian operators that were not used in our analysis, either because their contribution vanishes at tree-level or because their contribution is identical to that given by our operators at tree-level.  However, at loop level, these new operators contribute unique terms in the sum rules and potentially help unitarize the process at higher order.  Therefore, at loop level, there are new parameters that can, in principle, be adjusted in order to achieve unitarity.  Whether this is actually possible in practice is not clear to us.  However, it seems likely that any theory that is perturbatively unitary at the loop-level would require gauge invariance for the spin-1 sector.  For this reason, we now move on to gauge invariant theories.

\section{\label{sec:multiple spin-2}Trivially Gauge Invariant Theory}
In this section, we attempt to unitarize the spin-2 scattering in the absence of a spin-1 interaction.  We allow multiple spin-2 bosons and spin-0 bosons.  Since we do not have any interactions with a spin-1 boson, it is trivially gauge invariant.  As in the previous section, we consider the most general dimension-3 and -4 operators consistent with Lorentz invariance that make unique nonvanishing contributions to the tree-level process $h_{\mu_1\nu_1}(p_1)h_{\mu_2\nu_2}(p_2)\to h_{\mu_3\nu_3}(p_3)h_{\mu_4\nu_4}(p_4)$.   Our full set of operators along with their resulting Feynman rules are described in detail in Appendix~\ref{sec:Feynman Rules}.  We then use these Feynman rules to construct the full set of $2\to2$ scattering diagrams which includes the 4-point diagram as well as the S-, T- and U-channel diagrams which contain an intermediate spin-2 $h$ or $h_i$ (where $h_i$ is another spin-2 boson different than the one that is scattering) or a spin-0 $s_k$.  As in the previous section, we have explicitly checked that our calculation of the 625 helicity amplitudes satisfy the equivalencies outlined in Appendix~\ref{sec:equivalencies} with only 97 being linearly dependent.  

We series expand each helicity amplitude in large energy and demand that the coefficient of each positive power of energy and each unique term in $\cos\theta$ must vanish.  This gives us a set of sum rules that must be satisfied in order for unitarity to be valid at tree-level in this theory.  As in the previous section, we find that the amplitudes $\mathcal{M}_{0000}$ and $\mathcal{M}_{20\minus10}$ are sufficient to give us our result.  Among the sum rules that we find are
\begin{eqnarray}
\frac{g^{(1)\ 2}_{222}}{M_2^4} + \sum_i\frac{g^{(1)\ 2}_{222i}}{M_{2i}^4} &=& 0\ ,\nonumber\\
\frac{g^{(2)\ 2}_{222}}{M_2^4} + \frac{1}{9}\sum_i\frac{g^{(2)\ 2}_{222i}}{M_{2i}^4} &=& 0\ ,
\label{eq:trivial gauge invariance sum positives}
\end{eqnarray}
which can be obtained purely from $\mathcal{M}_{20\minus10}$.
Since this is the sum of positive numbers, the only way to satisfy these sum rules is to take
\begin{equation}
g^{(1)}_{222} = g^{(1)}_{222i} = g^{(2)}_{222} = g^{(2)}_{222i} = 0\ .
\end{equation}
This is in contrast to the previous section where we included spin-1 bosons and they contributed to these sum rules with the opposite sign (see, for example, the second equation of Eq.~\eqref{eq:sec1:sum rules}).
However, in the present absence of spin-1 fields, once the $hhh$ and $hhh_i$ vertices vanish, we have only the $hhhh$ vertices and the $hhs_k$ vertices and we find that they are not sufficient to unitarize this process nontrivially.  The only solution in this case is the trivial solution where all coupling constants are set to zero.
Therefore, we find that it is impossible to perturbatively unitarize massive spin-2 scattering with only dimension-3 and -4 operators in the absence of interactions with spin-1 bosons.  This is in interesting contrast to spin-1 scattering which can be unitizarized with only spin-1 bosons \cite{SekharChivukula:2001hz} and spin-$\frac{1}{2}$ fermions scattering to spin-1 bosons which can be unitarized with only spin-$\frac{1}{2}$ fermions and spin-1 bosons \cite{Sekhar Chivukula:2007mw}, both with no need for lower spin fields.

\section{\label{sec:adjoint rep}Single Gauge Theory}
In this section, we reintroduce the spin-1 field as a gauge boson and demand that the spin-2 boson interacts with it in a gauge invariant way.  Since we consider only real spin-2 bosons in this paper, we use the simplest real representation that is present in all compact Lie groups, namely the adjoint representation.  We consider the most general Lorentz and gauge invariant dimension-3 and -4 operators (see Appendix~\ref{sec:Feynman Rules}) that contribute to the process $h_{a_1\mu_1\nu_1}(p_1)h_{a_2\mu_2\nu_2}(p_2)\to h_{a_3\mu_3\nu_3}(p_3)h_{a_4\mu_4\nu_4}(p_4)$, where the subscript $a_i$ is the adjoint gauge index.  We then use these Feynman rules to construct the full set of $2\to2$ tree-level scattering diagrams which includes the 4-point diagram as well as the S-, T- and U-channel diagrams which contain an intermediate spin-2 $h$ or $h_i$, an intermediate spin-1 gauge boson $v$ or a spin-0 $s_k$.  We find that our helicity amplitudes do not depend on $F$ in the spin-1 propagator.  This is as we expect since our interactions are gauge invariant.  As in the previous two sections, we explicitly check that the 625 helicity combinations satisfy the equivalencies outlined in Appendix~\ref{sec:equivalencies} with only 97 being linearly independent.  

We also find that, in this case, the contribution from an intermediate spin-2 boson vanishes identically.  The reason for this is that the only operator with a product of three fields that each transform under the adjoint representation of the gauge group is totally antisymmetric in the gauge index.  That is, the gauge index is contracted by the structure constant which is totally antisymmetric.  The two possible terms we could have are
\begin{equation}
f_{abc} h_{a\mu\nu}(x)h_b^{\mu\nu}(x)h_{c\alpha}^{\ \ \alpha}(x)
\ ,\
 f_{abc} h_{a\mu\nu}(x)h_b^{\mu\alpha}(x)h_{c\alpha}^{\ \ \ \nu}(x)\ ,
 \label{eq:single gauge structure constant 1}
\end{equation}
for an intermediate $h$ and 
\begin{equation}
f_{abc} h_{a\mu\nu}(x)h_b^{\mu\nu}(x)h_{ic\alpha}^{\ \ \alpha}(x)
\ ,\
 f_{abc} h_{a\mu\nu}(x)h_b^{\mu\alpha}(x)h_{ic\alpha}^{\ \ \ \nu}(x)\ ,
  \label{eq:single gauge structure constant 2}
\end{equation}
for an intermediate $h_i$.
The first operator in each pair is identically zero due to the antisymmetry of the structure constant on the first two fields.  The second term is nonzero but gives a Feynman rule that is antisymmetric in the Lorentz indices of one of the three fields (see Eqs.~\eqref{eq:h^3:adjoint} and \eqref{eq:h^2hi:adjoint}).  However, when it is used in a Feynman diagram, since the external polarization vectors are symmetric in their Lorentz indices and since the spin-2 propagator is symmetric in each pair of its Lorentz indices (see Table~\ref{tab:fields}), the contribution from this vertex also vanishes.  This means that intermediate spin-2 bosons cannot help unitarize the spin-2 scattering amplitude when the spin-2 bosons transform under the adjoint representation of one gauge group.

After series expanding the remaining contributions to $\mathcal{M}_{0000}$ in large energy and setting the coefficients of each positive power of energy and $\cos\theta$ to zero, we get the following set of constraints that must be satisfied for tree-level unitarity
\begin{eqnarray}
\bar{g}_{2222}^{(A2)} &=& \frac{1}{4}\bar{g}^2\sum_j \bar{g}_{221j}^{(2)\ 2}\ ,\nonumber\\
\bar{g}_{2222}^{(B1)} &=& 0\ ,\nonumber\\
\sum_j \bar{g}_{221j}^{(1)\ 2}M_{1j}^2 &=& \sum_j \bar{g}_{221j}^{(2)\ 2}M_{1j}^2 = 0\ ,\nonumber\\
\sum_k \bar{g}_{220k}^2 &=& 0\ .
\label{eq:single adjoint sum rules}
\end{eqnarray}
The last two constraints set the sum of positive terms equal to zero.  Therefore, the only solution to these equations is the trivial one where all contributing couplings are zero
\begin{equation}
\bar{g}_{2222}^{(A2)} = \bar{g}_{2222}^{(B1)} = \bar{g}_{221j}^{(1)} = \bar{g}_{221j}^{(2)} = \bar{g}_{220k} = 0\ .
\end{equation}
On the other hand, the couplings $\bar{g}_{2222}^{(B2)}$, $\bar{g}_{222}^{(2)}$ and $\bar{g}_{222i}^{(2)}$ could be nonzero and still allow tree-level unitarity, since they do not contribute to $2\to2$ scattering at tree-level.  However, this case is very different than that of Sec.~\ref{sec:non gauge-invariant} with the standard propagators where a nontrivial cancellation happened between the diagrams.  In this section, each diagram vanishes independently and the $2\to2$ scattering is trivial.

\section{\label{sec:Symmetric cubic term}Double Gauge Theory}
In the last section, where the spin-2 bosons transformed under the adjoint representation of one gauge group, we found that diagrams with spin-2 intermediate states identically vanished and could not contribute to the unitarization of the tree-level scattering.  We found that this was due to the antisymmetry of the structure constant in the $hhh$ and $hhh_i$ Lagrangian operators, which produced Feynman rules that were antisymmetric in the Lorentz indices of at least one $h$ or $h_i$.  Combined with the symmetry of the Lorentz indices of the external polarization vector and the spin-2 propagator, these diagrams vanished individually.  In the present section, on the other hand, we would like to consider a theory whose interactions are gauge invariant and which also have non-vanishing contributions from intermediate spin-2 states.  We do this by constructing a theory with two gauge groups where the spin-2 bosons transform under the adjoint representation of both groups.  As a result, the operators with three spin-2 fields will have two structure constants as in
\begin{eqnarray}
&&f_{\bar{a}\bar{b}\bar{c}}f_{abc} h_{\bar{a}a\mu\nu}(x)h_{\bar{b}b}^{\mu\nu}(x)h_{\bar{c}c\alpha}^{\ \ \alpha}(x)\ ,\nonumber\\
&&f_{\bar{a}\bar{b}\bar{c}} f_{abc} h_{\bar{a}a\mu\nu}(x)h_{\bar{b}b}^{\mu\alpha}(x)h_{\bar{c}c\alpha}^{\ \ \ \nu}(x)\ ,
\label{eq:double adjoint operator 1}
\end{eqnarray}
for an intermediate $h$ and 
\begin{eqnarray}
&&f_{\bar{a}\bar{b}\bar{c}}f_{abc} h_{\bar{a}a\mu\nu}(x)h_{\bar{b}b}^{\mu\nu}(x)h_{i\bar{c}c\alpha}^{\ \ \alpha}(x)\ ,\nonumber\\
&&f_{\bar{a}\bar{b}\bar{c}} f_{abc} h_{\bar{a}a\mu\nu}(x)h_{\bar{b}b}^{\mu\alpha}(x)h_{i\bar{c}c\alpha}^{\ \ \ \nu}(x)\ ,
\label{eq:double adjoint operator 2}
\end{eqnarray}
for an intermediate $h_i$.  All of these operators are now nonzero and can contribute nontrivially to the amplitude.

As in previous sections, we consider the most general Lorentz and gauge invariant operators (see Appendix~\ref{sec:Feynman Rules}) that contribute to the process $h_{\bar{a}_1a_1\mu_1\nu_1}(p_1)h_{\bar{a}_2a_2\mu_2\nu_2}(p_2)\to h_{\bar{a}_3a_3\mu_3\nu_3}(p_3)h_{\bar{a}_4a_4\mu_4\nu_4}(p_4)$, where the subscripts $\bar{a}_i$ and $a_i$ are the adjoint gauge indices of the two gauge groups.  We then use these Feynman rules to construct the full set of $2\to2$ scattering diagrams which include the 4-point diagram as well as the S-, T- and U-channel diagrams which contain an intermediate spin-2 $h$ or $h_i$, an intermediate spin-1 gauge boson $v_1$ or $v_2$, or a spin-0 boson $s_k$.  We find that, as in the previous section, our results do not depend on $F$ in the spin-1 propagator, as we expect since our interactions are gauge invariant.  As in the previous three sections, we explicitly check that all 625 helicity combinations satisfy the equivalencies outlined in Appendix~\ref{sec:equivalencies} with only 97 being linearly independent.  

We again, as in previous sections, series expand the amplitudes in large energy and set the coefficient of each positive power of energy and each power of $\cos\theta$ to zero in order to ensure unitarity at high energy.  This gives us a set of constraints on the couplings and masses that must be satisfied.  	Among the constraints we get from the $\mathcal{M}_{0000}$ amplitude are
\begin{eqnarray}
\tilde{g}_{222}^{(1)\ 2} + \sum_i \tilde{g}_{222i}^{(1)\ 2}\frac{M_2^4}{M_{2i}^4} &=& 0\ ,\nonumber\\
9\tilde{g}_{222}^{(2)\ 2} + \sum_i \tilde{g}_{222i}^{(2)\ 2}\frac{M_2^4}{M_{2i}^4} &=& 0\ ,\nonumber\\
\tilde{g}_{2211}^{(1)\ 2} + \tilde{g}_{2212}^{(1)\ 2} &=& 0\ ,\nonumber\\
\tilde{g}_{2211}^{(2)\ 2} + \tilde{g}_{2212}^{(2)\ 2} &=& 0\ ,\nonumber\\
\sum_k \tilde{g}_{220k}^{(1)\ 2} &=& 0\ .
\label{eq:double gauge sum rules}
\end{eqnarray}
Each of these is the sum of positive terms which, therefore, requires that each of these coupling is zero.  Consequently, we see that, although we constructed a theory where the spin-2 bosons could, in principle, contribute as intermediate states to the $2\to2$ scattering, in practice they do not help.  All the $hhh$ and $hhh_i$ couplings must vanish in order to achieve perturbative unitarity.   Additionally, we also find that the couplings to gauge bosons and scalar bosons must vanish.  With these results, we find that the only couplings allowed to be nonzero must satisfy the following constraints
\begin{eqnarray}
2\tilde{g}_{2222}^{(A1)} - \tilde{g}_{2222}^{(B1)} - \tilde{g}_{2222}^{(C1)} &=& 0\nonumber\\
\tilde{g}_{2222}^{(B2)} + \tilde{g}_{2222}^{(C2)} - 2\tilde{g}_{2222}^{(D2)} &=& 0
\end{eqnarray}
These couplings are all $hhhh$ couplings.  There are no longer any S-, T- or U-channel diagrams.  The 4-point diagrams cancel among themselves and the amplitude is identically zero.  This case is very different than that in Section~\ref{sec:non gauge-invariant} where there was a nontrivial cancellation between 4-point and S-, T- and U-channel diagrams.  In the present case, each of these is zero individually and the theory is trivial.  A theory with gauge invariant operators and nontrivial contributions from intermediate spin-2 states, as formulated here, is not sufficient to achieve perturbative unitarity.

\section{\label{sec:conclusion}Conclusion}
We have analyzed the tree-level unitarity of $2\to2$ scattering in a set of illustrative quantum field theories which contain a massive spin-2 boson.  We have focused on theories with only dimension-3 and -4 operators and only spins 0, 1 and 2.  For convenience, we have taken all our fields to be real.  

In Section~\ref{sec:non gauge-invariant}, we considered a theory with a single spin-2 boson and as many spin-0 and -1 bosons as required to achieve unitarity, however, we did not force our interactions to be gauge invariant at this point, but allowed the operators to take on whatever Lorentz invariant form was required to achieve unitarity.  We found that tree-level unitarity was possible in this case as long as the coupling constants and masses satisfied a set of sum rules, which were presented in Eqs.~\eqref{eq:sec1:g=0} and \eqref{eq:sec1:sum rules}.  We further showed that, for general propagators, this set of constraints did not cause the amplitude to identically vanish for all energies and angles and demonstrated this by presenting the leading term in the amplitude for a particular choice of propagators in Eq.~\eqref{eq:nonzero E0 term}.  On the other hand, if the standard propagators were used, we showed that the constraints simplified further to those given in Eqs.~\eqref{eq:g220k=0} and \eqref{eq:secI simplified constraints} with the nontrivial Lagrangian given in Eq.~\eqref{eq:secI:Lint}.  With this theory and using the standard propagators, we noted that the amplitude vanished identically for all energies and angles.  However, we pointed out that this was due to a nontrivial cancellation between the various diagrams and that other scattering processes should still be nonzero.  This was followed by noting that a theory with one spin-2 boson and one spin-1 boson with the Lagrangian given in Eq.~\eqref{eq:secI:Lint: one v} satisfied tree-level unitarity in all $2\to2$ scattering processes.  Following this, we commented on the fact that the sum rules would likely be modified at the loop-level and require the further adjustment of new Lagrangian operators that were not required for the tree-level analysis.  Although it is possible that perturbative unitarity is achievable even at the loop-level, we consider it likely that perturbative unitarity at higher loops requires gauge invariance of the spin-1 sector.

In Section~\ref{sec:multiple spin-2}, we removed the spin-1 boson and concentrated on a theory with only spin-2 bosons and spin-0 bosons, but as many of each of these as required to achieve unitarity.  We noted that this theory is trivially gauge invariant.  In this case, we found that, among the sum rules for this theory, there are two, shown in Eq.~\eqref{eq:trivial gauge invariance sum positives}, that equate the sum of positive terms to zero which can only be satisfied if all the coupling constants for three spin-2 bosons vanish.  We then, further, found that the remaining interactions are not sufficient to unitarize the scattering unless all couplings are zero and the theory becomes trivial.  We noted that this was in contrast to what occurs in Higgsless theories where it has been shown that all the high energy growth of massive spin-1 scattering can be cancelled purely by other massive spin-1 bosons \cite{SekharChivukula:2001hz} and all the high energy growth of massive spin-$\frac{1}{2}$ scattering to massive spin-1 bosons can be cancelled purely by other massive spin-$\frac{1}{2}$ fermions and spin-1 bosons \cite{Sekhar Chivukula:2007mw}, both of which make no use of lower spin fields.

In Section~\ref{sec:adjoint rep}, we reintroduced a single spin-1 field as a gauge boson and required that all interactions were gauge invariant under this group.  We took the spin-2 boson to transform under the simplest real representation which is present in all gauge groups, the adjoint representation.  We discussed the fact that the Lagrangian operators with three spin-2 bosons must be constructed with the structure constant of the gauge group, as shown in Eqs.~\eqref{eq:single gauge structure constant 1} and \eqref{eq:single gauge structure constant 2}.  This lead to the vanishing of one of the Lagrangian operators.  The other Lagrangian operator did not vanish, but produced a Feynman rule that was antisymmetric in the Lorentz indices of one of its spin-2 fields.  When combined with the symmetry of the Lorentz indices in the polarization vectors and the propagator, this operator gave a vanishing contribution to the amplitude.  We then found the sum rules for the remaining couplings, presented in Eq.~\eqref{eq:single adjoint sum rules}, and found that only the trivial solution with all other couplings set to zero was allowed.  

In Section~\ref{sec:Symmetric cubic term}, we considered again a gauge invariant theory.  However, this time we constructed it in such a way that diagrams with intermediate spin-2 bosons did not identically vanish and therefore could, in principle, contribute to the unitarization of the scattering.  We did this by including two gauge groups and having the spin-2 bosons transform under the adjoint representation of both gauge groups.  With this definition, we found that the Lagrangian operators, presented in Eqs.~\eqref{eq:double adjoint operator 1} and \eqref{eq:double adjoint operator 2}, with three spin-2 bosons did not vanish and generated Feynman rules which were symmetric in the Lorentz indices on each field.  Nevertheless, we found that the sum rules for this theory, given in Eq.~\eqref{eq:double gauge sum rules}, required all these couplings to be zero, and the spin-2 bosons were still not able to help unitarize the scattering by contributing in the intermediate states.  As in the previous section, the sum rules only allowed the trivial solution where the amplitude identically vanished and no nontrivial cancellation between diagrams occurred.  

From our result, it is still not possible to determine whether all gauge invariant theories with real massive spin-2 bosons, only spins 2 and lower, and only relevant and marginal operators can be made perturbatively unitary.  For one thing, we did not consider arbitrary real representations under an arbitrary number of gauge groups.  We also did not consider complex spin-2 bosons transforming under complex representations of the gauge groups.  It is conceivable that there are special representations and special numbers of gauge groups, under which perturbative unitarity is achievable.  It seems to us that, it might be possible to construct a general argument either discovering a gauge invariant theory that is unitary or a general proof, perhaps by induction, that such a theory is impossible.  Additionally, we note that string theory predicts an infinite set of spins including all spins higher than 2 as well as those which are lower \cite{Sagnotti:2011qp}.  It is conceivable that the perturbative unitarity of scattering of particles with spin greater than 1 requires these fields to interact with an infinite set of fields with an infinite number of spins \cite{Bekaert:2009ud}.  Finally, it is also possible that nontrivial higher spin fields cannot be made perturbatively unitary at very high energies in quantum field theory and that this can only be accomplished in a more fundamental theory, such as string theory.  We hope this is not the case.

\appendix


\section{\label{app:no energy growth}Unitarity Bound}
In this appendix, we review how the unitarity of the $\mathcal{S}$-matrix implies that the scattering amplitude must remain below a constant and, therefore, can not grow indefinitely with energy.  Thus, at very high energy, all polynomial energy growth must vanish.  In App.~\ref{app:heuristic}, we give a heuristic proof based on finite dimensional spaces while in App.~\ref{app:optical}, we give a more complete proof in the case of an infinite dimensional Hilbert space, but only for $2\to2$ scattering.  
A bound on the amplitude was originally found by Froissart \cite{Froissart,Froissart:2010}.

\subsection{\label{app:heuristic}Heuristic Finite Dimensional Proof}
In this subsection, we give a heuristic review of the unitarity bound on scattering amplitudes that will only be strictly valid for finite dimensional spaces.  In the next subsection, we will consider infinite dimensional Hilbert space.  The $\mathcal{S}$-operator is defined to give the $\mathcal{S}$-matrix when acting on free-particle states
\begin{equation}
\mathcal{S}_{\beta\alpha} = \langle\beta|\mathcal{S}|\alpha\rangle\ ,
\end{equation}
where $\alpha$ and $\beta$ are a complete specification of the incoming and outgoing free-particle states.  
Since the probability of this transition is equal to the square of the $\mathcal{S}$-matrix and the sum of probabilities over the complete set of outgoing free-particle states must add to one
\begin{equation}
\sum_\beta P_{\alpha\beta} = \sum_\beta | \langle\beta|\mathcal{S}|\alpha\rangle |^2 = 1\ ,
\end{equation}
we have the relation
\begin{equation}
1 = \sum_\beta \langle\alpha|\mathcal{S}^\dagger|\beta\rangle \langle\beta|\mathcal{S}|\alpha\rangle 
= \langle\alpha|\mathcal{S}^\dagger\mathcal{S}|\alpha\rangle
\end{equation}
giving us the unitarity of the $\mathcal{S}$-operator.
We can form the scattering part of this operator by removing the identity as in
\begin{equation}
\mathcal{S}=1+i \mathcal{T}\ .
\end{equation}
Solving for $\mathcal{T}$, we have
\begin{equation}
\mathcal{T}=\frac{1}{i}\left(\mathcal{S}-1\right)\ .
\end{equation}
Since $\mathcal{S}$ is unitary, it can be written as $\exp(i\mathcal{H})$ where $\mathcal{H}$ is Hermitian and diagonalizable.  As a result, we can rewrite $\mathcal{T}$ as
\begin{eqnarray}
\mathcal{T}&=&\frac{1}{i}\left(e^{i\mathcal{H}_D}-1\right) \\
&=& e^{i\mathcal{H}_D/2}\frac{1}{i}\left(e^{i\mathcal{H}_D/2}-e^{-i\mathcal{H}_D/2}\right)\\
&=& 2e^{i\mathcal{H}_D/2}\sin\left(\mathcal{H}_D/2\right)
\end{eqnarray}
in the diagonal basis, where the exponential and the sin of an operator are defined in terms of their series expansion.
Inserting this operator between free particle states gives the amplitude
\begin{equation}
\mathcal{M}(\alpha\to\beta)=\langle\beta|\mathcal{T}|\alpha\rangle = 2\langle\beta|e^{i\mathcal{H}_D/2}\sin\left(\mathcal{H}_D/2\right)|\alpha\rangle\ .
\end{equation}
If we expand the initial and final free-particle states in terms of the eigenstates $\mathcal{H}_D|j\rangle=\Delta_{j}|j\rangle$,
\begin{equation}
|\alpha\rangle = \sum_j c_j|j\rangle \quad \mbox{and} \quad |\beta\rangle = \sum_j d_j|j\rangle\ ,
\end{equation}
where $\sum_j |c_j|^2=\sum_j |d_j|^2=1$, we have
\begin{equation}
\mathcal{M}(\alpha\to\beta) = 2\sum_j c_j d_j^* e^{i\Delta_j/2}\sin\left(\Delta_j/2\right)\ .
\end{equation}
The absolute value of this satisfies the triangle inequality
\begin{equation}
|\mathcal{M}(\alpha\to\beta)|\leq2\sum_j |c_j| |d_j| \ ,
\end{equation}
where we have dropped the phase $\exp(i\Delta_j/2)$ since it does not affect the absolute value and $|\sin(\Delta_j/2)|$ since it only strengthens this inequality.
This is the inner product of two vectors, so by the Cauchy-Schwartz inequality,
\begin{equation}
|\mathcal{M}(\alpha\to\beta)|\leq2\sqrt{\sum_j |c_j|^2} \sqrt{\sum_k |d_k|^2} \leq 2\ .
\end{equation}
Consequently, we see that in a finite dimensional space, the amplitude must always be below a constant.  This means that the amplitude can not grow indefinitely with energy, or in other words, any energy growth must be cancelled at very high energies.
We note that the same procedure could be followed for any finite-dimensional unitary operator with the same result.

\subsection{\label{app:optical}Partial Wave Unitarity Bound}
In this subsection, we review the partial wave unitarity bound on $2\to2$ scattering amplitudes.  
For simplicity, we assume spinless particles.
Because the $\mathcal{S}$-operator is unitary, we have
\begin{equation}
1 = \mathcal{S}^\dagger\mathcal{S} = \left(1-i\mathcal{T}^\dagger\right)\left(1+i\mathcal{T}\right)
\end{equation}
in terms of the scattering operator $\mathcal{T}$ which gives us
\begin{equation}
i\left(\mathcal{T}^\dagger-\mathcal{T}\right) = \mathcal{T}^\dagger\mathcal{T}\ .
\end{equation}
We can insert this between the states $|A_i\rangle$
 and $|A_f\rangle$
followed by the insertion of a complete set of states $|X\rangle$ on the right to obtain
\begin{equation}
i\left(\langle A_i|\mathcal{T}|A_f\rangle^*-\langle A_f|\mathcal{T}|A_i\rangle\right) 
= \sum_{X}\int_X \langle A_f | \mathcal{T}^\dagger |X \rangle \langle X| \mathcal{T} | A_i \rangle
\end{equation}
where the discrete sum $\sum_X$ is over the number of particles in state $|X\rangle$ as well as over all the discrete quantum numbers of the particles and the integral $\int_X$ is over the phase space of the particles in $|X\rangle$ such that $\sum_X\int_X |X\rangle\langle X|=1$.
The overall momentum conserving delta function can be removed from this expression by rewriting it in terms of the scattering amplitude
\begin{equation}
\langle \Psi_2|\mathcal{T}|\Psi_1\rangle = \left(2\pi\right)^4\delta^4\left(p_{\Psi_1}-p_{\Psi_2}\right)\mathcal{M}\left(\Psi_1\to \Psi_2\right)
\end{equation}
giving us
\begin{eqnarray}
i\left(\mathcal{M}\left(A_f\to A_i\right)^*-\mathcal{M}\left(A_i\to A_f\right)\right) &=&\nonumber\\
&& \hspace{-2.2in}\sum_X\int_X\left(2\pi\right)^4\delta^4\left(p_i-p_X\right)
\mathcal{M}\left(A_i\to X\right)\mathcal{M}(A_f \to X)^*\nonumber\ .\\
\end{eqnarray}
We next take $A_f=A_i$, in which case the final state is exactly the same as the initial state (forward elastic scattering), and find
\begin{eqnarray}
2\mbox{Im}\mathcal{M}\left(A_i\to A_i\right) &=&\nonumber\\
&&\hspace{-0.5in} \sum_X\int_X\left(2\pi\right)^4\delta^4\left(p_i-p_X\right)
|\mathcal{M}\left(A_i\to X\right)|^2\ .\nonumber\\
\end{eqnarray}
At this point, we could relate the right-hand side to the cross section to obtain the optical theorem.  
Instead, we will now restrict $A_i$ to a two-particle state (we will call it $A_2$) in its center of momentum (CM) frame 
and note that since the right-hand side is the sum of positive terms, one of these terms must be less than the total sum to obtain
\begin{equation}\label{eq:unit:beforePartial}
\mbox{Im}\mathcal{M}\left(A_2\to A_2\right) \geq \frac{1}{32\pi^2}\int d\Omega |\mathcal{M}\left(A_2\to X_2\right)|^2
\end{equation}
where $\int d\Omega$ is the integral over the solid angle ($d\Omega=\sin\theta d\theta d\phi$) and $X_2$ is a two-body final state.
We next expand each amplitude in partial waves as in
\begin{eqnarray}
\mathcal{M}\left(A_2\to X_2\right) &=& 8\pi\sum_n\left(2n+1\right)P_n\left(\cos\theta\right)b_{n}\nonumber\\
\mathcal{M}\left(A_2\to A_2\right) &=& 8\pi\sum_n\left(2n+1\right) a_{n}
\label{eq:amplitude decomposition}
\end{eqnarray}
where the $P_n(x)$ are Legendre polynomials and $\theta=0$ for $A_2\to A_2$.  
After using the orthonormality of the Legendre polynomials ($\int_{-1}^{-1} dx P_l(x)P_m(x)=\frac{2 \delta_{lm}}{2l + 1}$), we have
\begin{equation}
\sum_n\left(2n+1\right)\mbox{Im}\left(a_{n}\right)\geq \sum_m\left(2m+1\right)|b_{m}|^2
\end{equation}
However, since $|a_{n}|\geq \mbox{Im}\left(a_{n}\right)$, this can be rewritten as
\begin{equation}
\sum_n\left(2n+1\right)|a_{n}|\geq \sum_m\left(2m+1\right)|b_{m}|^2
\end{equation}

If we had started with individual angular momentum states from the beginning rather than plane wave states, we would have gotten each term in these sums individually where $n=m$ \cite{Itzykson:1980rh,Schwartz:2013pla}.  Then, we would have
\begin{equation}
|a_{n}|\geq|b_{n}|^2\ .
\label{eq:aAn>aXn}
\end{equation}
In particular, we can consider the case of elastic scattering where $b_n=a_n$, in which case, we have
\begin{equation}
|a_{n}|\geq|a_{n}|^2
\end{equation}
which can only be satisfied by
\begin{equation}
|a_{n}|\leq1\ .
\end{equation}
With this, we can also see from Eq.~\eqref{eq:aAn>aXn} that
\begin{equation}
|b_{n}|\leq1\ .
\end{equation}
In other words, each partial wave $2\to2$ scattering amplitude is bounded by 1.  Since it is bounded by a constant, it can not grow indefinitely with energy and, therefore, all polynomial energy growth in each partial wave amplitude must cancel at high energy in order to satisfy perturbative unitarity.  
Since it must cancel term by term in Eqs.~\eqref{eq:amplitude decomposition}, it must cancel in the sum.  As a result, all polynomial energy growth must cancel at high energy in the $2\to2$ scattering amplitude.


\section{\label{sec:Feynman Rules}Feynman Rules}
\begin{table*}
\begin{center}
\begin{tabular}{|cllll|}
\hline
Spin & Field & Mass & Propagator & \\
\hline\hline
$0$ & $s_k$ & $M_{0k}$ & $\Pi_k=\frac{i\Delta_k}{p^2-M_{0k}^2}$ & $\Delta_k=1$\\
$1$ & $v_{j\mu}$ & $M_{1j}$ & $\Pi_{j\mu\nu} = \frac{i\Delta_{j\mu\nu}}{p^2-M_{1j}^2}$ & $\Delta_{j\mu\nu}=-\eta_{\mu\nu}+F\frac{p_\mu p_\nu}{M_{1j}^2}$\\
$2$ & $h_{i\mu\nu}$ & $M_{2i}$ & $\Pi_{i\mu\nu,\alpha\beta} = \frac{i\Delta_{i\mu\nu,\alpha\beta}}{p^2-M_{2i}^2}$ & $\Delta_{i\mu\nu,\alpha\beta}=C_{12}\left(\eta_{\mu\alpha}\eta_{\nu\beta}+\eta_{\mu\beta}\eta_{\nu\alpha}\right)+C_{47}\left(\eta_{\mu\alpha}\frac{p_\nu p_\beta}{M_{2i}^2}+\eta_{\nu\beta}\frac{p_\mu p_\alpha}{M_{2i}^2}+\eta_{\mu\beta}\frac{p_\nu p_\alpha}{M_{2i}^2}+\eta_{\nu\alpha}\frac{p_\mu p_\beta}{M_{2i}^2}\right)$\\
&&&& $\hspace{0.6in}+C_3\eta_{\mu\nu}\eta_{\alpha\beta}+C_{89}\left(\eta_{\mu\nu}\frac{p_\alpha p_\beta}{M_{2i}^2}+\eta_{\alpha\beta}\frac{p_\mu p_\nu}{M_{2i}^2}\right)+C_{10}\frac{p_\mu p_\nu p_\alpha p_\beta}{M_{2i}^4}$\\
\hline
\end{tabular}
\end{center}
\caption{\label{tab:fields}Spin, field name, mass symbol and propagator used in this paper.  The subscripts $i,j$ and $k$ determine the member of the collection while $\mu, \nu, \alpha$ and $\beta$ are Lorentz indices.  The coefficients $F$ and $C_i$ are discussed in the text.}
\end{table*}
We consider a collection of spin-0, spin-1 and spin-2 fields which we will call $s_k$, $v_{j\mu}$ and $h_{i\mu\nu}$, respectively (see Table \ref{tab:fields}), where the subscripts $i,j$ and $k$ determine which of the collection is being referred to and $\mu$ and $\nu$ are Lorentz indices, although we will usually refer to the $h$ being scattered without the subscript $i$.

\subsection{\label{app:propagators}Propagators}
On physical grounds, the on-shell propagator numerator $\Delta$ (see Table \ref{tab:fields}) should be equal to a sum over products of the polarization vector times its conjugate \cite{Weinberg:1995mt} as in 
\begin{eqnarray}
\lim_{p^2\to M^2}\Delta\left(p\right) &=& \epsilon\left(p\right) \epsilon^*\left(p\right)\\
\lim_{p^2\to M^2}\Delta_{\mu\nu}\left(p\right) &=& \sum_{\sigma=-1}^1 \epsilon_{\sigma\mu}\left(p\right) \epsilon^*_{\sigma\nu}\left(p\right)\\
\lim_{p^2\to M^2}\Delta_{\mu\nu,\alpha\beta}\left(p\right) &=& \sum_{\sigma=-2}^2 \epsilon_{\sigma\mu\nu}\left(p\right) \epsilon^*_{\sigma\alpha\beta}\left(p\right)
\end{eqnarray}
for a spin-0, -1 and -2 field, respectively, where the subscripts $i,j$ and $k$ have been suppressed and $\sigma$ is the spin-z component or helicity (we will always use helicity in this article).  The reason this is true is that on-shell, the propagator numerator should be a projection operator which removes all spin components which are not physical.
In order to achieve this with the propagators listed in Table~\ref{tab:fields} requires 
\begin{equation}
F=1
\label{eq:standard spin-1 propagator}
\end{equation}
for spin-1 and 
\begin{equation}
C_{12}=\frac{1}{2}, C_3=-\frac{1}{3}, C_{47}=-\frac{1}{2}, C_{89}=\frac{1}{3},  C_{10}=\frac{2}{3}
\label{eq:standard spin-2 propagator}
\end{equation}
for spin-2.  We will often refer to the propagators with these coefficients as the standard propagators.

On the other hand, it is well-known that for spin-1 fields, $F=1$ is not the only consistent choice.  The Lagrangian can be ``gauge fixed" following the Faddeev-Popov method.  When this is done, the propagator takes a different form that no longer projects out the unphysical polarizations.  However, gauge fixing also introduces new unphysical fields, Faddeev-Popov ghosts and Goldstone bosons, which contribute in perturbation theory with the result that they cancel the effects of the unphysical polarizations of the vector field.  The Faddeev-Popov method is well-defined and based on a gauge symmetry and constrains the interactions to definite forms to accomplish this cancellation.  When the Faddeev-Popov method is followed in a theory where the vector bosons are gauge bosons and all operators are equal to or below mass-dimension four, renormalizability and perturbative unitarity are achieved at all orders in the perturbation series for any gauge choice.  One such popular gauge choice is the Feynman gauge where $F=0$.  In this paper, we only focus on the tree-level result where, as long as the external fields only include physical polarizations, the ghosts do not contribute.  The Goldstone bosons, on the other hand, can contribute at tree-level, depending on the process.

For spin-2, no well-defined method of achieving perturbative unitarity or renormalizability has been found.  In this paper, we take the point of view that this may be because we have not yet been sufficiently clever to find the appropriate procedure.  Since such a procedure may allow spin-2 propagators which do not remove unphysical polarizations, we will allow the propagator to vary from the projection operator by inserting unspecified coefficients for each unique term of the propagator as shown in Table \ref{tab:fields}.  We will, however, require that the propagator $\Pi_{\mu\nu,\alpha\beta}$ is symmetric in interchange of $\mu\nu\leftrightarrow\alpha\beta$ and also in $\mu\leftrightarrow\nu$ and $\alpha\leftrightarrow\beta$.  As a result of the propagator not removing unphysical polarizations, other fields analogous to the ghosts and Goldstone bosons of vector fields will be required to cancel the effects of the unphysical polarizations.  We assume that, as in the case of vector fields, any ghosts that are required do not contribute to tree-level amplitudes.  On the other hand, we include an unspecified number of scalar and vector fields along with the tensor field, some of which can take the analogous role of the Goldstone bosons, as required by tree-level unitarity.  Instead of starting with the symmetry which fixes the couplings in order to achieve unitarity (as is done for gauge bosons), we start with tree-level unitarity which fixes the couplings and hope to infer the symmetry afterwards.

\subsection{\label{app:h^4 interactions}$\mathbf{h^4}$ Interactions}
Since we are considering tree-level $2\to2$ scattering, a $h^4$ operator will only contribute to the diagram with all four $h$'s external and on-shell.  As a result, each $h_{\mu\nu}$ in this operator should be symmetric in $\mu$ and $\nu$ and traceless.  Furthermore, since we are only considering operators of dimension-4 or lower, there can be no derivatives in this operator.  

If $h$ does not transform under a gauge symmetry, the Lagrangian has two unique terms
\begin{eqnarray}
\mathcal{L}_{h^4} = &&
g^{(1)}_{2222} h_{\mu\nu}(x)h^{\mu\nu}(x)h_{\alpha\beta}(x)h^{\alpha\beta}(x)\nonumber\\
&&+ g^{(2)}_{2222} h_{\mu\nu}(x)h^{\mu\alpha}(x)h_{\alpha\beta}(x)h^{\nu\beta}(x)
\end{eqnarray}
The Feynman rules are obtained by Fourier transforming to momentum space and functionally differentiating with respect to $h$ four times.  This gives a total of 24 different contractions for each coupling, some of which might be identical by the symmetry of the metric or might have an identical contribution to $2\to2$ scattering due to the symmetry of the Lorentz indices on the external $h$.  This results in the following vertices
\\
\begin{minipage}{1.in}
\includegraphics[scale=0.33]{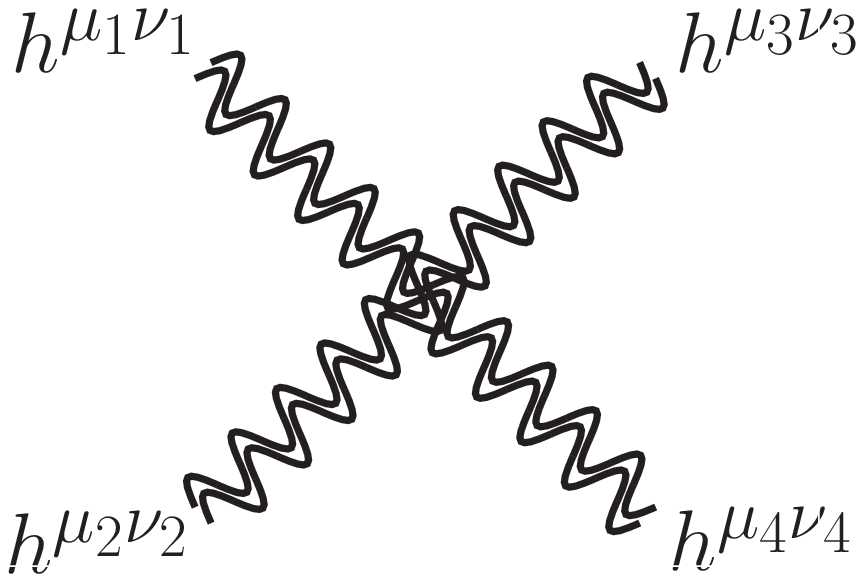}
\end{minipage}
\hfill
\begin{minipage}{2.35in}
\begin{eqnarray}
i8g_{2222}^{(1)}\big(&\eta_{\mu_1\mu_2}\eta_{\nu_1\nu_2}\eta_{\mu_3\mu_4}\eta_{\nu_3\nu_4}&\nonumber\\
&+\eta_{\mu_1\mu_3}\eta_{\nu_1\nu_3}\eta_{\mu_2\mu_4}\eta_{\nu_2\nu_4}&\nonumber\\
&+\eta_{\mu_1\mu_4}\eta_{\nu_1\nu_4}\eta_{\mu_3\mu_2}\eta_{\nu_3\nu_2}&\big)\nonumber\\\label{eq:g_2222^1}
\end{eqnarray}
\end{minipage}
\begin{minipage}{1.in}
\includegraphics[scale=0.33]{hhhh.eps}
\end{minipage}
\hfill
\begin{minipage}{2.35in}
\begin{eqnarray}
i8g_{2222}^{(2)}\big(&\eta_{\mu_1\mu_2}\eta_{\nu_1\nu_3}\eta_{\mu_3\mu_4}\eta_{\nu_2\nu_4}&\nonumber\\
&+\eta_{\mu_1\mu_3}\eta_{\nu_1\mu_4}\eta_{\mu_2\nu_3}\eta_{\nu_2\nu_4}&\nonumber\\
&+\eta_{\mu_1\mu_4}\eta_{\nu_1\nu_2}\eta_{\mu_3\mu_2}\eta_{\nu_3\nu_4}&\big)\nonumber\\\label{eq:g_2222^2}
\end{eqnarray}
\end{minipage}
where the second is simplified by using the symmetry of the Lorentz indices in the external states.  

If, on the other hand, $h$ does transform under a gauge symmetry, we will label it with an additional index $a$.  In this paper, we only consider the adjoint representation.  We have for the unique operators
\begin{eqnarray}
\mathcal{L}_{h^4} = &&
\bar{g}^{(B1)}_{2222} f_{ead}f_{ebc} h_{a\mu\nu}(x)h_b^{\mu\nu}(x)h_{c\alpha\beta}(x)h_d^{\alpha\beta}(x)\nonumber\\
&&+ \left(\bar{g}^{(A2)}_{2222} f_{eab}f_{ecd}+\bar{g}^{(B2)}_{2222} f_{ead}f_{ebc}\right)\times\nonumber\\
&& \quad h_{a\mu\nu}(x)h_b^{\mu\alpha}(x)h_{c\alpha\beta}(x)h_d^{\nu\beta}(x)
\end{eqnarray}
All others can be put in the form of one of these operators or do not contribute to the $2\to2$ scattering.
The Feynman rules are obtained by Fourier transforming to momentum space and functionally differentiating with respect to $h$ four times.  This gives a total of 24 different contractions for each coupling, some of which might be identical by the symmetry of the metric or might have an identical contribution to $2\to2$ scattering due to the symmetry of the Lorentz indices on the external $h$.  This results in the following vertices
\\
\begin{minipage}{1.in}
\includegraphics[scale=0.33]{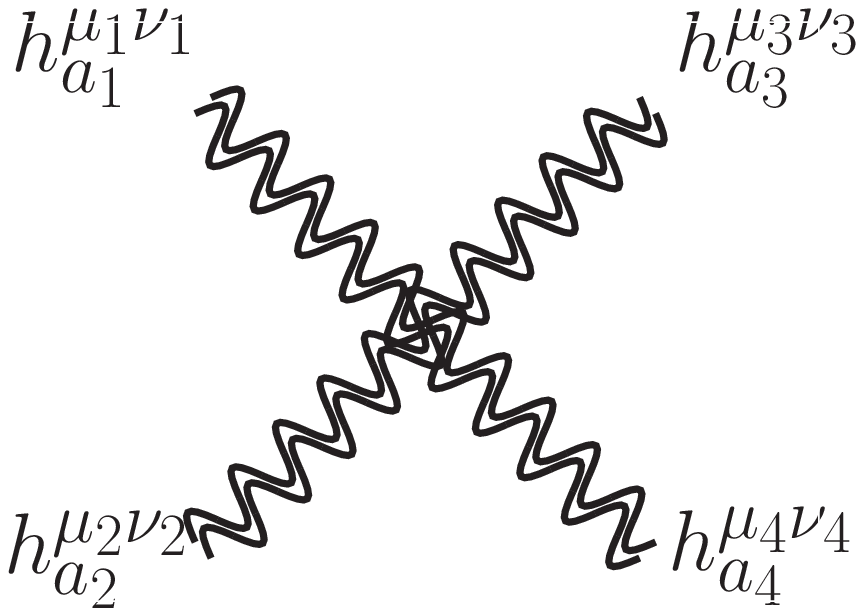}
\end{minipage}
\hfill
\begin{minipage}{2.35in}
\begin{eqnarray}
i4\bar{g}_{2222}^{(B1)}\big[
&\eta_{\mu_1\mu_2}\eta_{\nu_1\nu_2}\eta_{\mu_3\mu_4}\eta_{\nu_3\nu_4}\times\nonumber\\
&\left(f_{ba_1a_4}f_{ba_2a_3}+f_{ba_1a_3}f_{ba_2a_4}\right)&\nonumber\\
&+\eta_{\mu_1\mu_3}\eta_{\nu_1\nu_3}\eta_{\mu_2\mu_4}\eta_{\nu_2\nu_4}\times\nonumber\\
&\left(f_{ba_1a_2}f_{ba_3a_4}-f_{ba_1a_4}f_{ba_2a_3}\right)&\nonumber
\end{eqnarray}
\end{minipage}
\begin{eqnarray}
-\eta_{\mu_1\mu_4}\eta_{\nu_1\nu_4}\eta_{\mu_2\mu_3}\eta_{\nu_2\nu_3}
\left(f_{ba_1a_3}f_{ba_2a_4}+f_{ba_1a_2}f_{ba_3a_4}\right)\big]\nonumber\\
\label{eq:gf_2222^1}
\end{eqnarray}
\begin{minipage}{1.in}
\includegraphics[scale=0.33]{hahbhchd.eps}
\end{minipage}
\hfill
\begin{minipage}{2.35in}
\begin{eqnarray}
i4\bar{g}_{2222}^{(A2)}\big[
&f_{ba_1a_4}f_{ba_2a_3}\times\nonumber\\
&\big(\eta_{\mu_1\nu_2}\eta_{\mu_2\mu_3}\eta_{\mu_4\nu_3}\eta_{\nu_1\nu_4}&\nonumber\\
&-\eta_{\mu_1\nu_3}\eta_{\mu_2\mu_3}\eta_{\mu_4\nu_2}\eta_{\nu_1\nu_4}\big)&\nonumber
\end{eqnarray}
\end{minipage}
\begin{eqnarray}
+f_{ba_1a_3}f_{ba_2a_4}\big(
\eta_{\mu_1\nu_2}\eta_{\mu_2\mu_4}\eta_{\mu_3\nu_4}\eta_{\nu_1\nu_3}
-\eta_{\mu_1\nu_4}\eta_{\mu_2\mu_4}\eta_{\mu_3\nu_2}\eta_{\nu_1\nu_3}\big)\nonumber\\
+f_{ba_1a_2}f_{ba_3a_4}\big(
\eta_{\mu_1\nu_3}\eta_{\mu_2\nu_4}\eta_{\mu_3\mu_4}\eta_{\nu_1\nu_2}
-\eta_{\mu_1\nu_4}\eta_{\mu_2\nu_3}\eta_{\mu_3\mu_4}\eta_{\nu_1\nu_2}\big)\big]
\nonumber\\
\label{eq:gf_2222^2}
\end{eqnarray}
where the symmetry of the Lorentz indices on the external states was used to simplify this vertex and
\begin{minipage}{1.in}
\includegraphics[scale=0.33]{hahbhchd.eps}
\end{minipage}
\hfill
\begin{minipage}{2.35in}
\begin{equation}
0
\end{equation}
\end{minipage}
where the symmetry of the Lorentz indices of the external states was used to simplify this vertex.  We see that $\bar{g}^{(B2)}_{2222}$ does not contribute to our $2\to2$ process.

If $h$ transforms under the adjoint representation of two gauge groups, we will label the first group index by letters from the beginning of the alphabet ($a,b,c,d,e$) and the second group index as barred letters from the beginning of the alphabet ($\bar{a},\bar{b},\bar{c},\bar{d},\bar{e}$).  The unique operators are then given by
\begin{eqnarray}
\mathcal{L}_{h^4} &=& 
\Big(\tilde{g}^{(A1)}_{2222}f_{\bar{e}\bar{a}\bar{b}}f_{\bar{e}\bar{c}\bar{d}}f_{eab}f_{ecd} + \tilde{g}^{(B1)}_{2222}f_{\bar{e}\bar{a}\bar{b}}f_{\bar{e}\bar{c}\bar{d}}f_{ead}f_{ebc}\nonumber\\
&&+ \tilde{g}^{(C1)}_{2222}f_{\bar{e}\bar{a}\bar{d}}f_{\bar{e}\bar{b}\bar{c}}f_{eab}f_{ecd} + \tilde{g}^{(D1)}_{2222}f_{\bar{e}\bar{a}\bar{d}}f_{\bar{e}\bar{b}\bar{c}}f_{ead}f_{ebc} \Big)\nonumber\\
&&\times h_{\bar{a}a\mu\nu}(x)h_{\bar{b}b}^{\mu\nu}(x)h_{\bar{c}c\alpha\beta}(x)h_{\bar{d}d}^{\alpha\beta}(x)+\nonumber\\
&&\Big(\tilde{g}^{(A2)}_{2222}f_{\bar{e}\bar{a}\bar{b}}f_{\bar{e}\bar{c}\bar{d}}f_{eab}f_{ecd} + \tilde{g}^{(B2)}_{2222}f_{\bar{e}\bar{a}\bar{b}}f_{\bar{e}\bar{c}\bar{d}}f_{ead}f_{ebc}\nonumber\\
&&+ \tilde{g}^{(C2)}_{2222}f_{\bar{e}\bar{a}\bar{d}}f_{\bar{e}\bar{b}\bar{c}}f_{eab}f_{ecd} + \tilde{g}^{(D2)}_{2222}f_{\bar{e}\bar{a}\bar{d}}f_{\bar{e}\bar{b}\bar{c}}f_{ead}f_{ebc} \Big)\nonumber\\
&&\times h_{\bar{a}a\mu\nu}(x)h_{\bar{b}b}^{\mu\alpha}(x)h_{\bar{c}c\alpha\beta}(x)h_{\bar{d}d}^{\nu\beta}(x)\nonumber\\
\end{eqnarray}
where both $f_{\bar{e}\bar{a}\bar{c}}f_{\bar{e}\bar{b}\bar{d}}$ and $f_{eac}f_{ebd}$ are replaced using the Jacobi identity.

The Feynman rules are obtained by Fourier transforming to momentum space and functionally differentiating with respect to $h$ four times.  This gives a total of 24 different contractions for each coupling, some of which might be identical by the symmetry of the metric or might have an identical contribution to $2\to2$ scattering due to the symmetry of the Lorentz indices on the external $h$.  The resulting Feynman rule is too long to fully reproduce here.

\subsection{\label{app:h^3 interactions}$\mathbf{h^3}$ Interactions}
Since we are considering tree-level $2\to2$ scattering, a $h^3$ operator will only contribute to the diagram with at least two $h$'s on-shell.  As a result, at least two of the $h_{\mu\nu}$ in this operator should be symmetric in $\mu$ and $\nu$ and traceless.  Furthermore, since we are only considering operators of dimension-4 or lower, there can be no derivatives in this operator.  

If $h$ does not transform under a gauge symmetry, the Lagrangian has the two independent contributions
\begin{eqnarray}
\mathcal{L}_{h^3} = &&
g^{(1)}_{222} h_{\mu\nu}(x)h^{\mu\nu}(x)h_{\alpha}^{\ \ \alpha}(x)\nonumber\\
&&+ g^{(2)}_{222} h_{\mu\nu}(x)h^{\mu\alpha}(x)h_{\alpha}^{\ \ \nu}(x)
\end{eqnarray}
The Feynman rules are obtained by Fourier transforming to momentum space and functionally differentiating with respect to $h$ three times.  This gives a total of 6 different contractions for each coupling, some of which might be identical by the symmetry of the metric or might have an identical contribution to $2\to2$ scattering due to the symmetry of the Lorentz indices on the external $h$.  Furthermore, only one $h$ can be off-shell.  We have the Feynman vertices
\\
\begin{minipage}{1.2in}
\includegraphics[scale=0.33]{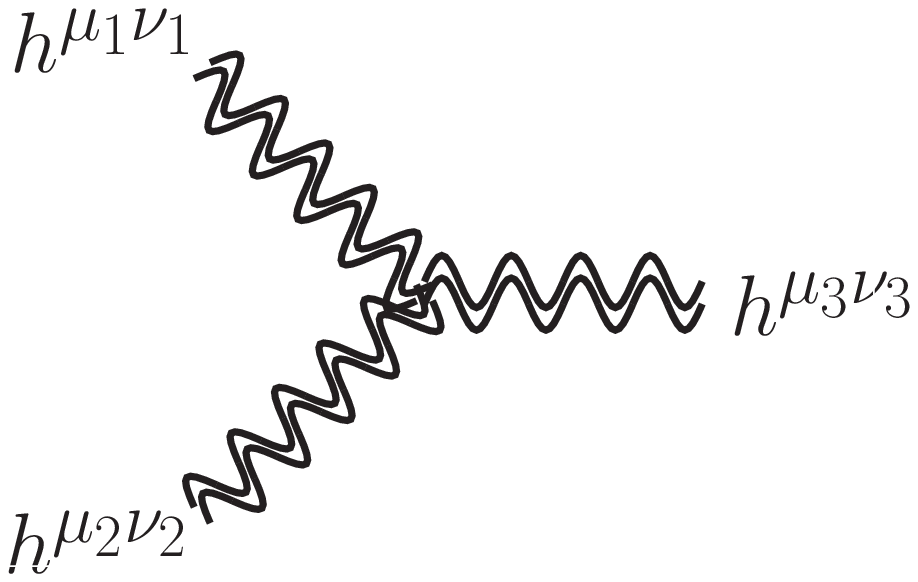}
\end{minipage}
\hfill
\begin{minipage}{2.15in}
\begin{eqnarray}
&&i2g_{222}^{(1)}\big(\eta_{\mu_1\mu_2}\eta_{\nu_1\nu_2}\eta_{\mu_3\nu_3}\nonumber\\
&&+\eta_{\mu_1\mu_3}\eta_{\nu_1\nu_3}\eta_{\mu_2\nu_2}+\eta_{\mu_2\mu_3}\eta_{\nu_2\nu_3}\eta_{\mu_1\nu_1}\big)\nonumber\\
\label{eq:g_222^1}
\end{eqnarray}
\end{minipage}
\begin{minipage}{1.2in}
\includegraphics[scale=0.33]{hhh.eps}
\end{minipage}
\hfill
\begin{minipage}{2.15in}
\begin{eqnarray}
&&i 3g_{222}^{(2)}\times\nonumber\\
&&\big(\eta_{\mu_1\mu_2}\eta_{\nu_1\nu_3}\eta_{\nu_2\mu_3}+\eta_{\mu_1\mu_2}\eta_{\nu_1\mu_3}\eta_{\nu_2\nu_3}\big)\nonumber\\
\label{eq:g_222^2}
\end{eqnarray}
\end{minipage}
The second is simplified by using the symmetry of the Lorentz indices in the external states.

If, on the other hand, $h$ does transform under the adjoint representation, we have the unique operator
\begin{equation}
\mathcal{L}_{h^3} = 
\bar{g}^{(2)}_{222} f_{abc} h_{a\mu\nu}(x)h_b^{\mu\alpha}(x)h_{c\alpha}^{\ \ \ \nu}(x)
\end{equation}
Again, Fourier transforming to momentum space and functionally differentiating gives us six terms some of which give equal contributions due to the symmetry of the external $h$\\
\begin{minipage}{1.2in}
\includegraphics[scale=0.33]{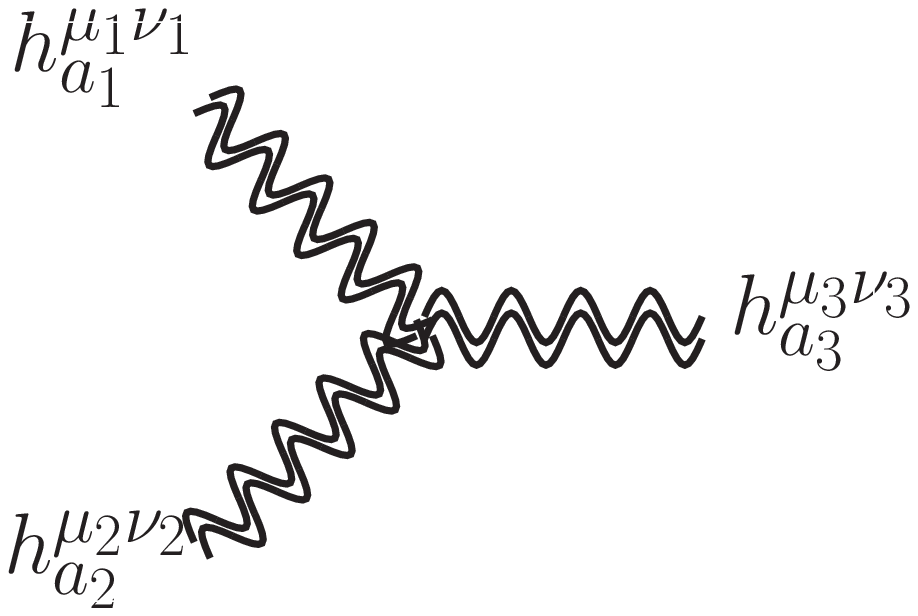}
\end{minipage}
\hfill
\begin{minipage}{2.15in}
\begin{eqnarray}
&&i \bar{g}_{222}^{(2)}f_{a_1a_2a_3}\times\nonumber\\
&&\big(\eta_{\mu_1\mu_2}\eta_{\nu_1\mu_3}\eta_{\nu_2\nu_3}-\eta_{\mu_1\mu_3}\eta_{\nu_1\mu_2}\eta_{\nu_2\nu_3}\nonumber\\
&&+\eta_{\mu_2\mu_3}\eta_{\nu_2\mu_1}\eta_{\nu_3\nu_1}-\eta_{\mu_2\mu_1}\eta_{\nu_2\mu_3}\eta_{\nu_1\nu_3}\nonumber\\
&&+\eta_{\mu_3\mu_1}\eta_{\nu_3\mu_2}\eta_{\nu_1\nu_2}-\eta_{\mu_3\mu_2}\eta_{\nu_3\mu_1}\eta_{\nu_1\nu_2}\big)\nonumber\\
\label{eq:h^3:adjoint}
\end{eqnarray}
\end{minipage}
which will give a vanishing contribution to the $2\to2$ scattering amplitude when the external fields are taken on-shell.

If $h$ transforms under the adjoint representation of two gauge groups, we must contract both sets of indices with the structure constant, so the allowed operators are given by
\begin{eqnarray}
\mathcal{L}_{h^3} &=& 
\tilde{g}^{(1)}_{222}f_{\bar{a}\bar{b}\bar{c}}f_{abc}h_{\bar{a}a\mu\nu}(x)h_{\bar{b}b}^{\mu\nu}(x)h_{\bar{c}c\alpha}^{\ \ \ \alpha}(x)\nonumber\\
&&+\tilde{g}^{(2)}_{222}f_{\bar{a}\bar{b}\bar{c}}f_{abc}h_{\bar{a}a\mu\nu}(x)h_{\bar{b}b}^{\mu\alpha}(x)h_{\bar{c}c\alpha}^{\ \ \ \nu}(x)
\end{eqnarray}
Again, Fourier transforming to momentum space and functionally differentiating gives us six terms some of which give equal contributions due to the symmetry of the external $h$\\
\begin{minipage}{1.2in}
\includegraphics[scale=0.33]{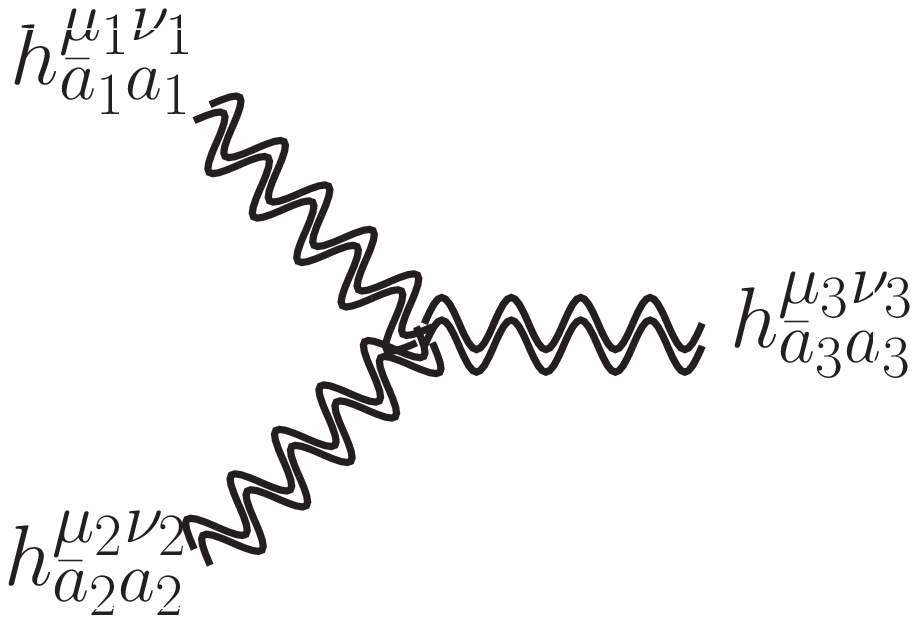}
\end{minipage}
\hfill
\begin{minipage}{2.15in}
\begin{eqnarray}
&&i 2\tilde{g}_{222}^{(1)}f_{\bar{a}_1\bar{a}_2\bar{a}_3}f_{a_1a_2a_3}\times\nonumber\\
&&\big(\eta_{\mu_1\mu_2}\eta_{\nu_1\nu_2}\eta_{\mu_3\nu_3}+\eta_{\mu_2\mu_3}\eta_{\nu_2\nu_3}\eta_{\mu_1\nu_1}\nonumber\\
&&+\eta_{\mu_3\mu_1}\eta_{\nu_3\nu_1}\eta_{\mu_2\nu_2}\big)\nonumber\\
\end{eqnarray}
\end{minipage}
\begin{minipage}{1.2in}
\includegraphics[scale=0.33]{haahbbhcc.eps}
\end{minipage}
\hfill
\begin{minipage}{2.15in}
\begin{eqnarray}
&&i \tilde{g}_{222}^{(2)}f_{\bar{a}_1\bar{a}_2\bar{a}_3}f_{a_1a_2a_3}\times\nonumber\\
&&\big(\eta_{\mu_1\mu_2}\eta_{\nu_1\mu_3}\eta_{\nu_2\nu_3}+\eta_{\mu_1\mu_3}\eta_{\nu_1\mu_2}\eta_{\nu_2\nu_3}\nonumber\\
&&+\eta_{\mu_2\mu_3}\eta_{\nu_2\mu_1}\eta_{\nu_3\nu_1}+\eta_{\mu_2\mu_1}\eta_{\nu_2\mu_3}\eta_{\nu_1\nu_3}\nonumber\\
&&+\eta_{\mu_3\mu_1}\eta_{\nu_3\mu_2}\eta_{\nu_1\nu_2}+\eta_{\mu_3\mu_2}\eta_{\nu_3\mu_1}\eta_{\nu_1\nu_2}\big)\nonumber\\
\end{eqnarray}
\end{minipage}

\subsection{$\mathbf{h^2h_i}$ Interactions}
These are interactions where the two external $h$ interact with another spin-2 object called $h_i$.  Since we are considering tree-level $2\to2$ scattering, a $h^2h_i$ operator will only contribute to the diagram with the two $h$'s on-shell.  As a result, the two $h_{\mu\nu}$ in this operator should be symmetric in $\mu$ and $\nu$ and traceless.  Furthermore, since we are only considering operators of dimension-4 or lower, there can be no derivatives in this operator.  

If $h$ does not transform under a gauge symmetry, the Lagrangian has two independent contributions
\begin{eqnarray}
\mathcal{L}_{h^2h_i} = &&
g^{(1)}_{222i} h_{\mu\nu}(x)h^{\mu\nu}(x)h_{i\alpha}^{\ \ \alpha}(x)\nonumber\\
&&+ g^{(2)}_{222i} h_{\mu\nu}(x)h^{\mu\alpha}(x)h_{i\alpha}^{\ \ \nu}(x)
\end{eqnarray}
The Feynman rules are obtained by Fourier transforming to momentum space and functionally differentiating with respect to $h$ two times and $h_i$ once.  This gives a total of 2 different contractions for each coupling, some of which might be identical by the symmetry of the metric or might have an identical contribution to $2\to2$ scattering due to the symmetry of the Lorentz indices on the external $h$.  Furthermore, only $h_i$ can be off-shell.  We have the Feynman vertices
\\
\begin{minipage}{1.2in}
\includegraphics[scale=0.33]{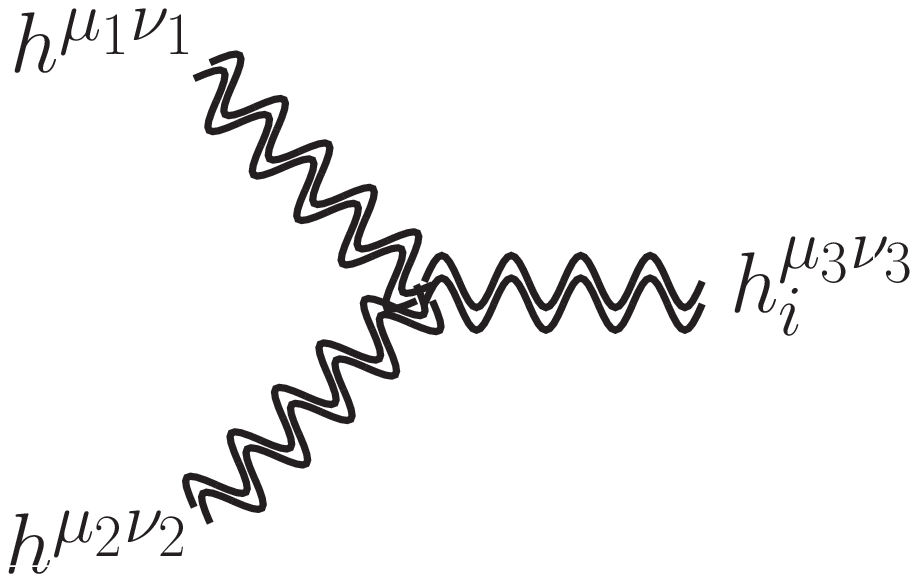}
\end{minipage}
\hfill
\begin{minipage}{2.15in}
\begin{equation}
i2g_{222i}^{(1)}\eta_{\mu_1\mu_2}\eta_{\nu_1\nu_2}\eta_{\mu_3\nu_3}
\label{eq:g_222i^1}
\end{equation}
\end{minipage}
\begin{minipage}{1.2in}
\includegraphics[scale=0.33]{hhhi.eps}
\end{minipage}
\hfill
\begin{minipage}{2.15in}
\begin{eqnarray}
&&i g_{222i}^{(2)}\times\nonumber\\
&&\big(\eta_{\mu_1\mu_2}\eta_{\nu_1\nu_3}\eta_{\nu_2\mu_3}+\eta_{\mu_1\mu_2}\eta_{\nu_1\mu_3}\eta_{\nu_2\nu_3}\big)\nonumber\\
\label{eq:g_222i^2}
\end{eqnarray}
\end{minipage}

If, on the other hand, $h$ does transform under the adjoint representation, we have the unique operator
\begin{equation}
\mathcal{L}_{h^2h_i} =
\bar{g}^{(2)}_{222i} f_{a b c} h_{a\mu\nu}(x)h_b^{\mu\alpha}(x)h_{i c \alpha}^{\ \ \ \nu}(x)
\end{equation}
Again, Fourier transforming to momentum space and functionally differentiating gives us two terms\\
\begin{minipage}{1.2in}
\includegraphics[scale=0.33]{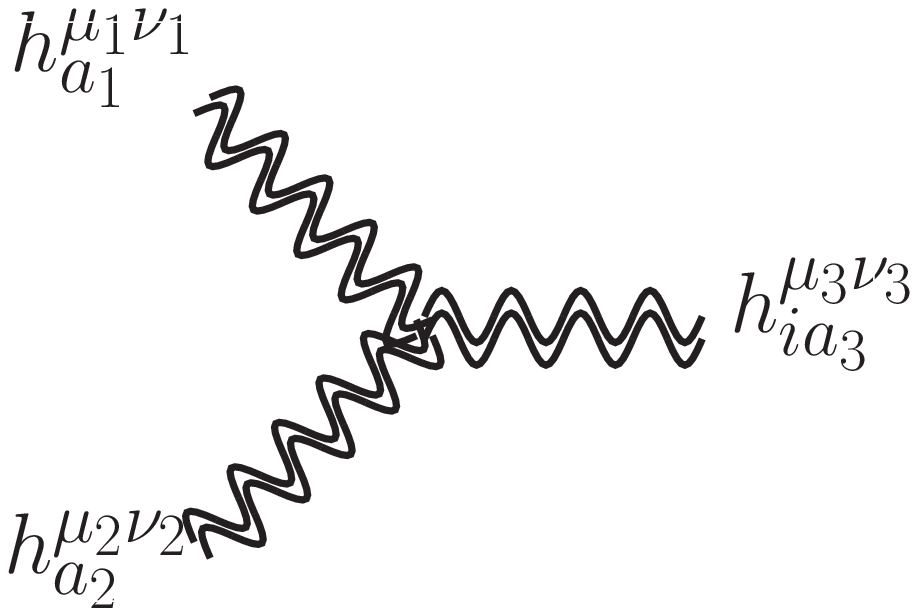}
\end{minipage}
\hfill
\begin{minipage}{2.15in}
\begin{eqnarray}
&&i \bar{g}_{222i}^{(2)}f_{a_1a_2a_3}\times\nonumber\\
&&\big(\eta_{\mu_1\mu_2}\eta_{\nu_1\mu_3}\eta_{\nu_2\nu_3}-\eta_{\mu_1\mu_2}\eta_{\nu_1\nu_3}\eta_{\nu_2\mu_3}\big)\nonumber\\
\label{eq:h^2hi:adjoint}
\end{eqnarray}
\end{minipage}

If $h$ transforms under the adjoint representation of two gauge gropus, we have
\begin{eqnarray}
\mathcal{L}_{h^2h_i} &=&
\tilde{g}^{(1)}_{222i} f_{\bar{a}\bar{b}\bar{c}}f_{abc} h_{\bar{a}a\mu\nu}(x)h_{\bar{b}b}^{\mu\nu}(x)h_{i\bar{c}c\alpha}^{\ \ \ \ \alpha}(x)+\nonumber\\
&&\tilde{g}^{(2)}_{222i} f_{\bar{a}\bar{b}\bar{c}}f_{abc} h_{\bar{a}a\mu\nu}(x)h_{\bar{b}b}^{\mu\alpha}(x)h_{i \bar{c}c \alpha}^{\ \ \ \ \nu}(x)
\end{eqnarray}
Again, Fourier transforming to momentum space and functionally differentiating gives us the Feynman rules\\
\begin{minipage}{1.2in}
\includegraphics[scale=0.33]{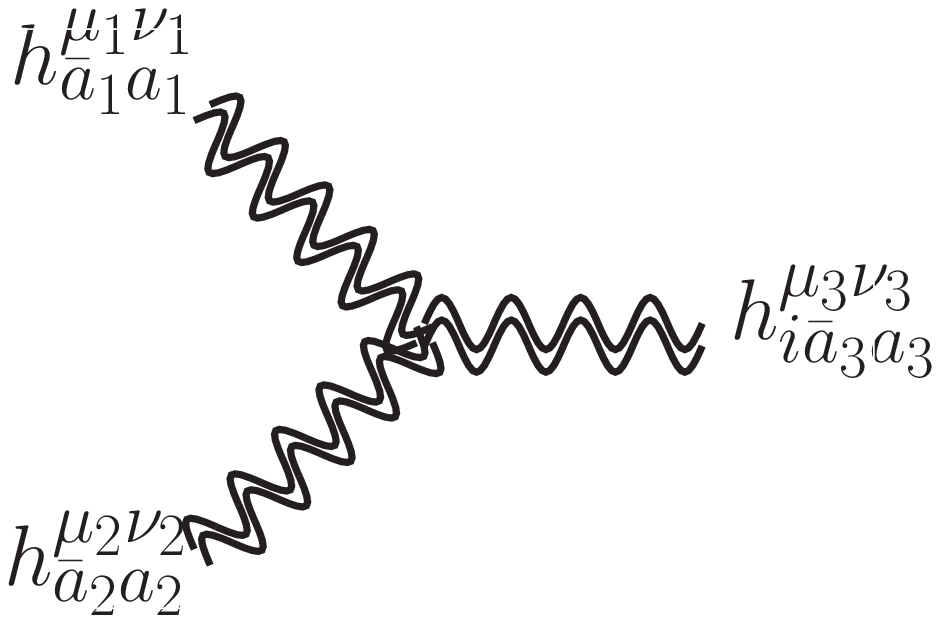}
\end{minipage}
\hfill
\begin{minipage}{2.15in}
\begin{eqnarray}
&&i 2\tilde{g}_{222i}^{(1)}f_{\bar{a}_1\bar{a}_2\bar{a}_3}f_{a_1a_2a_3}\times\nonumber\\
&&\quad\eta_{\mu_1\mu_2}\eta_{\nu_1\nu_2}\eta_{\mu_3\nu_3}\nonumber\\
\end{eqnarray}
\end{minipage}
and\\
\begin{minipage}{1.2in}
\includegraphics[scale=0.33]{haahbbhcci.eps}
\end{minipage}
\hfill
\begin{minipage}{2.15in}
\begin{eqnarray}
&&i \tilde{g}_{222i}^{(2)}f_{\bar{a}_1\bar{a}_2\bar{a}_3}f_{a_1a_2a_3}\times\nonumber\\
&&\big(\eta_{\mu_1\mu_2}\eta_{\nu_1\mu_3}\eta_{\nu_2\nu_3}+\eta_{\mu_1\mu_2}\eta_{\nu_1\nu_3}\eta_{\nu_2\mu_3}\big)\nonumber\\
\end{eqnarray}
\end{minipage}

\subsection{\label{sec:h^2v_j interactions}$\mathbf{h^2v_j}$ Interactions}
Since we are considering tree-level $2\to2$ scattering, a $h^2v_j$ operator will only contribute to the diagram with at least one $h$ external and on-shell.  As a result, at least one $h_{a\mu\nu}$ in this operator should be symmetric in $\mu$ and $\nu$ and traceless.  Furthermore, since we are only considering operators of dimension-4 or lower, there must be exactly one derivative in this operator.  

If we consider the vector field to not be a gauge boson for the moment, the operators are
\begin{eqnarray}
\mathcal{L}_{h^2v_j} = &
i g_{221j}^{(1)} h^{\mu\nu}(x)h_{\alpha\nu}(x)\partial_\mu v_{j}^{\alpha}(x)&\nonumber\\
&+i g_{221j}^{(2)} h^{\mu\nu}(x)h_{\mu\nu}(x)\partial_\alpha v_{j}^{\alpha}(x)&\nonumber\\
\label{eq:op:g_221j}
\end{eqnarray}
The Feynman rules are obtained by Fourier transforming to momentum space and functionally differentiating with respect to $h$ two times and $v_j$ one time.  This gives a total of 2 different contractions for each coupling, some of which are identical by the symmetry of the Lorentz indices on the external state.  We have the Feynman vertices
\\
\begin{minipage}{1.in}
\includegraphics[scale=0.33]{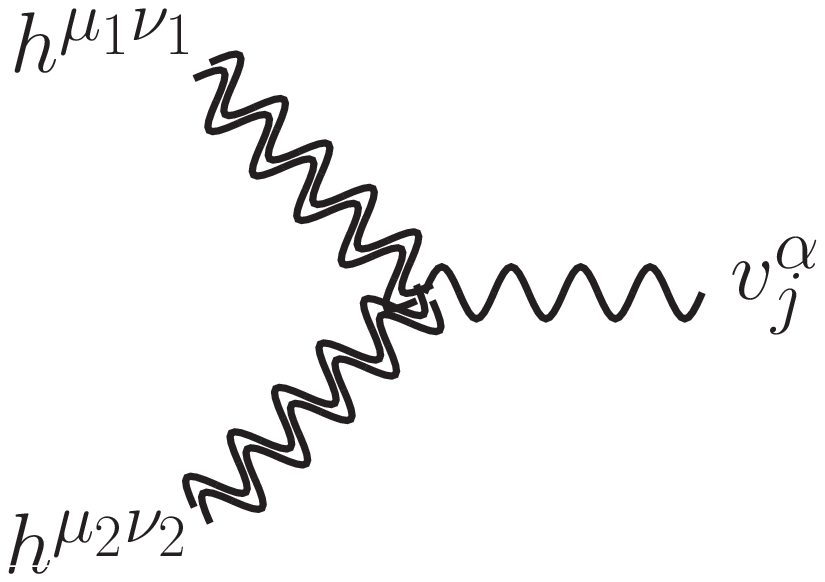}
\end{minipage}
\hfill
\begin{minipage}{2.35in}
\begin{equation}
ig^{(1)}_{221j}\eta_{\mu_1\mu_2}\left(p_{3\nu_1}\eta_{\nu_2\alpha}+p_{3\nu_2}\eta_{\nu_1\alpha}\right)
\label{eq:g_221j^1}
\end{equation}
\end{minipage}
\begin{minipage}{1.in}
\includegraphics[scale=0.33]{hhvj.eps}
\end{minipage}
\hfill
\begin{minipage}{2.35in}
\begin{equation}
i2g^{(2)}_{221j}p_{3\alpha}\eta_{\mu_1\mu_2}\eta_{\nu_1\nu_2}
\label{eq:g_221j^2}
\end{equation}
\end{minipage}
where all the momenta are ingoing.

If, on the other hand, $h$ transforms under the gauge transformation where $v$ is the gauge boson, the interactions come from the covariant derivative terms.  There are two terms that potentially contribute
\begin{eqnarray}
\mathcal{L}_{D} =&
\frac{1}{2}\bar{g}^{(1)}_{221j}\left(D_\alpha h_{\mu\nu}(x)\right)_a \left(D^\alpha h^{\mu\nu}(x)\right)_a\nonumber\\
& + \frac{1}{2}\bar{g}^{(2)}_{221j}\left(D_\alpha h_{\mu\nu}(x)\right)_a \left(D^\mu h^{\alpha\nu}(x)\right)_a
\end{eqnarray}
where
\begin{equation}
\left(D_\alpha h_{\mu\nu}(x)\right)_a = \partial_\alpha h_{a\mu\nu}(x)+\bar{g} f_{abc} v_{j b \alpha} h_{c\mu\nu}
\end{equation}
and $\bar{g}^{(1)}_{221j}$ and $\bar{g}^{(2)}_{221j}$ are normalization constants determined by the inverse of the propagator.
Expanding and extracting the three-point operators gives
\begin{eqnarray}
\mathcal{L}_{h^2v_j} = &
-\bar{g}\bar{g}^{(1)}_{221j}f_{abc}v_{ja}^\alpha(x) \partial_\alpha h_{b\mu\nu}(x) h_c^{\mu\nu}(x) \nonumber\\
&-\bar{g}\bar{g}^{(2)}_{221j}f_{abc}v_{ja}^\mu(x) \partial_\alpha h_{b\mu\nu}(x) h_c^{\alpha\nu}(x) 
\end{eqnarray}
Fourier transforming and functionally differentiating gives the Feynman rules
\\
\begin{minipage}{1.in}
\includegraphics[scale=0.33]{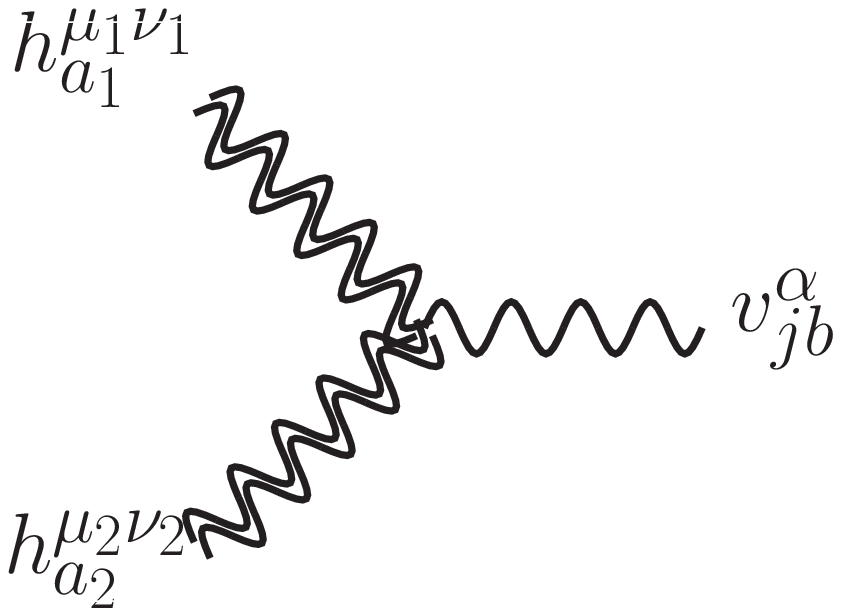}
\end{minipage}
\hfill
\begin{minipage}{2.35in}
\begin{equation}
i\bar{g}\bar{g}^{(1)}_{221j}f_{a_1a_2b}\eta_{\mu_1\mu_2}\eta_{\nu_1\nu_2}\left(p_{1\alpha}-p_{2\alpha}\right)
\label{eq:gf_221j^1}
\end{equation}
\end{minipage}
\begin{minipage}{1.in}
\includegraphics[scale=0.33]{hahbvcj.eps}
\end{minipage}
\hfill
\begin{minipage}{2.35in}
\begin{eqnarray}
&&i\bar{g}\bar{g}^{(2)}_{221j}f_{a_1a_2b}\eta_{\nu_1\nu_2}\times\nonumber\\
&&\left(\eta_{\alpha\mu_1}p_{1\mu_2}-\eta_{\alpha\mu_2}p_{2\mu_1}\right)
\label{eq:gf_221j^2}
\end{eqnarray}
\end{minipage}

If $h$ transforms under the adjoint representation of two gauge groups where the gauge bosons are $\bar{v}$ and $v$, the interactions come from the covariant derivative terms.  There are two terms that potentially contribute
\begin{eqnarray}
\mathcal{L}_{D} =&
\frac{1}{2}\tilde{g}^{(1)}_{221j}\left(D_\alpha h_{\mu\nu}(x)\right)_{\bar{a}a} \left(D^\alpha h^{\mu\nu}(x)\right)_{\bar{a}a}\nonumber\\
& + \frac{1}{2}\tilde{g}^{(2)}_{221j}\left(D_\alpha h_{\mu\nu}(x)\right)_{\bar{a}a} \left(D^\mu h^{\alpha\nu}(x)\right)_{\bar{a}a}
\end{eqnarray}
where
\begin{eqnarray}
\left(D_\alpha h_{\mu\nu}(x)\right)_{\bar{a}a} = \partial_\alpha h_{\bar{a}a\mu\nu}(x)
&&+\bar{g} f_{\bar{a}\bar{b}\bar{c}} \bar{v}_{\bar{b} \alpha} h_{\bar{c}a\mu\nu}\nonumber\\
&&+g f_{abc} v_{b \alpha} h_{\bar{a}c\mu\nu}
\end{eqnarray}
and $\tilde{g}^{(1)}_{221j}$ and $\tilde{g}^{(2)}_{221j}$ are normalization constants determined by the inverse of the propagator.
Expanding and extracting the three-point operators gives
\begin{eqnarray}
\mathcal{L}_{h^2v_j} &= &
-\bar{g}\tilde{g}^{(1)}_{221j}f_{\bar{a}\bar{b}\bar{c}}\bar{v}_{\bar{a}}^\alpha(x) \partial_\alpha h_{\bar{b}a\mu\nu}(x) h_{\bar{c}a}^{\mu\nu}(x) \nonumber\\
&&-\bar{g}\tilde{g}^{(2)}_{221j}f_{\bar{a}\bar{b}\bar{c}}\bar{v}_{\bar{a}}^\mu(x) \partial_\alpha h_{\bar{b}a\mu\nu}(x) h_{\bar{c}a}^{\alpha\nu}(x)\nonumber\\
&&-g\tilde{g}^{(1)}_{221j}f_{abc}v_{a}^\alpha(x) \partial_\alpha h_{\bar{a}b\mu\nu}(x) h_{\bar{a}c}^{\mu\nu}(x) \nonumber\\
&&-g\tilde{g}^{(2)}_{221j}f_{abc}v_{a}^\mu(x) \partial_\alpha h_{\bar{a}b\mu\nu}(x) h_{\bar{a}c}^{\alpha\nu}(x) 
\end{eqnarray}
Fourier transforming and functionally differentiating gives the Feynman rules
\\
\begin{minipage}{1.in}
\includegraphics[scale=0.33]{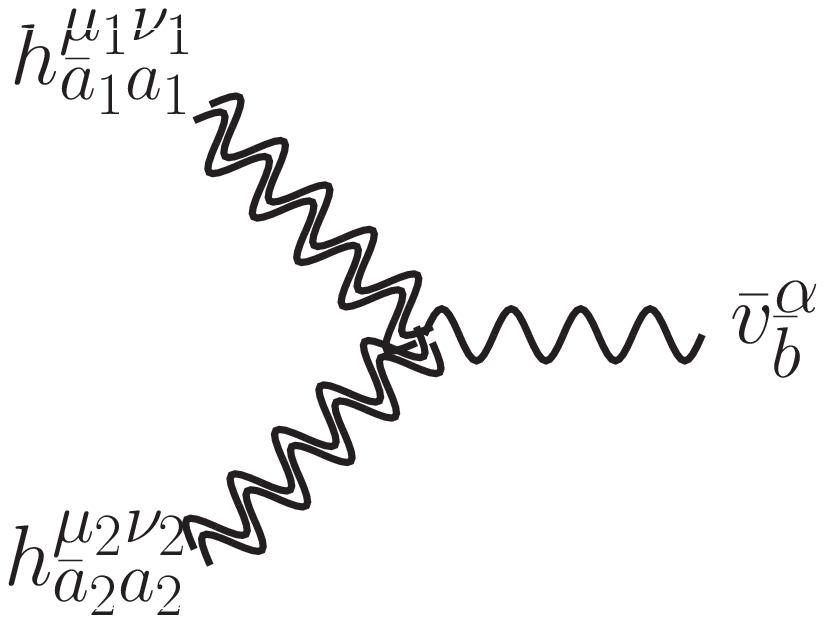}
\end{minipage}
\hfill
\begin{minipage}{2.35in}
\begin{eqnarray}
&&i \bar{g}\tilde{g}^{(1)}_{221j}\delta_{a_1a_2}f_{\bar{a}_1\bar{a}_2\bar{b}}\times\nonumber\\
&&\eta_{\mu_1\mu_2}\eta_{\nu_1\nu_2}\left(p_{1\alpha}-p_{2\alpha}\right)
\label{eq:gff_221j^1}
\end{eqnarray}
\end{minipage}
\begin{minipage}{1.in}
\includegraphics[scale=0.33]{haahbbvbc.eps}
\end{minipage}
\hfill
\begin{minipage}{2.35in}
\begin{eqnarray}
&&i \bar{g}\tilde{g}^{(2)}_{221j}\delta_{a_1a_2}f_{\bar{a}_1\bar{a}_2\bar{b}}\times\nonumber\\
&&\eta_{\nu_1\nu_2}\left(\eta_{\alpha\mu_1}p_{1\mu_2}-\eta_{\alpha\mu_2}p_{2\mu_1}\right)
\label{eq:gff_221j^2}
\end{eqnarray}
\end{minipage}
\begin{minipage}{1.in}
\includegraphics[scale=0.33]{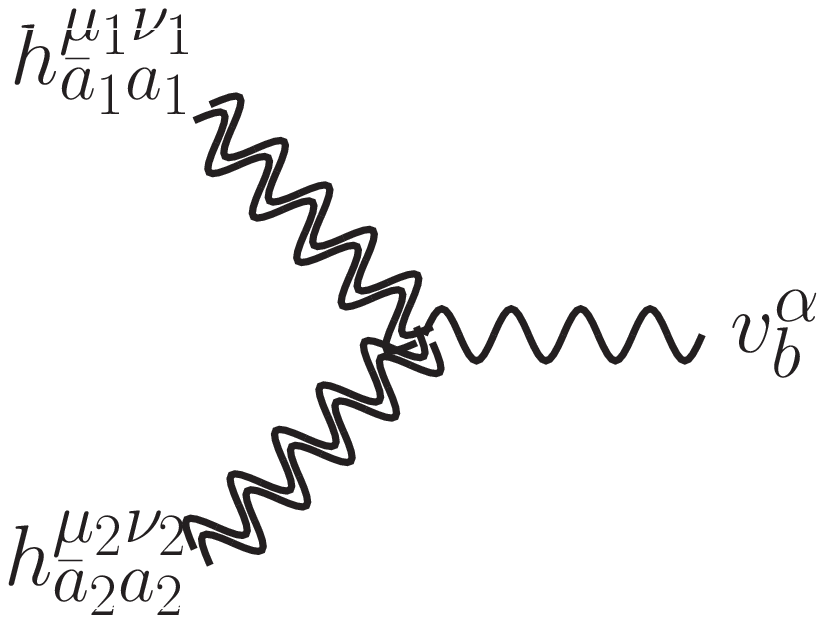}
\end{minipage}
\hfill
\begin{minipage}{2.35in}
\begin{eqnarray}
&&i g\tilde{g}^{(1)}_{221j}\delta_{\bar{a}_1\bar{a}_2}f_{a_1a_2b}\times\nonumber\\
&&\eta_{\mu_1\mu_2}\eta_{\nu_1\nu_2}\left(p_{1\alpha}-p_{2\alpha}\right)
\label{eq:gff_221j^1}
\end{eqnarray}
\end{minipage}
\begin{minipage}{1.in}
\includegraphics[scale=0.33]{haahbbvc.eps}
\end{minipage}
\hfill
\begin{minipage}{2.35in}
\begin{eqnarray}
&&i g\tilde{g}^{(2)}_{221j}\delta_{\bar{a}_1\bar{a}_2}f_{a_1a_2b}\times\nonumber\\
&&\eta_{\nu_1\nu_2}\left(\eta_{\alpha\mu_1}p_{1\mu_2}-\eta_{\alpha\mu_2}p_{2\mu_1}\right)
\label{eq:gff_221j^2}
\end{eqnarray}
\end{minipage}

\subsection{\label{app:hhsk interaction}$\mathbf{h^2s_k}$ Interactions}
Since we are considering tree-level $2\to2$ scattering, a $h^2s_k$ operator will only contribute to the diagram with at least one $h$ external and on-shell.  As a result, at least one $h_{\mu\nu}$ in this operator should be symmetric in $\mu$ and $\nu$ and traceless.  Furthermore, since we are only considering operators of dimension-4 or lower, there can be no derivatives in this operator.  

If $h$ does not transform under a gauge symmetry, the general operator is of the form
\begin{equation}
\mathcal{L}_{h^2s_k} = 
g_{220k} h_{\mu\nu}(x)h^{\mu\nu}(x)s_{k}(x)
\label{eq:g_220k}
\end{equation}
The Feynman rules are obtained by Fourier transforming to momentum space and functionally differentiating with respect to $h$ two times and $s_k$ one time.  This gives a total of 2 different contractions for this coupling, both of which are identical by the symmetry of the Lorentz indices on the external state.  We have the Feynman vertices
\\
\begin{minipage}{1.in}
\includegraphics[scale=0.33]{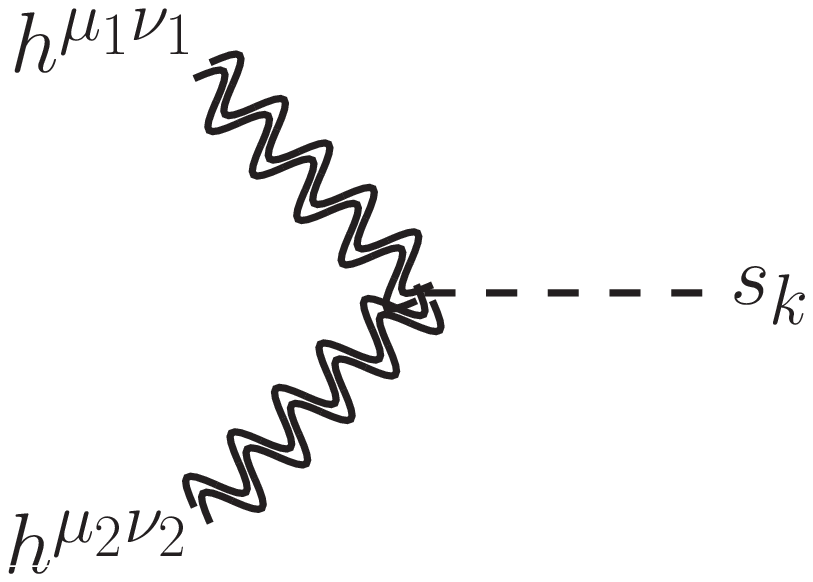}
\end{minipage}
\hfill
\begin{minipage}{2.35in}
\begin{equation}
i2g_{220k}\eta_{\mu_1\mu_2}\eta_{\nu_1\nu_2}
\label{eq:g_220k}
\end{equation}
\end{minipage}

If, on the other hand, $h$ does transform under the adjoint representation, $s$ can not transform under the adjoint representation in order to get a nonzero operator.  This results in 
\begin{equation}
\mathcal{L}_{h^2s_k} = 
\bar{g}_{220k} h_{a\mu\nu}(x)h_a^{\mu\nu}(x)s_{k}(x)
\end{equation}
Again, Fourier transforming and functionally differentiating gives the Feynman rules
\\
\begin{minipage}{1.in}
\includegraphics[scale=0.33]{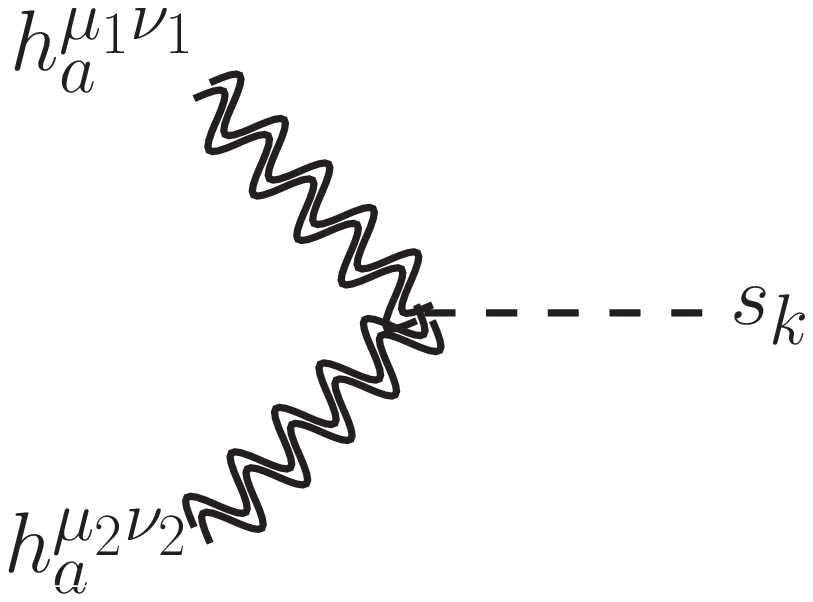}
\end{minipage}
\hfill
\begin{minipage}{2.35in}
\begin{equation}
i2\bar{g}_{220k}\eta_{\mu_1\mu_2}\eta_{\nu_1\nu_2}
\end{equation}
\end{minipage}

If $h$ transforms under the adjoint representation of two gauge groups, $s$ can either be a singlet under both gauge groups or transform under the adjoint representation under both gauge groups in order to get a nonzero operator.  This results in 
\begin{eqnarray}
\mathcal{L}_{h^2s_k} &=& 
\tilde{g}^{(1)}_{220k} h_{\bar{a}a\mu\nu}(x)h_{\bar{a}a}^{\mu\nu}(x)s_{k}(x)\nonumber\\
&&+\tilde{g}^{(2)}_{220k} f_{\bar{a}\bar{b}\bar{c}}f_{abc} h_{\bar{a}a\mu\nu}(x)h_{\bar{b}b}^{\mu\nu}(x)\bar{s}_{k\bar{c}c}(x)\nonumber\\
\end{eqnarray}
Again, Fourier transforming and functionally differentiating gives the Feynman rules
\\
\begin{minipage}{1.in}
\includegraphics[scale=0.33]{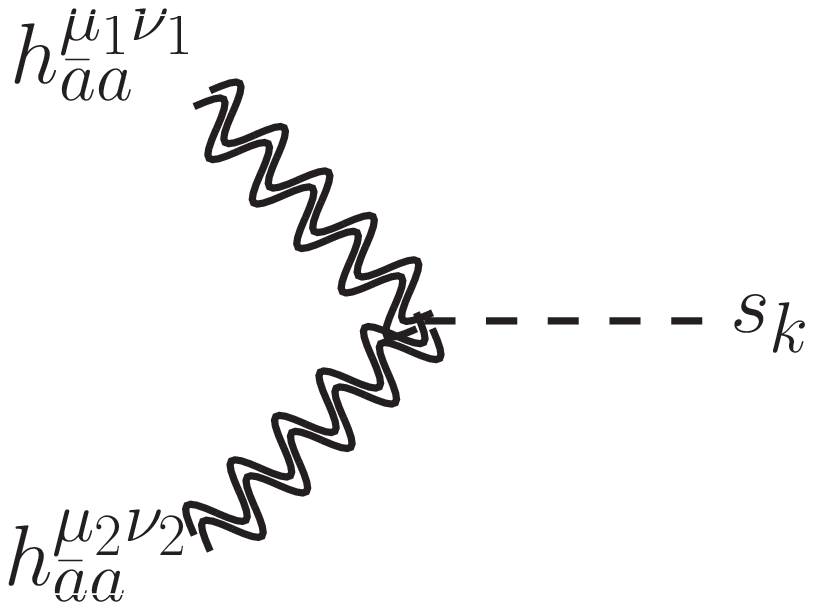}
\end{minipage}
\hfill
\begin{minipage}{2.35in}
\begin{equation}
i2\tilde{g}^{(1)}_{220k}\eta_{\mu_1\mu_2}\eta_{\nu_1\nu_2}
\end{equation}
\end{minipage}
\begin{minipage}{1.in}
\includegraphics[scale=0.33]{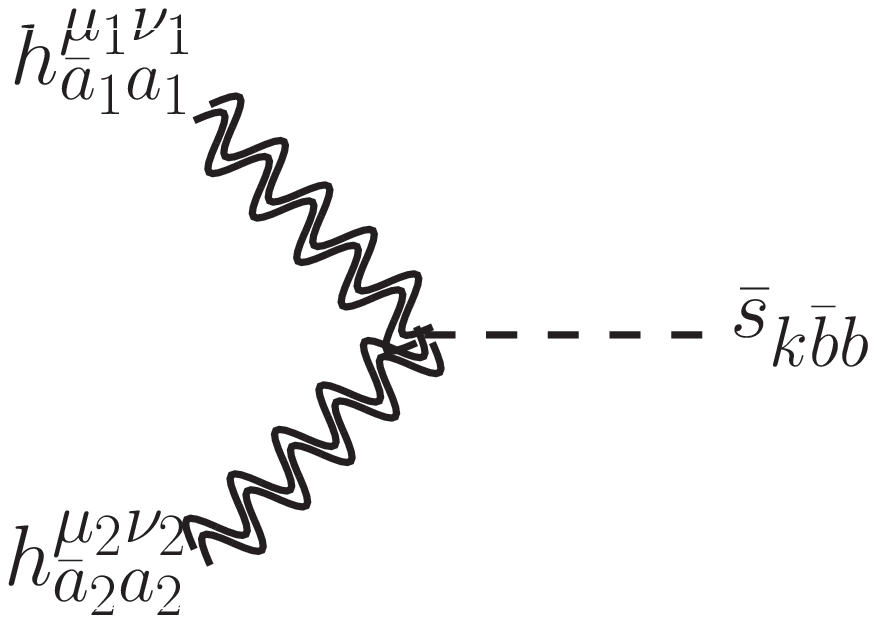}
\end{minipage}
\hfill
\begin{minipage}{2.35in}
\begin{equation}
i2\tilde{g}^{(2)}_{220k}f_{\bar{a}_1\bar{a}_2\bar{b}}f_{a_1a_2b}\eta_{\mu_1\mu_2}\eta_{\nu_1\nu_2}
\end{equation}
\end{minipage}

\section{\label{app:Explicit Diagrams}Diagrams}

\subsection{\label{app:polarization vectors}Momenta and Polarization Vectors}
\begin{figure}
\begin{center}
\includegraphics[scale=0.5]{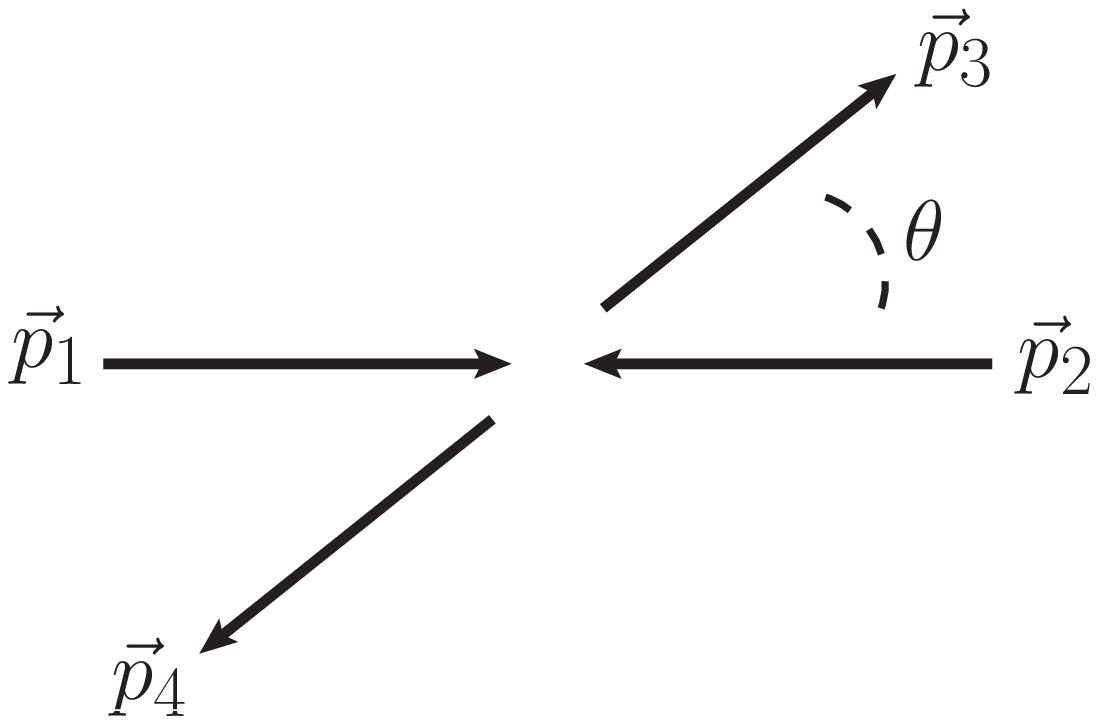}
\end{center}
\caption{\label{fig:Momentum Diagram}Momenta of the $2\to2$ scattering process.}
\end{figure}

For convenience, we will take the momenta to be
\begin{eqnarray}
p_1^\mu &=& \left(E,0,0,k\right)\\
p_2^\mu &=& \left(E,0,0,-k\right)\\
p_3^\mu &=& \left(E,k\sin\theta,0,k\cos\theta\right)\\
p_4^\mu &=& \left(E,-k\sin\theta,0,-k\cos\theta\right)
\end{eqnarray}
where $k=\sqrt{E^2-M_2^2}$, as shown in Fig.~\ref{fig:Momentum Diagram}.
The polarization vectors for spin-1 are then
\begin{eqnarray}
\epsilon_{\pm1}^\mu\left(p_1\right) &=& \frac{1}{\sqrt{2}}\left(0,\mp1,-i,0\right)\\
\epsilon_0^\mu\left(p_1\right) &=& \frac{1}{M_2}\left(k,0,0,E\right)\\
\epsilon_{\pm1}^\mu\left(p_2\right) &=& \frac{1}{\sqrt{2}}\left(0,\pm1,-i,0\right)\\
\epsilon_0^\mu\left(p_2\right) &=& \frac{1}{M_2}\left(k,0,0,-E\right)\\
\epsilon_{\pm1}^{*\mu}\left(p_3\right) &=& \frac{1}{\sqrt{2}}\left(0,\mp\cos\theta,i,\pm\sin\theta\right)\\
\epsilon_0^{*\mu}\left(p_3\right) &=& \frac{1}{M_2}\left(k,E\sin\theta,0,E\cos\theta\right)\\
\epsilon_{\pm1}^{*\mu}\left(p_4\right) &=& \frac{1}{\sqrt{2}}\left(0,\pm\cos\theta,i,\mp\sin\theta\right)\\
\epsilon_0^{*\mu}\left(p_4\right) &=& \frac{1}{M_2}\left(k,-E\sin\theta,0,-E\cos\theta\right)
\end{eqnarray}
The polarization vectors for spin-2 are then given by
\begin{eqnarray}
\epsilon_{\pm2}^{\mu\nu}\left(p\right) &=& \epsilon_{\pm1}^{\mu}\left(p\right)\epsilon_{\pm1}^{\nu}\left(p\right)\\
\epsilon_{\pm1}^{\mu\nu}\left(p\right) &=& \frac{1}{\sqrt{2}}\left[\epsilon_{\pm1}^{\mu}\left(p\right)\epsilon_{0}^{\nu}\left(p\right)+\epsilon_{0}^{\mu}\left(p\right)\epsilon_{\pm1}^{\nu}\left(p\right)\right]\\
\epsilon_{0}^{\mu\nu}\left(p\right) &=& \frac{1}{\sqrt{6}}\big[\epsilon_{+1}^{\mu}\left(p\right)\epsilon_{-1}^{\nu}\left(p\right)+2\epsilon_{0}^{\mu}\left(p\right)\epsilon_{0}^{\nu}\left(p\right)\nonumber\\
&&\quad \quad +\epsilon_{-1}^{\mu}\left(p\right)\epsilon_{+1}^{\nu}\left(p\right)\big]
\end{eqnarray}

\subsection{\label{sec:equivalencies}Enumeration and Equivalencies}
\begin{figure}
\begin{center}
\includegraphics[scale=0.5]{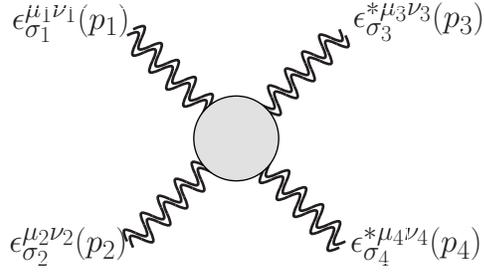}
\end{center}
\caption{\label{fig:General Diagram}Feynman diagrams contributing to the $2\to2$ scattering process.  The polarizations of the external states are given by $\epsilon_\sigma^{\mu\nu}(p)$.}
\end{figure}
We are interested in finding the constraints on the Lagrangian in order to achieve a fully tree-level unitary theory that includes massive spin-2 bosons.  The worst potential violation of unitarity is in the scattering process of the spin-2 bosons, namely $h^{\mu_1\nu_1}(p_1),h^{\mu_2\nu_2}(p_2)\to h^{\mu_3\nu_3}(p_3),h^{\mu_4\nu_4}(p_4)$.  Each external massive spin-2 boson contains 5 helicity states, labeled as $-2,-1,0,1$ and $2$.  Since there are four external spin-2 bosons, this gives us a total of $5^4=625$ different helicity combinations.  The energy growth in each of these helicity combinations must be cancelled.  We illustrate in Fig.~\ref{fig:General Diagram} the process we are interested in, where we have replaced the external fields with their polarization vectors.  
\begin{table*}
\begin{center}
\begin{tabular}{|llllll||llllll|}
\hline\hline
& \multicolumn{4}{c}{$E^8$} &&& \multicolumn{4}{c}{$E^3$} &\\
\hline
1: & A: & $\mathcal{M}_{0,0,0,0}$ & & & & 
43: & D2: & $\mathcal{M}_{2,2,1,0}$ & $=\mathcal{M}_{2,2,0,1}$ & $=-\mathcal{M}_{1,0,2,2}$ & $=-\mathcal{M}_{0,1,2,2}$ \\
\cline{1-6}
& \multicolumn{4}{c}{$E^7$} & &
44: & D2: & $\mathcal{M}_{2,1,2,0}$ & $=\mathcal{M}_{1,2,0,2}$ & $=-\mathcal{M}_{2,0,2,1}$ & $=-\mathcal{M}_{0,2,1,2}$ \\
\cline{1-6}
2: & B: & $\mathcal{M}_{1,0,0,0}$ & $=\mathcal{M}_{0,1,0,0}$ & $=-\mathcal{M}_{0,0,1,0}$ & $=-\mathcal{M}_{0,0,0,1}$ &
45: & D2: & $\mathcal{M}_{2,1,0,2}$ & $=\mathcal{M}_{1,2,2,0}$ & $=-\mathcal{M}_{0,2,2,1}$ & $=-\mathcal{M}_{2,0,1,2}$\\
\cline{1-6}
& \multicolumn{4}{c}{$E^6$} & &
46: & D2: & $\mathcal{M}_{2,2,0,\minus1}$ & $=\mathcal{M}_{2,2,\minus1,0}$ & $=-\mathcal{M}_{0,\minus1,2,2}$ & $=-\mathcal{M}_{\minus1,0,2,2}$\\
\cline{1-6}
3: & B: & $\mathcal{M}_{2,0,0,0}$ & $=\mathcal{M}_{0,2,0,0}$ & $=\mathcal{M}_{0,0,2,0}$ & $=\mathcal{M}_{0,0,0,2}$ &
47: & D2: & $\mathcal{M}_{2,0,2,\minus1}$ & $=\mathcal{M}_{0,2,\minus1,2}$ & $=-\mathcal{M}_{2,\minus1,2,0}$ & $=-\mathcal{M}_{\minus1,2,0,2}$\\
4: & C: & $\mathcal{M}_{1,1,0,0}$ && $=\mathcal{M}_{0,0,1,1}$ & &
48: & D2: & $\mathcal{M}_{2,0,\minus1,2}$ & $=\mathcal{M}_{0,2,2,\minus1}$ & $=-\mathcal{M}_{\minus1,2,2,0}$ & $=-\mathcal{M}_{2,\minus1,0,2}$\\
5: & C: & $\mathcal{M}_{1,0,1,0}$ & $=\mathcal{M}_{0,1,0,1}$ & & &
49: & B: & $\mathcal{M}_{2,1,1,1}$ & $=\mathcal{M}_{1,2,1,1}$ & $=-\mathcal{M}_{1,1,2,1}$ & $=-\mathcal{M}_{1,1,1,2}$\\
6: & C: & $\mathcal{M}_{1,0,0,1}$ & $=\mathcal{M}_{0,1,1,0}$ & & &
50: & D1: & $\mathcal{M}_{2,1,1,\minus1}$ & $=\mathcal{M}_{1,2,\minus1,1}$ & $=-\mathcal{M}_{1,\minus1,2,1}$ & $=-\mathcal{M}_{\minus1,1,1,2}$\\
7: & D0: & $\mathcal{M}_{1,0,0,\minus1}$ & $=\mathcal{M}_{0,1,\minus1,0}$ & & &
51: & D1: & $\mathcal{M}_{2,1,\minus1,1}$ & $=\mathcal{M}_{1,2,1,\minus1}$ & $=-\mathcal{M}_{\minus1,1,2,1}$ & $=-\mathcal{M}_{1,\minus1,1,2}$\\
8: & D0: & $\mathcal{M}_{1,0,\minus1,0}$ & $=\mathcal{M}_{0,1,0,\minus1}$ & & &
52: & D1: & $\mathcal{M}_{2,\minus1,1,1}$ & $=\mathcal{M}_{\minus1,2,1,1}$ & $=-\mathcal{M}_{1,1,2,\minus1}$ & $=-\mathcal{M}_{1,1,\minus1,2}$\\
9: & D0: & $\mathcal{M}_{1,\minus1,0,0}$ && $=\mathcal{M}_{0,0,1,\minus1}$ & &
53: & E: & $\mathcal{M}_{2,1,0,\minus2}$ & $=\mathcal{M}_{1,2,\minus2,0}$ & $=-\mathcal{M}_{0,\minus2,2,1}$ & $=-\mathcal{M}_{\minus2,0,1,2}$\\
\cline{1-6}
& \multicolumn{4}{c}{$E^5$} & &
54: & E: & $\mathcal{M}_{2,1,\minus2,0}$ & $=\mathcal{M}_{1,2,0,\minus2}$ & $=-\mathcal{M}_{\minus2,0,2,1}$ & $=-\mathcal{M}_{0,\minus2,1,2}$\\
\cline{1-6}
10: & D0: & $\mathcal{M}_{2,1,0,0}$ & $=\mathcal{M}_{1,2,0,0}$ & $=-\mathcal{M}_{0,0,2,1}$ & $=-\mathcal{M}_{0,0,1,2}$ &
55: & E: & $\mathcal{M}_{2,0,1,\minus2}$ & $=\mathcal{M}_{0,2,\minus2,1}$ & $=-\mathcal{M}_{1,\minus2,2,0}$ & $=-\mathcal{M}_{\minus2,1,0,2}$\\
11: & D0: & $\mathcal{M}_{2,0,1,0}$ & $=\mathcal{M}_{0,2,0,1}$ & $=-\mathcal{M}_{1,0,2,0}$ & $=-\mathcal{M}_{0,1,0,2}$ &
56: & E: & $\mathcal{M}_{2,0,\minus2,1}$ & $=\mathcal{M}_{0,2,1,\minus2}$ & $=-\mathcal{M}_{\minus2,1,2,0}$ & $=-\mathcal{M}_{1,\minus2,0,2}$\\
12: & D0: & $\mathcal{M}_{2,0,0,1}$ & $=\mathcal{M}_{0,2,1,0}$ & $=-\mathcal{M}_{0,1,2,0}$ & $=-\mathcal{M}_{1,0,0,2}$ &
57: & E: & $\mathcal{M}_{2,\minus2,1,0}$ & $=\mathcal{M}_{\minus2,2,0,1}$ & $=-\mathcal{M}_{1,0,2,\minus2}$ & $=-\mathcal{M}_{0,1,\minus2,2}$\\
13: & D0: & $\mathcal{M}_{2,0,0,\minus1}$ & $=\mathcal{M}_{0,2,\minus1,0}$ & $=-\mathcal{M}_{0,\minus1,2,0}$ & $=-\mathcal{M}_{\minus1,0,0,2}$ &
58: & E: & $\mathcal{M}_{2,\minus2,0,1}$ & $=\mathcal{M}_{\minus2,2,1,0}$ & $=-\mathcal{M}_{0,1,2,\minus2}$ & $=-\mathcal{M}_{1,0,\minus2,2}$\\
14: & D0: & $\mathcal{M}_{2,0,\minus1,0}$ & $=\mathcal{M}_{0,2,0,\minus1}$ & $=-\mathcal{M}_{\minus1,0,2,0}$ & $=-\mathcal{M}_{0,\minus1,0,2}$ &
59: & D1: & $\mathcal{M}_{\minus2,\minus1,1,1}$ & $=\mathcal{M}_{\minus1,\minus2,1,1}$ & $=-\mathcal{M}_{1,1,\minus2,\minus1}$ & $=-\mathcal{M}_{1,1,\minus1,\minus2}$\\
15: & D0: & $\mathcal{M}_{2,\minus1,0,0}$ & $=\mathcal{M}_{\minus1,2,0,0}$ & $=-\mathcal{M}_{0,0,2,\minus1}$ & $=-\mathcal{M}_{0,0,\minus1,2}$ &
60: & D1: & $\mathcal{M}_{\minus2,1,\minus1,1}$ & $=\mathcal{M}_{1,\minus2,1,\minus1}$ & $=-\mathcal{M}_{\minus1,1,\minus2,1}$ & $=-\mathcal{M}_{1,\minus1,1,\minus2}$\\
16: & B: & $\mathcal{M}_{0,1,1,1}$ & $=\mathcal{M}_{1,0,1,1}$ & $=-\mathcal{M}_{1,1,0,1}$ & $=-\mathcal{M}_{1,1,1,0}$ &
61: & D1: & $\mathcal{M}_{\minus2,1,1,\minus1}$ & $=\mathcal{M}_{1,\minus2,\minus1,1}$ & $=-\mathcal{M}_{1,\minus1,\minus2,1}$ & $=-\mathcal{M}_{\minus1,1,1,\minus2}$\\
17: & D1: & $\mathcal{M}_{1,1,0,\minus1}$ & $=\mathcal{M}_{1,1,\minus1,0}$ & $=-\mathcal{M}_{0,\minus1,1,1}$ & $=-\mathcal{M}_{\minus1,0,1,1}$ &
62: & B: & $\mathcal{M}_{\minus2,1,1,1}$ & $=\mathcal{M}_{1,\minus2,1,1}$ & $=-\mathcal{M}_{1,1,\minus2,1}$ & $=-\mathcal{M}_{1,1,1,\minus2}$\\
\cline{7-12}
18: & D1: & $\mathcal{M}_{1,0,1,\minus1}$ & $=\mathcal{M}_{0,1,\minus1,1}$ & $=-\mathcal{M}_{1,\minus1,1,0}$ & $=-\mathcal{M}_{\minus1,1,0,1}$ &
& \multicolumn{4}{c}{$E^2$}&\\
\cline{7-12}
19: & D1: & $\mathcal{M}_{1,0,\minus1,1}$ & $=\mathcal{M}_{0,1,1,\minus1}$ & $=-\mathcal{M}_{\minus1,1,1,0}$ & $=-\mathcal{M}_{1,\minus1,0,1}$ &
63: & B: & $\mathcal{M}_{0,2,2,2}$ & $=\mathcal{M}_{2,0,2,2}$ & $=\mathcal{M}_{2,2,0,2}$ & $=\mathcal{M}_{2,2,2,0}$\\
\cline{1-6}
& \multicolumn{4}{c}{$E^4$} & &
64: & C: & $\mathcal{M}_{2,2,1,1}$ && $=\mathcal{M}_{1,1,2,2}$ & \\
\cline{1-6}
20: & C: & $\mathcal{M}_{2,2,0,0}$ && $=\mathcal{M}_{0,0,2,2}$ & &
65: & C: & $\mathcal{M}_{2,1,2,1}$ & $=\mathcal{M}_{1,2,1,2}$ & &\\
21: & C: & $\mathcal{M}_{2,0,2,0}$ & $=\mathcal{M}_{0,2,0,2}$ & & &
66: & C: & $\mathcal{M}_{2,1,1,2}$ & $=\mathcal{M}_{1,2,2,1}$ & &\\
22: & C: & $\mathcal{M}_{2,0,0,2}$ & $=\mathcal{M}_{0,2,2,0}$ & & &
67: & D2: & $\mathcal{M}_{2,2,1,\minus1}$ & $=\mathcal{M}_{2,2,\minus1,1}$ & $=\mathcal{M}_{1,\minus1,2,2}$ & $=\mathcal{M}_{\minus1,1,2,2}$\\
23: & D1: & $\mathcal{M}_{2,1,1,0}$ & $=\mathcal{M}_{1,2,0,1}$ & $=\mathcal{M}_{1,0,2,1}$ & $=\mathcal{M}_{0,1,1,2}$ &
68: & D2: & $\mathcal{M}_{2,1,2,\minus1}$ & $=\mathcal{M}_{1,2,\minus1,2}$ & $=\mathcal{M}_{2,\minus1,2,1}$ & $=\mathcal{M}_{\minus1,2,1,2}$\\
24: & D1: & $\mathcal{M}_{2,1,0,1}$ & $=\mathcal{M}_{1,2,1,0}$ & $=\mathcal{M}_{0,1,2,1}$ & $=\mathcal{M}_{1,0,1,2}$ &
69: & D2: & $\mathcal{M}_{2,1,\minus1,2}$ & $=\mathcal{M}_{1,2,2,\minus1}$ & $=\mathcal{M}_{\minus1,2,2,1}$ & $=\mathcal{M}_{2,\minus1,1,2}$\\
25: & D1: & $\mathcal{M}_{2,0,1,1}$ & $=\mathcal{M}_{0,2,1,1}$ & $=\mathcal{M}_{1,1,2,0}$ & $=\mathcal{M}_{1,1,0,2}$ &
70: & D2: & $\mathcal{M}_{2,2,0,\minus2}$ & $=\mathcal{M}_{2,2,\minus2,0}$ & $=\mathcal{M}_{0,\minus2,2,2}$ & $=\mathcal{M}_{\minus2,0,2,2}$\\
26: & E: & $\mathcal{M}_{2,1,0,\minus1}$ & $=\mathcal{M}_{1,2,\minus1,0}$ & $=\mathcal{M}_{0,\minus1,2,1}$ & $=\mathcal{M}_{\minus1,0,1,2}$ &
71: & D2: & $\mathcal{M}_{2,0,2,\minus2}$ & $=\mathcal{M}_{0,2,\minus2,2}$ & $=\mathcal{M}_{2,\minus2,2,0}$ & $=\mathcal{M}_{\minus2,2,0,2}$\\
27: & E: & $\mathcal{M}_{2,1,\minus1,0}$ & $=\mathcal{M}_{1,2,0,\minus1}$ & $=\mathcal{M}_{\minus1,0,2,1}$ & $=\mathcal{M}_{0,\minus1,1,2}$ &
72: & D2: & $\mathcal{M}_{2,0,\minus2,2}$ & $=\mathcal{M}_{0,2,2,\minus2}$ & $=\mathcal{M}_{\minus2,2,2,0}$ & $=\mathcal{M}_{2,\minus2,0,2}$\\
28: & E: & $\mathcal{M}_{2,0,1,\minus1}$ & $=\mathcal{M}_{0,2,\minus1,1}$ & $=\mathcal{M}_{1,\minus1,2,0}$ & $=\mathcal{M}_{\minus1,1,0,2}$ &
73: & C: & $\mathcal{M}_{2,2,\minus1,\minus1}$ && $=\mathcal{M}_{\minus1,\minus1,2,2}$ & \\
29: & E: & $\mathcal{M}_{2,0,\minus1,1}$ & $=\mathcal{M}_{0,2,1,\minus1}$ & $=\mathcal{M}_{\minus1,1,2,0}$ & $=\mathcal{M}_{1,\minus1,0,2}$ &
74: & C: & $\mathcal{M}_{2,\minus1,2,\minus1}$ & $=\mathcal{M}_{\minus1,2,\minus1,2}$ & &\\
30: & E: & $\mathcal{M}_{2,\minus1,1,0}$ & $=\mathcal{M}_{\minus1,2,0,1}$ & $=\mathcal{M}_{1,0,2,\minus1}$ & $=\mathcal{M}_{0,1,\minus1,2}$ &
75: & C: & $\mathcal{M}_{2,\minus1,\minus1,2}$ & $=\mathcal{M}_{\minus1,2,2,\minus1}$ & &\\
31: & E: & $\mathcal{M}_{2,\minus1,0,1}$ & $=\mathcal{M}_{\minus1,2,1,0}$ & $=\mathcal{M}_{0,1,2,\minus1}$ & $=\mathcal{M}_{1,0,\minus1,2}$ &
76: & D1: & $\mathcal{M}_{2,1,1,\minus2}$ & $=\mathcal{M}_{1,2,\minus2,1}$ & $=\mathcal{M}_{1,\minus2,2,1}$ & $=\mathcal{M}_{\minus2,1,1,2}$\\
32: & D0: & $\mathcal{M}_{2,0,0,\minus2}$ & $=\mathcal{M}_{0,2,\minus2,0}$ & & &
77: & D1: & $\mathcal{M}_{2,1,\minus2,1}$ & $=\mathcal{M}_{1,2,1,\minus2}$ & $=\mathcal{M}_{\minus2,1,2,1}$ & $=\mathcal{M}_{1,\minus2,1,2}$\\
33: & D0: & $\mathcal{M}_{2,0,\minus2,0}$ & $=\mathcal{M}_{0,2,0,\minus2}$ & & &
78: & D1: & $\mathcal{M}_{2,\minus2,1,1}$ & $=\mathcal{M}_{\minus2,2,1,1}$ & $=\mathcal{M}_{1,1,2,\minus2}$ & $=\mathcal{M}_{1,1,\minus2,2}$\\
34: & D0: & $\mathcal{M}_{2,\minus2,0,0}$ && $=\mathcal{M}_{0,0,2,\minus2}$ & & 
79: & E: & $\mathcal{M}_{2,1,\minus1,\minus2}$ & $=\mathcal{M}_{1,2,\minus2,\minus1}$ & &\\
35: & D1: & $\mathcal{M}_{\minus2,0,1,1}$ & $=\mathcal{M}_{0,\minus2,1,1}$ & $=\mathcal{M}_{1,1,\minus2,0}$ & $=\mathcal{M}_{1,1,0,\minus2}$ &
80: & E: & $\mathcal{M}_{2,1,\minus2,\minus1}$ & $=\mathcal{M}_{1,2,\minus1,\minus2}$ & &\\
36: & D1: & $\mathcal{M}_{\minus2,1,0,1}$ & $=\mathcal{M}_{1,\minus2,1,0}$ & $=\mathcal{M}_{0,1,\minus2,1}$ & $=\mathcal{M}_{1,0,1,\minus2}$ &
81: & E: & $\mathcal{M}_{2,\minus1,1,\minus2}$ & $=\mathcal{M}_{\minus1,2,\minus2,1}$ & &\\
37: & D1: & $\mathcal{M}_{\minus2,1,1,0}$ & $=\mathcal{M}_{1,\minus2,0,1}$ & $=\mathcal{M}_{1,0,\minus2,1}$ & $=\mathcal{M}_{0,1,1,\minus2}$ &
82: & E: & $\mathcal{M}_{2,\minus1,\minus2,1}$ & $=\mathcal{M}_{\minus1,2,1,\minus2}$ & &\\
38: & A: & $\mathcal{M}_{1,1,1,1}$ & & & &
83: & E: & $\mathcal{M}_{2,\minus2,1,\minus1}$ && $=\mathcal{M}_{1,\minus1,2,\minus2}$ & \\
39: & B: & $\mathcal{M}_{\minus1,1,1,1}$ & $=\mathcal{M}_{1,\minus1,1,1}$ & $=\mathcal{M}_{1,1,\minus1,1}$ & $=\mathcal{M}_{1,1,1,\minus1}$ &
84: & E: & $\mathcal{M}_{2,\minus2,\minus1,1}$ && $=\mathcal{M}_{\minus1,1,2,\minus2}$ & \\
\cline{7-12}
40: & C: & $\mathcal{M}_{1,1,\minus1,\minus1}$ & & & &
& \multicolumn{4}{c}{$E^1$}&\\
\cline{7-12}
41: & C: & $\mathcal{M}_{1,\minus1,1,\minus1}$ & & & &
85: & B: & $\mathcal{M}_{1,2,2,2}$ & $=\mathcal{M}_{2,1,2,2}$ & $=-\mathcal{M}_{2,2,1,2}$ & $=-\mathcal{M}_{2,2,2,1}$\\
42: & C: & $\mathcal{M}_{1,\minus1,\minus1,1}$ & & & &
86: & B: & $\mathcal{M}_{\minus1,2,2,2}$ & $=\mathcal{M}_{2,\minus1,2,2}$ & $=-\mathcal{M}_{2,2,\minus1,2}$ & $=-\mathcal{M}_{2,2,2,\minus1}$\\
\cline{1-6}
& \multicolumn{4}{c}{$E^0$} & &
87: & D2: & $\mathcal{M}_{2,1,2,\minus2}$ & $=\mathcal{M}_{1,2,\minus2,2}$ & $=-\mathcal{M}_{2,\minus2,2,1}$ & $=-\mathcal{M}_{\minus2,2,1,2}$\\
\cline{1-6}
93: & A: & $\mathcal{M}_{2,2,2,2}$ & & & &
88: & D2: & $\mathcal{M}_{2,2,1,\minus2}$ & $=\mathcal{M}_{2,2,\minus2,1}$ & $=-\mathcal{M}_{1,\minus2,2,2}$ & $=-\mathcal{M}_{\minus2,1,2,2}$\\
94: & B: & $\mathcal{M}_{\minus2,2,2,2}$ & $=\mathcal{M}_{2,\minus2,2,2}$ & $=\mathcal{M}_{2,2,\minus2,2}$ & $=\mathcal{M}_{2,2,2,\minus2}$ &
89: & D2: & $\mathcal{M}_{2,1,\minus2,2}$ & $=\mathcal{M}_{1,2,2,\minus2}$ & $=-\mathcal{M}_{\minus2,2,2,1}$ & $=-\mathcal{M}_{2,\minus2,1,2}$\\
95: & C: & $\mathcal{M}_{2,2,\minus2,\minus2}$ & & & &
90: & D2: & $\mathcal{M}_{2,2,\minus1,\minus2}$ & $=\mathcal{M}_{2,2,\minus2,\minus1}$ & $=-\mathcal{M}_{\minus1,\minus2,2,2}$ & $=-\mathcal{M}_{\minus2,\minus1,2,2}$\\
96: & C: & $\mathcal{M}_{2,\minus2,2,\minus2}$ & & & &
91: & D2: & $\mathcal{M}_{2,\minus1,2,\minus2}$ & $=\mathcal{M}_{\minus1,2,\minus2,2}$ & $=-\mathcal{M}_{2,\minus2,2,\minus1}$ & $=-\mathcal{M}_{\minus2,2,\minus1,2}$\\
97: & C: & $\mathcal{M}_{2,\minus2,\minus2,2}$ & & & &
92: & D2: & $\mathcal{M}_{2,\minus1,\minus2,2}$ & $=\mathcal{M}_{\minus1,2,2,\minus2}$ & $=-\mathcal{M}_{\minus2,2,2,\minus1}$ & $=-\mathcal{M}_{2,\minus2,\minus1,2}$\\
\hline\hline
\end{tabular}
\end{center}
\caption{\label{tab:97 amps}A list of the 97 linearly independent amplitudes along with the amplitudes that are equivalent to them by Eq.~\eqref{eq:flip equivalence}.}
\end{table*}

However, these $5^4=625$ helicity combinations are not all unique.  They satisfy two types of equalities:
\begin{equation}
\mathcal{M}_{\sigma_1\sigma_2\sigma_3\sigma_4} = \left(-1\right)^{\sigma_1+\sigma_2+\sigma_3+\sigma_4}\mathcal{M}_{-\sigma_1-\sigma_2-\sigma_3-\sigma_4}
\label{eq:opposite helicity equivalence}
\end{equation}
and
\begin{eqnarray}
\mathcal{M}_{\sigma_1\sigma_2\sigma_3\sigma_4} &=& \mathcal{M}_{\sigma_2\sigma_1\sigma_4\sigma_3}\nonumber\\
&=& \left(-1\right)^{\sigma_1+\sigma_2+\sigma_3+\sigma_4}\mathcal{M}_{\sigma_3\sigma_4\sigma_1\sigma_2}\nonumber\\
&=& \left(-1\right)^{\sigma_1+\sigma_2+\sigma_3+\sigma_4}\mathcal{M}_{\sigma_4\sigma_3\sigma_2\sigma_1}
\label{eq:flip equivalence}
\end{eqnarray}
where $\sigma_1,\sigma_2$ and $\sigma_3,\sigma_4$ are the two incoming and two outgoing helicities, respectively.  These equivalencies are given by the symmetries of the $\mathcal{S}$-matrix under Lorentz transformations, including parity and time-reversal\footnote{For a complete discussion of the symmetries of the $\mathcal{S}$-matrix under Lorentz transformations, see Section 3.3 of \cite{Weinberg:1995mt}.}, since these symmetries are satisfied by our operators (see App.~\ref{sec:Feynman Rules}).  To begin with, we consider ordinary spatial rotations.  In the CM frame of our collision, a plane is defined by the momentum of the incoming and outgoing particles.  If we rotate by 180$^\circ$ around the axis perpendicular to this plane, it brings the process into another process that looks exactly the same except that helicities 1 and 2 are switched and separately helicities 3 and 4 are switched.  This gives the first equality of Eq.~\eqref{eq:flip equivalence}.  This also relates the second and third lines of Eq.~\eqref{eq:flip equivalence}.

We next consider time-reversal, under which the three-momenta and the spin components flip sign but the helicity remains the same.  The direction of the collision also reverses so that particles 3 and 4 are colliding to create particles 1 and 2.  If we rotate around the axis perpendicular to the collision plane by $180^\circ-\theta$ followed by a rotation of $180^\circ$ around $\vec{p}_3$, the process looks exactly the same but with helicities 1 and 3 switched and separately 2 and 4 switched.  This gives the equality between the first and second line of Eq.~\eqref{eq:flip equivalence}.  

We finally consider parity, under which the three-momenta of the particles flip sign but the spin-components do not change.  As a result, the helicities flip sign.  Furthermore, since we are flipping the three-momenta, the process looks the same except that we have interchanged helicities 1 and 2 and separately 3 and 4.  However, since we have already established in Eq.~\eqref{eq:flip equivalence} that such a switch of helicities does not effect the amplitude, we can move them back to their original positions.  So, combining these, gives Eq.~\eqref{eq:opposite helicity equivalence}.  

These two sets of equalities reduce the initial $625$ helicity combinations down to a total of $97$ linearly independent amplitudes, which are enumerated in Table~\ref{tab:97 amps}.  Along with the 97 linearly independent amplitudes, we list the amplitudes that are equivalent to them by Eq.~\eqref{eq:flip equivalence}, but not those related by Eq.~\eqref{eq:opposite helicity equivalence}.  There are a total of 313 amplitudes shown in Table~\ref{tab:97 amps}.  All of them except $\mathcal{M}_{0,0,0,0}$ are related to another amplitude by Eq.~\eqref{eq:opposite helicity equivalence} giving 625 ($312\times2+1$) amplitudes as expected.   

The 313 amplitudes shown in Table~\ref{tab:97 amps} are organized according to the energy growth of the polarization vectors in the amplitude.  As seen in App.~\ref{app:polarization vectors}, the polarization vectors $\epsilon^{\mu\nu}_\sigma(p)$ grow like $E^2$ if $\sigma=0$, like $E^1$ if $\sigma=\pm1$ and like $E^0$ if $\sigma=\pm2$.  They are further categorized according to the number of helicities that are equal as we will now explain:
\begin{enumerate}[A:]
\item If all four helicities are the same ($\sigma=\sigma_1=\sigma_2=\sigma_3=\sigma_4$), there are only 3 linearly independent values for $\sigma$ (say $2,1$ and $0$).  These are labeled by A in Table~\ref{tab:97 amps}.  The other two choices for $\sigma$ ($-1$ and $-2$) are related to these by use of Eq.~\eqref{eq:opposite helicity equivalence}. 
\item If three of the helicities are the same, then Eq.~\eqref{eq:flip equivalence} can be used to move the helicity that is unique to the fourth position which we will call $\sigma_4$ ($\sigma=\sigma_1=\sigma_2=\sigma_3\neq\sigma_4$).  If $\sigma>0$, there are two choices (2 or 1) for $\sigma$ and four remaining choices for $\sigma_4$ for a total of 8 ($2\times4$) helicity combinations.  If $\sigma=0$, there are two choices with $\sigma_4>0$.  All other possibilities are the same as one of these by use of Eq.~\eqref{eq:opposite helicity equivalence}.  This gives a total of 10 linearly independent helicity combinations, which are labeled with a B in Table~\ref{tab:97 amps}.
\item If the helicities split into two pairs of identical helicities (call them $\sigma_a>\sigma_b$) we need to analyze the possibilities for each pair.  If $\sigma_a=2$, then $\sigma_b$ can be any of the remaining helicities, so there are 4 possibilities.  If $\sigma_a=1$, then $\sigma_b$ can be 0 or -1, so there are 2 possibilities.  The other possibilities are covered by Eq.~\eqref{eq:opposite helicity equivalence}.  This gives a total of $4+2=6$ different choices for $\sigma_a$ and $\sigma_b$.  Now, we concern ourselves with the positions of the helicities.  Using Eq.~\eqref{eq:flip equivalence}, one of the $\sigma_a$ can be moved to position 1.  Then, there are three linearly independent orders given by $\mathcal{M}_{\sigma_a\sigma_a\sigma_b\sigma_b}$, $\mathcal{M}_{\sigma_a\sigma_b\sigma_a\sigma_b}$ and $\mathcal{M}_{\sigma_a\sigma_b\sigma_b\sigma_a}$ which can not be related to each other using Eq.~\eqref{eq:flip equivalence}.  This gives a total of 18 ($6\times3$) linearly independent amplitudes, which are marked with a C in Table~\ref{tab:97 amps}.   
\item If there is one pair of identical helicities with the other two helicities being unique, there are three choices for the identical pair.  Let's call it $\sigma_a$ and it can be $2, 1$ or $0$.  The other two helicities for $\sigma_a$, namely $-1$ and $-2$, are related to these by Eq.~\eqref{eq:opposite helicity equivalence}.  We will call the other unique helicities $\sigma_b$ and $\sigma_c$, where $\sigma_b>\sigma_c$.  Once these are chosen, there are three linearly independent orders that can not be related to one another by use of Eq.~\eqref{eq:flip equivalence}, namely $\mathcal{M}_{\sigma_a,\sigma_a,\sigma_b,\sigma_c}$, $\mathcal{M}_{\sigma_a,\sigma_b,\sigma_a,\sigma_c}$ and $\mathcal{M}_{\sigma_a,\sigma_b,\sigma_c,\sigma_a}$.  We describe each possibility in turn:
\begin{enumerate}[1:]
\item[2:] If $\sigma_a=2$, then $\sigma_b$ can be $1, 0$ or $-1$ with $\sigma_c$ being smaller for a total of 6 ($3+2+1$) possibilities.  Combining this with the three independent orderings gives a total of 18 ($6\times3$) linearly independent amplitudes, which are marked with a D2 in Table~\ref{tab:97 amps}.
\item[1:] If $\sigma_a=1$, then $\sigma_b$ can be $2, 0$ or $-1$ with the final helicity being smaller for a total of 6 ($3+2+1$) possibilities.  Combining this with the three independent orderings gives a total of 18 ($6\times3$) linearly independent amplitudes, which are marked with a D1 in Table~\ref{tab:97 amps}.
\item[0:] If $\sigma_a=0$, then $\sigma_b$ can be either $2$ or $1$ ($-1$ and $-2$ are related to these by use of Eq.~\eqref{eq:opposite helicity equivalence}).  If $\sigma_b=2$, then $\sigma_c$ can be $1, -1$ or $-2$.  If $\sigma_b=1$, then, $\sigma_c$ can only be $-1$.  This gives a total of 4 ($3+1$) choices for the other two helicities.  Combining this with the three independent orderings gives a total of 12 ($4\times3$) linearly independent amplitudes.  These amplitudes are marked with a D0 in Table~\ref{tab:97 amps}.
\end{enumerate}  
\item If all four helicities are unique, we label them in decreasing order as $\sigma_a>\sigma_b>\sigma_c>\sigma_d$.  The helicity that is not included could be $0, -1$ or $-2$, which gives 3 choices.  The other possibilities ($2$ and $1$) can be related to these by use of Eq.~\eqref{eq:opposite helicity equivalence}.  Once the helicities are chosen, we can always move $\sigma_a$ to position 1 using Eq.~\eqref{eq:flip equivalence}.  Then, there are 6 ($3!$) ways of placing the other three helicities.  This gives a total of 18 ($3\times6$) linearly independent amplitudes, which are marked with an E in Table~\ref{tab:97 amps}.
\end{enumerate}

\begin{widetext}
\subsection{\label{app:M0000}Expanded $\mathbf{\mathcal{M}_{0000}}$ Diagrams}
In this section, we use the Feynman rules given in App.~\ref{sec:Feynman Rules} to explicitly calculate the $2\to2$ scattering amplitude with the helicities $\sigma_1=\sigma_2=\sigma_3=\sigma_4=0$.  These are the diagrams that give the largest energy growth at high energy and, therefore, require the greatest cancellations.  We explicitly expand each diagram in energy and use the standard propagators ($F=1$, $C_{12}=\frac{1}{2}$, $C_3=-\frac{1}{3}$, $C_{47}=-\frac{1}{2}$, $C_{89}=\frac{1}{3}$ and $C_{10}=\frac{2}{3}$) for compactness and clarity in these expressions.

Using the Feynman rules given in Eqs.~\eqref{eq:g_2222^1} and \eqref{eq:g_2222^2} gives the 4-point diagram\\
\begin{minipage}{7in}
\begin{minipage}{1.75in}
\includegraphics[scale=0.4]{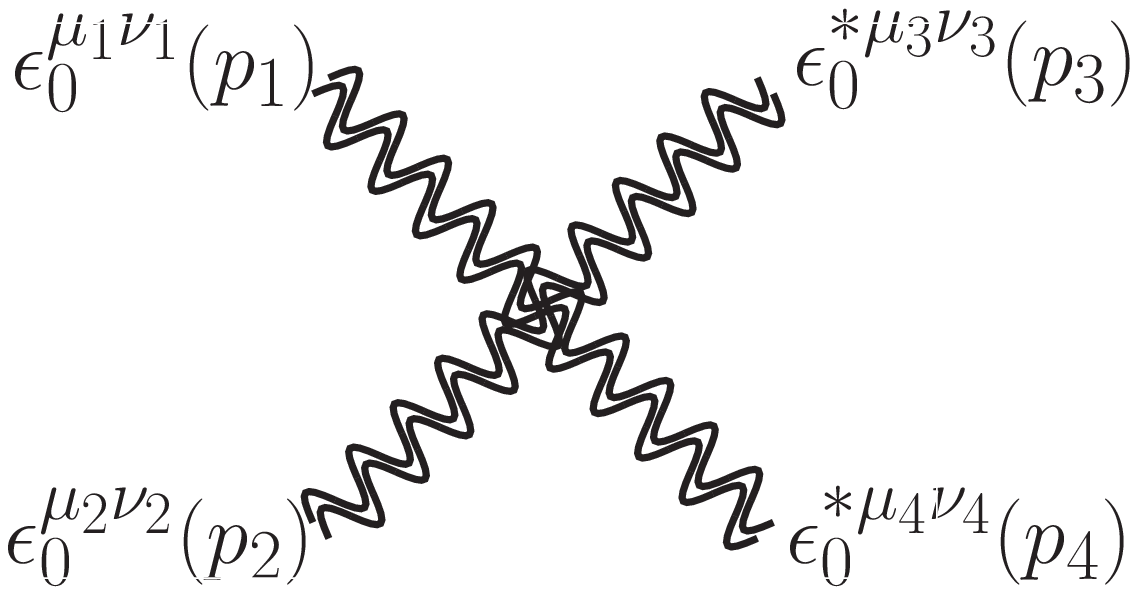}
\end{minipage}
\hfill
\begin{minipage}{5.25in}
\begin{eqnarray}
&&\mathcal{M}^4_{0000} =
\frac{E^8}{M_2^8}\left(2g^{(1)}_{2222}+g^{(2)}_{2222}\right)
\left(\frac{32}{9}\cos^4\theta+\frac{64}{3}\cos^2+32\right)\nonumber
\end{eqnarray}
\end{minipage}
\begin{eqnarray}
&&+\frac{E^6}{M_2^6}\Bigg[\frac{64}{9}\left(2g^{(1)}_{2222}+g^{(2)}_{2222}\right)\cos^4\theta
-\frac{256}{9}\left(3g^{(1)}_{2222}+g^{(2)}_{2222}\right)\cos^2\theta
-\frac{64}{9}\left(22g^{(1)}_{2222}+13g^{(2)}_{2222}\right)\Bigg]\nonumber\\
&&+\frac{E^4}{M_2^4}\Bigg[\frac{16}{3}\left(2g^{(1)}_{2222}+g^{(2)}_{2222}\right)\cos^4\theta
+\frac{64}{9}g^{(1)}_{2222}\cos^2\theta
+\frac{16}{9}\left(102g^{(1)}_{2222}+61g^{(2)}_{2222}\right)\Bigg]\nonumber\\
&&+\frac{E^2}{M_2^2}\Bigg[\frac{16}{9}\left(2g^{(1)}_{2222}+g^{(2)}_{2222}\right)\cos^4\theta
+\frac{16}{9}\left(4g^{(1)}_{2222}-3g^{(2)}_{2222}\right)\cos^2\theta
-\frac{32}{3}\left(9g^{(1)}_{2222}+5g^{(2)}_{2222}\right)\Bigg]\nonumber\\
&&+\Bigg[
\frac{2}{9}\cos^4\theta\left(2g^{(1)}_{2222}+g^{(2)}_{2222}\right)
+\frac{4}{9}\cos^2\theta\left(10g^{(1)}_{2222}+g^{(2)}_{2222}\right)
+\frac{2}{9}\left(86g^{(1)}_{2222}+51g^{(2)}_{2222}\right)
\Bigg]
\end{eqnarray}
\end{minipage}

Using the Feynman rules given in Eqs.~\eqref{eq:g_222^1} and \eqref{eq:g_222^2} gives the S-channel diagram\\
\begin{minipage}{7in}
\begin{minipage}{2in}
\includegraphics[scale=0.4]{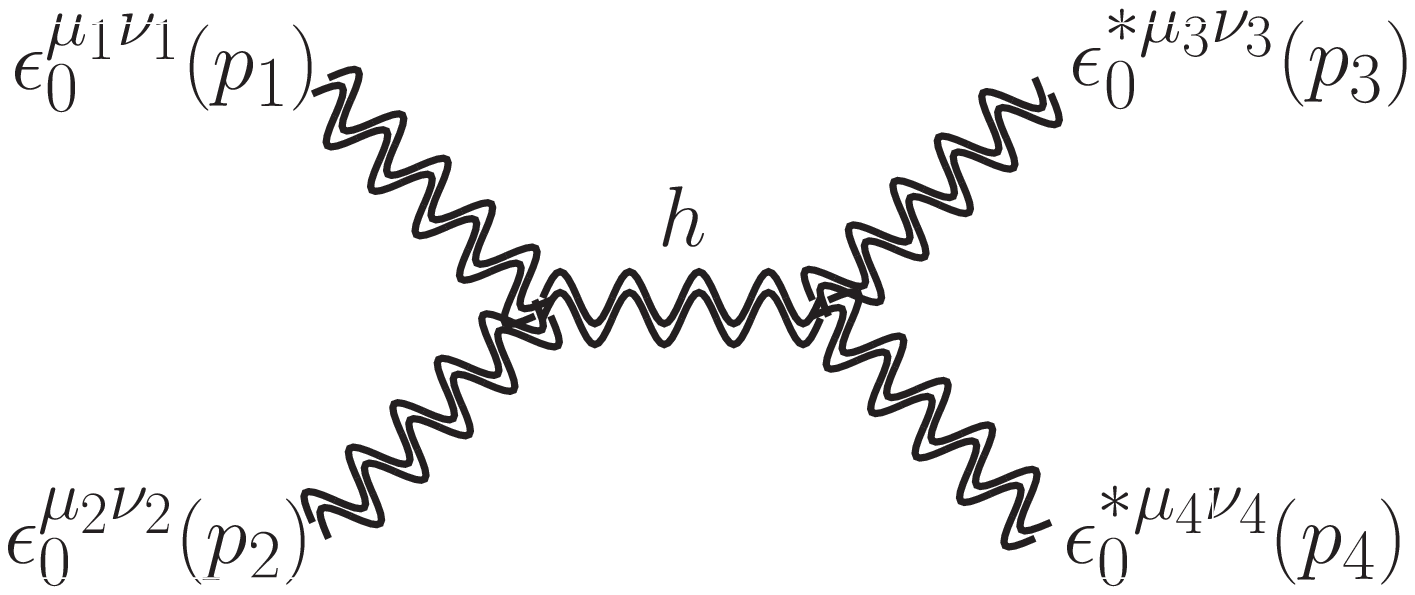}
\end{minipage}
\hfill
\begin{minipage}{5in}
\begin{eqnarray}
&&\mathcal{M}^{Sh}_{0000} =
-\frac{E^{10}}{M_2^{12}}\frac{512}{27}\left(2g^{(1)}_{222}+3g^{(2)}_{222}\right)^2
+\frac{E^8}{M_2^{10}}\frac{512}{9}\left(2g^{(1)}_{222}+3g^{(2)}_{222}\right)\left(g^{(1)}_{222}+3g^{(2)}_{222}\right)\nonumber\\
&&\quad\quad\quad+\frac{E^6}{M_2^8}\Bigg[-16g^{(2)\ 2}_{222}\cos^2\theta
-\frac{16}{9}\left(32g^{(1)\ 2}_{222}+240g^{(1)}_{222}g^{(2)}_{222}+333g^{(2)\ 2}_{222}\right)\Bigg]\nonumber
\end{eqnarray}
\end{minipage}
\begin{eqnarray}
&&+\frac{E^4}{M_2^6}\Bigg[12g^{(2)\ 2}_{222}\cos^2\theta
-\frac{4}{27}\left(64g^{(1)\ 2}_{222}-960g^{(1)}_{222}g^{(2)}_{222}-2205g^{(2)\ 2}_{222}\right)\Bigg]
+\frac{E^2}{M_2^4}\Bigg[3g^{(2)\ 2}_{222}\cos^2\theta+\frac{1}{9}\left(160g^{(1)\ 2}_{222}-729g^{(2)\ 2}_{222}\right)\Bigg]\nonumber\\
&&-\Bigg[
\frac{5}{4}g^{(2)\ 2}_{222}\cos^2\theta
+\frac{1}{12}\left(64g^{(1)\ 2}_{222}+96g^{(1)}_{222}g^{(2)}_{222}-69g^{(2)\ 2}_{222}\right)
\Bigg]
+\mathcal{O}\left(E^{-2}\right)
\end{eqnarray}
\end{minipage}
and the T+U-channel diagrams\\
\begin{minipage}{7in}
\begin{minipage}{2in}
\includegraphics[scale=0.4]{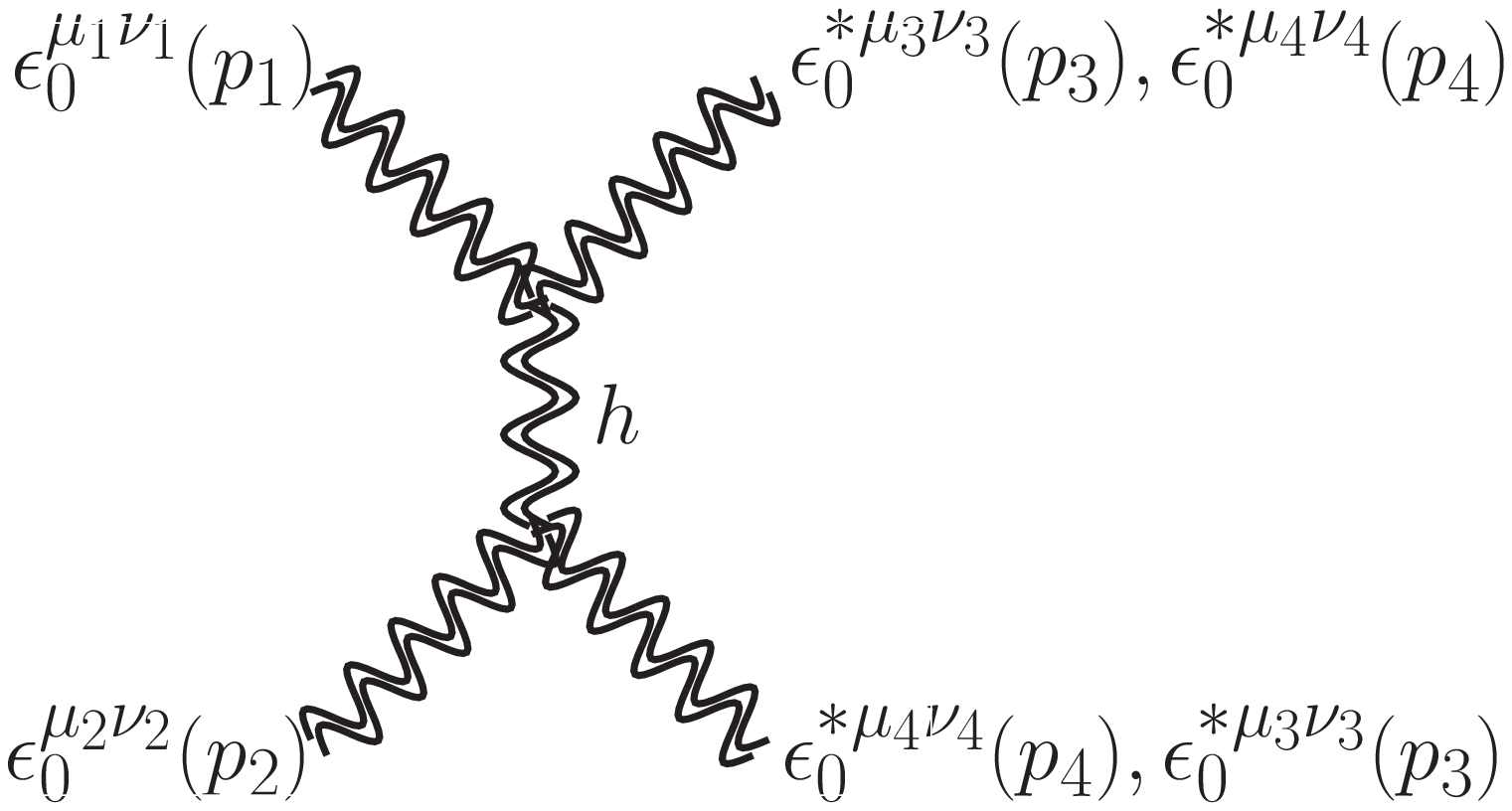}
\end{minipage}
\hfill
\begin{minipage}{4.75in}
\begin{eqnarray}
&&\mathcal{M}^{TUh}_{0000} =
\frac{E^{10}}{M_2^{12}}\frac{32}{27}\left(2g^{(1)}_{222}+3g^{(2)}_{222}\right)^2\left(5\cos^4\theta+10\cos^2\theta+1\right)\nonumber\\
&&-\frac{E^8}{M_2^{10}}\frac{32}{27}\left(2g^{(1)}_{222}+3g^{(2)}_{222}\right)\Bigg[\left(8g^{(1)}_{222}+3g^{(2)}_{222}\right)\cos^4\theta\nonumber\\
&&\hspace{1in}+2\left(44g^{(1)}_{222}+39g^{(2)}_{222}\right)\cos^2\theta
+\left(16g^{(1)}_{222}+15g^{(2)}_{222}\right)\Bigg]\nonumber
\end{eqnarray}
\end{minipage}
\begin{eqnarray}
&&+\frac{E^6}{M_2^8}\frac{8}{27}\Bigg[-4\left(2g^{(1)}_{222}+3g^{(2)}_{222}\right)\left(11g^{(1)}_{222}+12g^{(2)}_{222}\right)\cos^4\theta
+\left(928g^{(1)\ 2}_{222}+1704g^{(1)}_{222}g^{(2)}_{222}+603g^{(2)\ 2}_{222}\right)\cos^2\theta\nonumber\\
&&\hspace{0.5in}+\left(376g^{(1)\ 2}_{222}+732g^{(1)}_{222}g^{(2)}_{222}+441g^{(2)\ 2}_{222}\right)\Bigg]\nonumber\\
&&-\frac{E^4}{M_2^6}\frac{4}{27}\Bigg[4\left(2g^{(1)}_{222}+3g^{(2)}_{222}\right)^2\cos^4\theta
+\left(736g^{(1)\ 2}_{222}+768g^{(1)}_{222}g^{(2)}_{222}+333g^{(2)\ 2}_{222}\right)\cos^2\theta\nonumber\\
&&\hspace{0.5in}+\left(976g^{(1)\ 2}_{222}+1488g^{(1)}_{222}g^{(2)}_{222}+981g^{(2)\ 2}_{222}\right)
\Bigg]\nonumber\\
&&+\frac{E^2}{M_2^4}\frac{2}{3\left(\cos^2\theta-1\right)}\Bigg[
\frac{1}{9}\left(2g^{(1)}_{222}+3g^{(2)}_{222}\right)\left(2g^{(1)}_{222}+21g^{(2)}_{222}\right)\cos^6\theta
+\frac{2}{3}\left(38g^{(1)\ 2}_{222}-24g^{(1)}_{222}g^{(2)}_{222}-117g^{(2)\ 2}_{222}\right)\cos^4\theta\nonumber\\
&&\hspace{0.5in}+4\left(23g^{(1)\ 2}_{222}+36g^{(1)}_{222}g^{(2)}_{222}-6g^{(2)\ 2}_{222}\right)\cos^2\theta
-\frac{1}{9}\left(1060g^{(1)\ 2}_{222}+1200g^{(1)}_{222}g^{(2)}_{222}+1089g^{(2)\ 2}_{222}\right)
\Bigg]\nonumber\\
&&-\frac{1}{M_2^2}\frac{1}{27\left(\cos^2\theta-1\right)^2}\Bigg[
2\left(2g^{(1)}_{222}+3g^{(2)}_{222}\right)\left(4g^{(1)}_{222}-3g^{(2)}_{222}\right)\cos^8\theta
+4\left(32g^{(1)\ 2}_{222}+24g^{(1)}_{222}g^{(2)}_{222}+261g^{(2)\ 2}_{222}\right)\cos^6\theta\nonumber\\
&&\hspace{0.5in}+\left(96g^{(1)\ 2}_{222}+72g^{(1)}_{222}g^{(2)}_{222}+1269g^{(2)\ 2}_{222}\right)\cos^4\theta
+\left(-640g^{(1)\ 2}_{222}-480g^{(1)}_{222}g^{(2)}_{222}+1206g^{(2)\ 2}_{222}\right)\cos^2\theta\nonumber\\
&&\hspace{0.5in}+\left(400g^{(1)\ 2}_{222}+300g^{(1)}_{222}g^{(2)}_{222}+387g^{(2)\ 2}_{222}\right)
\Bigg]
+\mathcal{O}\left(E^{-2}\right)
\end{eqnarray}
\end{minipage}

Using the Feynman rules given in Eqs.~\eqref{eq:g_222i^1} and \eqref{eq:g_222i^2} gives the S-channel diagram\\
\begin{minipage}{7in}
\begin{minipage}{2in}
\includegraphics[scale=0.4]{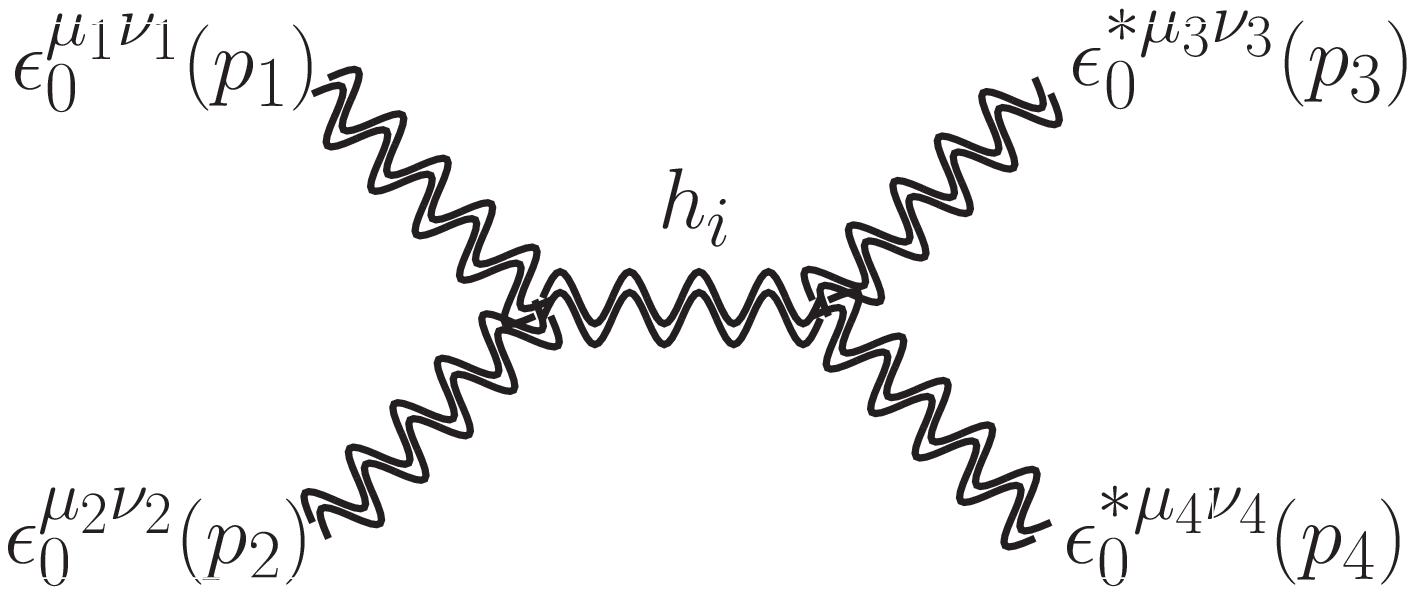}
\end{minipage}
\begin{minipage}{4in}
\begin{eqnarray}
&&\mathcal{M}^{Shi}_{0000} =
-\frac{E^{10}}{M_2^{12}}\frac{512M_2^4}{27M_{2i}^4}\left(2g^{(1)}_{222i}+g^{(2)}_{222i}\right)^2\nonumber\\
&&\hspace{0.5in}+\frac{E^8}{M_2^{10}}\frac{512}{27}\left(2g^{(1)}_{222i}+g^{(2)}_{222i}\right)\Bigg[
\frac{M_2^4}{M_{2i}^4}\left(4g^{(1)}_{222i}+3g^{(2)}_{222i}\right)
-\frac{M_2^2}{M_{2i}^2}g^{(1)}_{222i}\Bigg]\nonumber
\end{eqnarray}
\end{minipage}
\begin{eqnarray}
&&+\frac{E^6}{M_2^8}\frac{16}{27}\Bigg[
-3g^{(2)\ 2}_{222i}\cos^2\theta-8\frac{M_2^4}{M_{2i}^4}\left(28g^{(1)\ 2}_{222i}+38g^{(1)}_{222i}g^{(2)}_{222i}+13g^{(2)\ 2}_{222i}\right)
+8\frac{M_2^2}{M_{2i}^2}\left(16g^{(1)\ 2}_{222i}+8g^{(1)}_{222i}g^{(2)}_{222i}-g^{(2)\ 2}_{222i}\right)\nonumber\\
&&\hspace{0.5in}+g^{(2)\ 2}_{222i}\Bigg]\nonumber\\
&&+\frac{E^4}{M_2^6}\frac{4}{27}\Bigg[
3g^{(2)\ 2}_{222i}\cos^2\theta\left(4-\frac{M_{2i}^2}{M_2^2}\right)
+\frac{32M_2^4}{M_{2i}^4}\left(3g^{(1)}_{222i}+2g^{(2)}_{222i}\right)\left(4g^{(1)}_{222i}+3g^{(2)}_{222i}\right)\nonumber\\
&&\hspace{0.5in}-\frac{56M_2^2}{M_{2i}^2}\left(8g^{(1)\ 2}_{222i}+4g^{(1)}_{222i}g^{(2)}_{222i}-g^{(2)\ 2}_{222i}\right)
-4g^{(2)\ 2}_{222i}
+\frac{M_{2i}^2}{M_2^2}g^{(2)\ 2}_{222i}
\Bigg]\nonumber\\
&&+\frac{E^2}{M_2^4}\frac{1}{27}\Bigg[
3g^{(2)\ 2}_{222i}\cos^2\theta\left(4\frac{M_{2i}^2}{M_2^2}-3\frac{M_{2i}^4}{M_2^4}\right)
-32\frac{M_2^4}{M_{2i}^4}\left(3g^{(1)}_{222i}+2g^{(2)}_{222i}\right)^2
+16\frac{M_2^2}{M_{2i}^2}\left(48g^{(1)\ 2}_{222i}+24g^{(1)}_{222i}g^{(2)}_{222i}-7g^{(2)\ 2}_{222i}\right)\nonumber\\
&&\hspace{0.5in}-4\frac{M_{2i}^2}{M_2^2}g^{(2)\ 2}_{222i}
+\frac{M_{2i}^4}{M_2^4}g^{(2)\ 2}_{222i}
\Bigg]\nonumber\\
&&+\frac{1}{M_2^2}\frac{1}{108}\Bigg[
-3g^{(2)\ 2}_{222i}\cos^2\theta\left(8-4\frac{M_{2i}^4}{M_2^4}+\frac{M_{2i}^6}{M_2^6}\right)
-32\frac{M_2^2}{M_{2i}^2}\left(6g^{(1)}_{222i}-g^{(2)}_{222i}\right)\left(3g^{(1)}_{222i}+2g^{(2)}_{222i}\right)\nonumber\\
&&\hspace{0.5in}+g^{(2)\ 2}_{222i}\left(8-4\frac{M_{2i}^4}{M_2^4}+\frac{M_{2i}^6}{M_2^6}\right)
\Bigg]
+\mathcal{O}\left(E^{-2}\right)
\end{eqnarray}
\end{minipage}
and the T+U-channel diagrams\\
\begin{minipage}{7in}
\begin{minipage}{2in}
\includegraphics[scale=0.4]{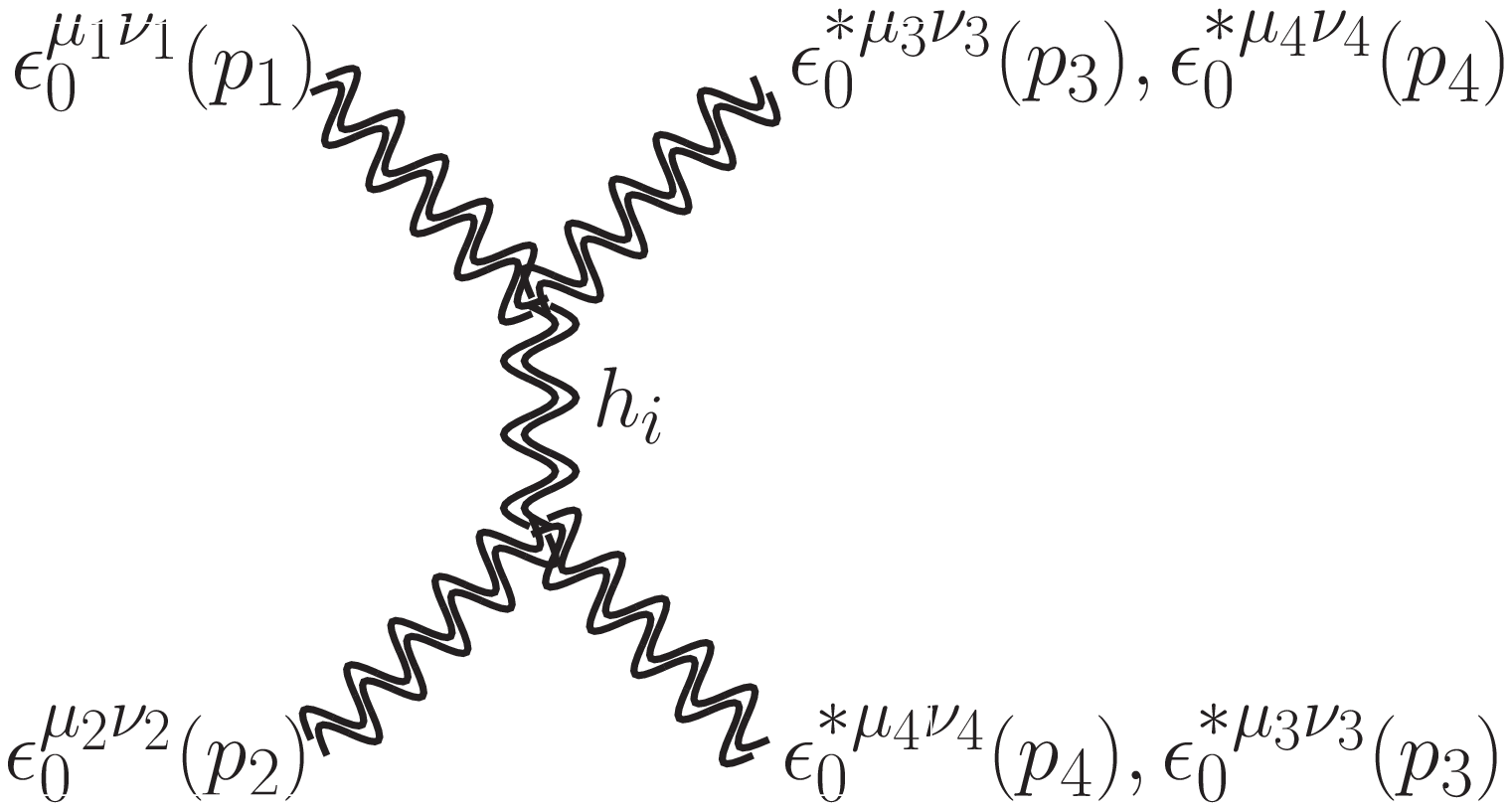}
\end{minipage}
\hfill
\begin{minipage}{4.5in}
\begin{eqnarray}
&&\mathcal{M}^{TUhi}_{0000} =
\frac{E^{10}}{M_2^{12}}\frac{32}{27}\frac{M_2^4}{M_{2i}^4}\left(2g^{(1)}_{222i}+g^{(2)}_{222i}\right)^2\left(5\cos^4\theta+10\cos^2\theta+1\right)\nonumber\\
&&+\frac{E^8M_{2i}^2}{M_2^{8}}\frac{32}{27}\left(2g^{(1)}_{222i}+g^{(2)}_{222i}\right)\Bigg[
\cos^4\theta\left(-\frac{M_2^2}{M_{2i}^2}\left(6g^{(1)}_{222i}+g^{(2)}_{222i}\right)-2g^{(1)}_{222i}\right)\nonumber\\
&&\hspace{0.5in}-2\cos^2\theta\left(\frac{M_2^2}{M_{2i}^2}\left(38g^{(1)}_{222i}+13g^{(2)}_{222i}\right)-6g^{(1)}_{222i}\right)\nonumber\\
&&\hspace{0.5in}-\frac{M_2^2}{M_{2i}^2}\left(14g^{(1)}_{222i}+5g^{(2)}_{222i}\right)-2g^{(1)}_{222i}
\Bigg]\nonumber
\end{eqnarray}
\end{minipage}
\begin{eqnarray}
&&+\frac{E^6}{M_2^8}\frac{8}{27}\Bigg[
-4\cos^4\theta\left(2g^{(1)}_{222i}+g^{(2)}_{222i}\right)\left(\frac{M_2^4}{M_{2i}^4}\left(7g^{(1)}_{222i}+4g^{(2)}_{222i}\right)+4\frac{M_2^2}{M_{2i}^2}g^{(1)}_{222i}\right)\nonumber\\
&&\hspace{0.5in}+\cos^2\theta\left(8\frac{M_2^4}{M_{2i}^4}\left(92g^{(1)\ 2}_{222i}+59g^{(1)}_{222i}g^{(2)}_{222i}+8g^{(2)\ 2}_{222i}\right)+12\frac{M_2^2}{M_{2i}^2}\left(4g^{(1)}_{222i}+g^{(2)}_{222i}\right)^2-9g^{(2)\ 2}_{222i}\right)\nonumber\\
&&\hspace{0.5in}+4\frac{M_2^4}{M_{2i}^4}\left(70g^{(1)\ 2}_{222i}+49g^{(1)}_{222i}g^{(2)}_{222i}+8g^{(2)\ 2}_{222i}\right)
+4\frac{M_2^2}{M_{2i}^2}\left(24g^{(1)\ 2}_{222i}+12g^{(1)}_{222i}g^{(2)}_{222i}+g^{(2)\ 2}_{222i}\right)
+13g^{(2)\ 2}_{222i}
\Bigg]\nonumber\\
&&+\frac{E^4}{M_2^6}\frac{4}{27}\Bigg[
4\cos^4\theta\left(2g^{(1)}_{222i}+g^{(2)}_{222i}\right)\left(\frac{M_2^4}{M_{2i}^4}\left(4g^{(1)}_{222i}-g^{(2)}_{222i}\right)-6\frac{M_2^2}{M_{2i}^2}g^{(1)}_{222i}\right)\nonumber\\
&&\hspace{0.in}-\cos^2\theta\left(
4\frac{M_2^4}{M_{2i}^4}\left(176g^{(1)\ 2}_{222i}+60g^{(1)}_{222i}g^{(2)}_{222i}-g^{(2)\ 2}_{222i}\right)
+2\frac{M_2^2}{M_{2i}^2}\left(16g^{(1)\ 2}_{222i}+8g^{(1)}_{222i}g^{(2)}_{222i}+7g^{(2)\ 2}_{222i}\right)
+26g^{(2)\ 2}_{222i}
+\frac{M_{2i}^2}{M_2^2}g^{(2)}_{222i}
\right)\nonumber\\
&&\hspace{0.5in}+\frac{16}{3}\frac{M_2^4}{M_{2i}^4}\left(17g^{(1)}_{222i}+12g^{(2)}_{222i}\right)
+\frac{56}{3}\frac{M_2^2}{M_{2i}^2}\left(-2g^{(1)}_{222i}+g^{(2)}_{222i}\right)
-\frac{4}{3}g^{(2)}_{222i}
+\frac{1}{3}\frac{M_{2i}^2}{M_2^2}g^{(2)}_{222i}
\Bigg]\nonumber\\
&&+\frac{E^2}{M_2^4}\frac{2}{27}\frac{1}{\left(\cos^2\theta-1\right)}\Bigg[
\cos^6\theta\left(2g^{(1)}_{222i}+g^{(2)}_{222i}\right)\left(
\frac{M_2^4}{M_{2i}^4}\left(18g^{(1)}_{222i}+7g^{(2)}_{222i}\right)
-16\frac{M_2^2}{M_{2i}^2}g^{(1)}_{222i}\right)\nonumber\\
&&\hspace{0.35in}+2\cos^4\theta\left(
\frac{M_2^4}{M_{2i}^4}\left(130g^{(1)\ 2}_{222i}-16g^{(1)}_{222i}g^{(2)}_{222i}-7g^{(2)\ 2}_{222i}\right)
-8\frac{M_2^2}{M_{2i}^2}\left(2g^{(1)\ 2}_{222i}+g^{(1)}_{222i}g^{(2)}_{222i}+g^{(2)\ 2}_{222i}\right)
-20g^{(2)\ 2}_{222i}-4\frac{M_{2i}^2}{M_2^2}
\right)\nonumber\\
&&\hspace{0.5in}+\cos^2\theta\Bigg(
\frac{M_2^4}{M_{2i}^4}\left(284g^{(1)\ 2}_{222i}+160g^{(1)}_{222i}g^{(2)}_{222i}+7g^{(2)\ 2}_{222i}\right)
+8\frac{M_2^2}{M_{2i}^2}\left(68g^{(1)\ 2}_{222i}+34g^{(1)}_{222i}g^{(2)}_{222i}+7g^{(2)\ 2}_{222i}\right)\nonumber\\
&&\hspace{1in}-76\frac{M_{2i}^2}{M_2^2}g^{(2)\ 2}_{222i}
-11\frac{M_{2i}^4}{M_2^4}g^{(2)\ 2}_{222i}
\Bigg)\nonumber\\
&&\hspace{0.5in}-20\frac{M_2^4}{M_{2i}^4}g^{(1)}_{222i}\left(29g^{(1)}_{222i}+8g^{(2)}_{222i}\right)
-40\frac{M_2^2}{M_{2i}^2}\left(12g^{(1)\ 2}_{222i}+6g^{(1)}_{222i}g^{(2)}_{222i}+g^{(2)\ 2}_{222i}\right)
-56g^{(2)\ 2}_{222i}
-12\frac{M_{2i}^2}{M_2^2}g^{(2)\ 2}_{222i}\nonumber\\
&&\hspace{1in}-13\frac{M_{2i}^4}{M_2^4}g^{(2)\ 2}_{222i}
\Bigg]\nonumber\\
&&+\frac{1}{M_2^2}\frac{1}{27}\frac{1}{\left(\cos^2\theta-1\right)^2}\Bigg[
2\cos^8\theta\Bigg(
\frac{M_2^4}{M_{2i}^4}\left(-4g^{(1)\ 2}_{222i}+g^{(2)\ 2}_{222i}\right)
-2\frac{M_2^2}{M_{2i}^2}g^{(1)}_{222i}\left(2g^{(1)}_{222i}+g^{(2)}_{222i}\right)
\Bigg)\nonumber\\
&&\hspace{0.5in}-2\cos^6\theta\Bigg(
\frac{M_2^4}{M_{2i}^4}\left(32g^{(1)\ 2}_{222i}+3g^{(2)\ 2}_{222i}\right)
+4\frac{M_2^2}{M_{2i}^2}\left(8g^{(1)\ 2}_{222i}+4g^{(1)}_{222i}g^{(2)}_{222i}+g^{(2)\ 2}_{222i}\right)
+34g^{(2)\ 2}_{222i}
+17\frac{M_{2i}^2}{M_2^2}g^{(2)\ 2}_{222i}
\Bigg)\nonumber\\
&&\hspace{0.5in}+\cos^4\theta\Bigg(
6\frac{M_2^4}{M_{2i}^4}\left(-8g^{(1)\ 2}_{222i}+g^{(2)\ 2}_{222i}\right)
-8\frac{M_2^2}{M_{2i}^2}\left(6g^{(1)\ 2}_{222i}+3g^{(1)}_{222i}g^{(2)}_{222i}-2g^{(2)\ 2}_{222i}\right)\nonumber\\
&&\hspace{1in}+g^{(2)\ 2}_{222i}\left(124-204\frac{M_{2i}^2}{M_2^2}-82\frac{M_{2i}^4}{M_2^4}-\frac{M_{2i}^6}{M_2^6}\right)\Bigg)\nonumber
\end{eqnarray}
\end{minipage}

\begin{eqnarray}
&&\hspace{0.in}+2\cos^2\theta\Bigg(
\frac{M_2^4}{M_{2i}^4}\left(160g^{(1)\ 2}_{222i}-g^{(2)\ 2}_{222i}\right)
+4\frac{M_2^2}{M_{2i}^2}\left(40g^{(1)\ 2}_{222i}+20g^{(1)}_{222i}g^{(2)}_{222i}-g^{(2)\ 2}_{222i}\right)\nonumber\\
&&\hspace{0.5in}+g^{(2)\ 2}_{222i}\left(-44+78\frac{M_{2i}^2}{M_2^2}-124\frac{M_{2i}^4}{M_2^4}-34\frac{M_{2i}^6}{M_2^6}\right)
\Bigg)\nonumber\\
&&\hspace{0.in}-200\frac{M_2^4}{M_{2i}^4}g^{(1)\ 2}_{222i}
-100\frac{M_2^2}{M_{2i}^2}g^{(1)}_{222i}\left(2g^{(1)}_{222i}+g^{(2)}_{222i}\right)
-g^{(2)\ 2}_{222i}\left(12+32\frac{M_{2i}^2}{M_2^2}-14\frac{M_{2i}^4}{M_2^4}+13\frac{M_{2i}^6}{M_2^6}\right)
\Bigg]
+\mathcal{O}\left(E^{-2}\right)
\end{eqnarray}

Using the Feynman rules given in Eqs.~\eqref{eq:g_221j^1} and \eqref{eq:g_221j^2} gives the S-channel diagram\\
\begin{minipage}{7in}
\begin{minipage}{2in}
\includegraphics[scale=0.4]{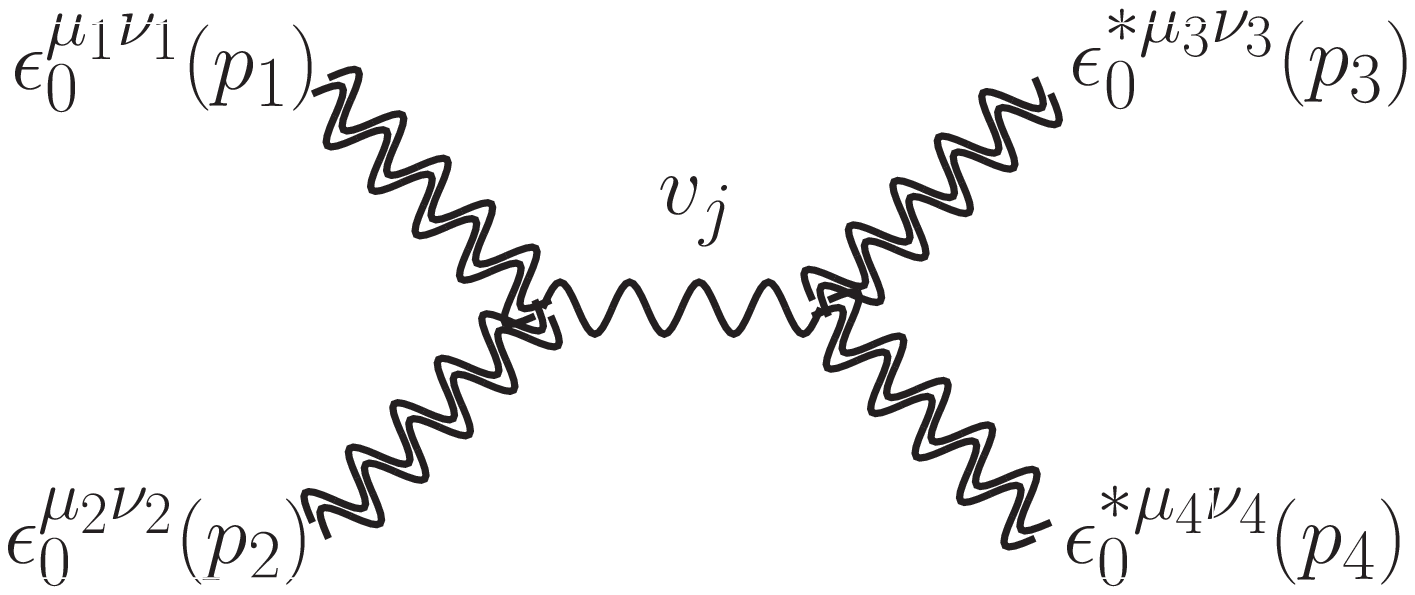}
\end{minipage}
\begin{minipage}{4in}
\begin{eqnarray}
&&\mathcal{M}^{Svj}_{0000} =
\frac{E^{10}}{M_2^{8}M_{1j}^2}\frac{256}{9}\left(g^{(1)}_{221j}+2g^{(2)}_{221j}\right)^2\nonumber\\
&&-\frac{E^8}{M_2^{6}M_{1j}^2}\frac{256}{9}\left(g^{(1)}_{221j}+2g^{(2)}_{221j}\right)\left(3g^{(1)}_{221j}+4g^{(2)}_{221j}\right)\nonumber
\end{eqnarray}
\end{minipage}
\begin{eqnarray}
&&+\frac{E^6}{M_2^4M_{1j}^2}\frac{64}{9}
\left(13g^{(1)\ 2}_{221j}+38g^{(1)}_{221j}g^{(2)}_{221j}+28g^{(2)\ 2}_{221j}\right)
-\frac{E^4}{M_2^2M_{1j}^2}\frac{64}{9}
\left(2g^{(1)}_{221j}+3g^{(2)}_{221j}\right)
\left(3g^{(1)}_{221j}+4g^{(2)}_{221j}\right)\nonumber\\
&&+\frac{E^2}{M_{1j}^2}\frac{16}{9}
\left(2g^{(1)}_{221j}+3g^{(2)}_{221j}\right)^2
\end{eqnarray}
\end{minipage}
and the T+U-channel diagrams\\
\begin{minipage}{7in}
\begin{minipage}{2in}
\includegraphics[scale=0.4]{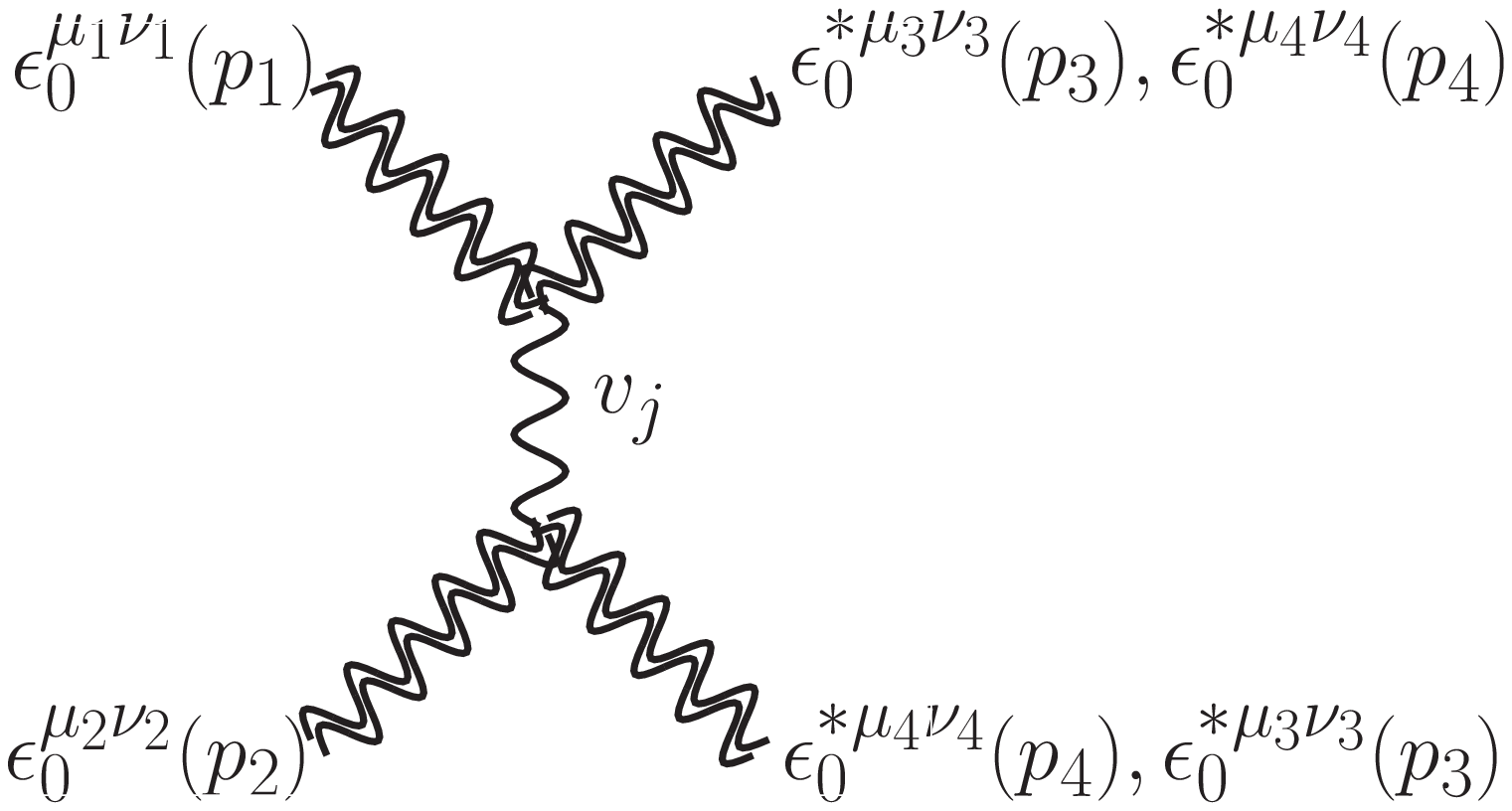}
\end{minipage}
\hfill
\begin{minipage}{4.75in}
\begin{eqnarray}
&&\mathcal{M}^{TUvj}_{0000} =
-\frac{E^{10}}{M_2^{8}M_{1j}^2}\frac{16}{9}\left(g^{(1)}_{221j}+2g^{(2)}_{221j}\right)^2
\left(5\cos^4\theta+10\cos^2+1\right)\nonumber\\
&&+\frac{E^8}{M_2^{6}M_{1j}^2}\frac{16}{9}\left(g^{(1)}_{221j}+2g^{(2)}_{221j}\right)\Bigg[
\cos^4\theta\left(g^{(1)}_{221j}+6g^{(2)}_{221j}\right)\nonumber\\
&&\hspace{0.75in}+2\cos^2\theta\left(13g^{(2)}_{221j}+38g^{(2)}_{221j}\right)
+\left(5g^{(1)}_{221j}+14g^{(2)}_{221j}\right)
\Bigg]\nonumber
\end{eqnarray}
\end{minipage}
\begin{eqnarray}
&&+\frac{E^6}{M_2^4M_{1j}^2}\frac{16}{9}\Bigg[
\cos^4\theta\left(g^{(1)}_{221j}+2g^{(2)}_{221j}\right)\left(4g^{(1)}_{221j}+7g^{(2)}_{221j}\right)
-2\cos^2\theta\left(8g^{(1)\ 2}_{221j}+59g^{(1)}_{221j}g^{(2)}_{221j}+92g^{(2)\ 2}_{221j}\right)\nonumber\\
&&\hspace{0.5in}-\left(8g^{(1)\ 2}_{221j}+49g^{(1)}_{221j}g^{(2)}_{221j}+70g^{(2)}_{221j}\right)
\Bigg]\nonumber\\
&&+\frac{E^4}{M_2^2M_{1j}^2}\frac{8}{9}\Bigg[
\cos^4\theta\left(g^{(1)}_{221j}+2g^{(2)}_{221j}\right)\left(g^{(1)}_{221j}-4g^{(2)}_{221j}\right)
+\cos^2\theta\left(-g^{(1)\ 2}_{221j}+60g^{(1)}_{221j}g^{(2)}_{221j}+176g^{(2)\ 2}_{221j}\right)\nonumber\\
&&\hspace{0.5in}+2\left(4g^{(1)\ 2}_{221j}+39g^{(1)}_{221j}g^{(2)}_{221j}+76g^{(2)\ 2}_{221j}\right)
\Bigg]\nonumber\\
&&-\frac{E^2}{M_{1j}^2}\frac{1}{9}\Bigg[
\cos^4\theta\left(g^{(1)}_{221j}+2g^{(2)}_{221j}\right)\left(7g^{(1)}_{221j}+18g^{(2)}_{221j}\right)
-\cos^2\theta\left(7g^{(1)\ 2}_{221j}-296g^{(2)\ 2}_{221j}\right)\nonumber\\
&&\hspace{0.5in}+20g^{(2)}_{221j}\left(8g^{(1)}_{221j}+29g^{(2)}_{221j}\right)
\Bigg]\nonumber\\
&&+\frac{M_2^2}{M_{1j}^2}\frac{1}{9}\Bigg[
-\cos^4\theta\left(g^{(1)\ 2}_{221j}-4g^{(2)\ 2}_{221j}\right)
+\cos^2\theta\left(g^{(1)\ 2}_{221j}+40g^{(2)\ 2}_{221j}\right)+100g^{(2)\ 2}_{221j}
\Bigg]\nonumber\\
\end{eqnarray}
\end{minipage}

Using the Feynman rules given in Eq.~\eqref{eq:g_220k} gives the S-channel diagram\\
\begin{minipage}{7in}
\begin{minipage}{2in}
\includegraphics[scale=0.4]{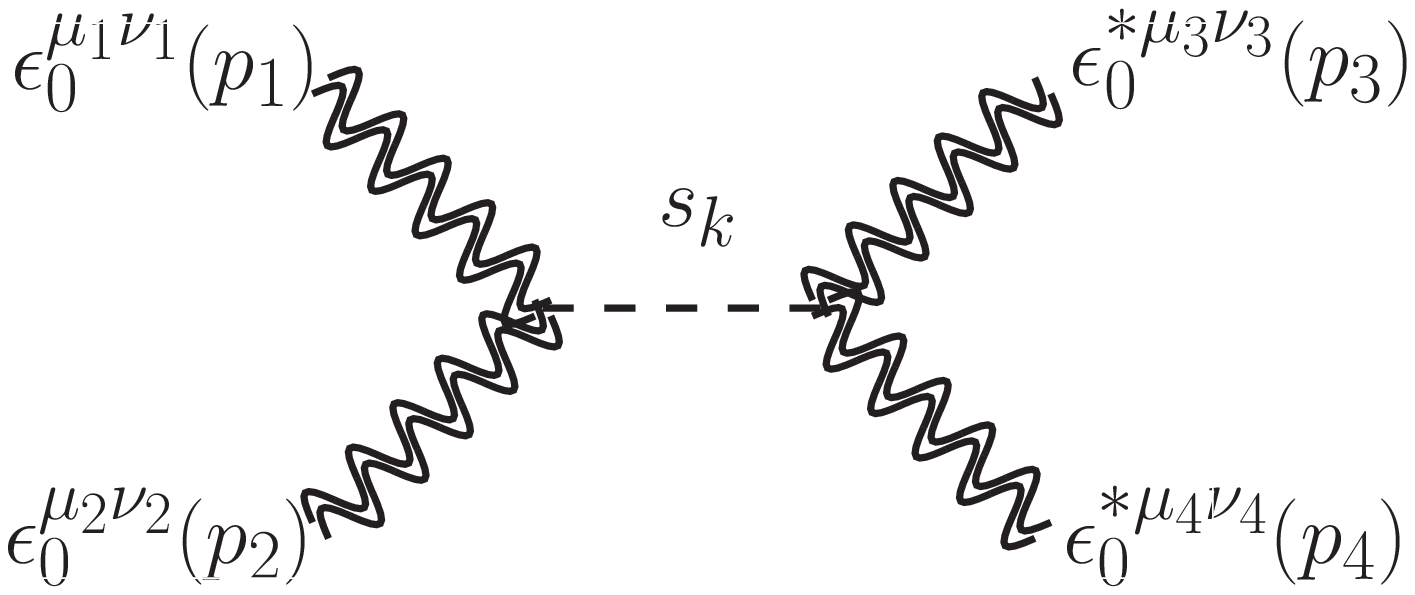}
\end{minipage}
\begin{minipage}{5in}
\begin{eqnarray}
&&\mathcal{M}^{Ssk}_{0000} =
-\frac{E^{6}}{M_2^{8}}\frac{64}{9}g_{220k}^2
+\frac{E^4}{M_2^{6}}\frac{16}{9}g_{220k}^2\left(8-\frac{M_{0k}^2}{M_2^2}\right)\nonumber\\
&&-\frac{E^2}{M_2^4}\frac{4}{9}g_{220k}^2\left(28-8\frac{M_{0k}^2}{M_2^2}+\frac{M_{0k}^4}{M_2^4}\right)\nonumber
\end{eqnarray}
\end{minipage}
\begin{eqnarray}
&&+\frac{1}{M_2^2}\frac{1}{9}g^2_{220k}\Bigg[
48-28\frac{M_{0k}^2}{M_2^2}+8\frac{M_{0k}^4}{M_2^4}-\frac{M_{0k}^6}{M_2^6}
\Bigg]
+\mathcal{O}\left(E^{-2}\right)
\end{eqnarray}
\end{minipage}\\
and the T+U-channel diagrams\\
\begin{minipage}{7in}
\begin{minipage}{2in}
\includegraphics[scale=0.4]{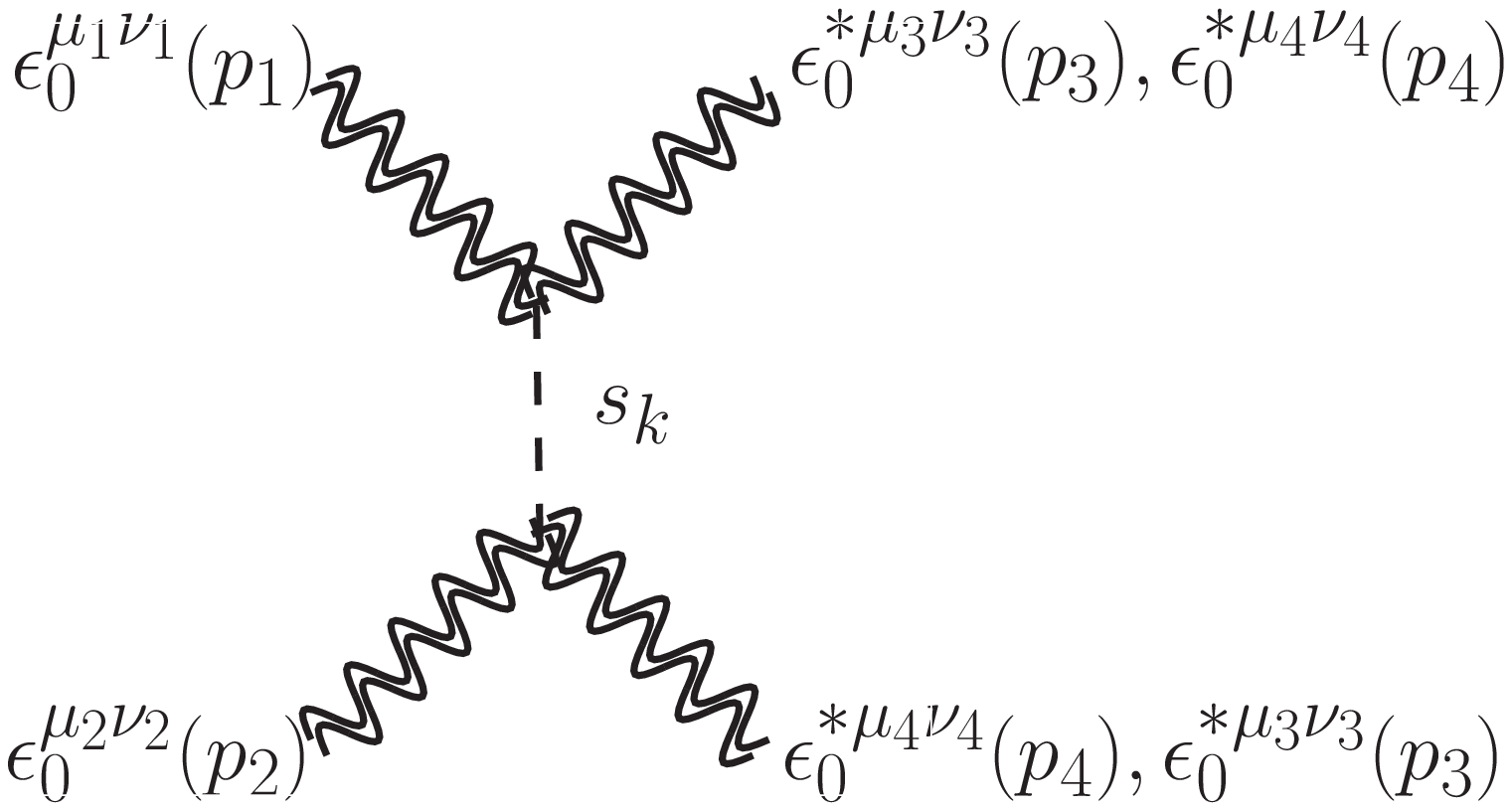}
\end{minipage}
\hfill
\begin{minipage}{4.75in}
\begin{eqnarray}
&&\mathcal{M}^{TUsk}_{0000} =
\frac{E^{6}}{M_2^{8}}\frac{16}{3}g_{220k}^2\left(\cos^2\theta+\frac{1}{3}\right)\nonumber\\
&&+\frac{E^4}{M_2^{6}}\frac{8}{9}g_{220k}^2\Bigg[
\cos^2\theta\left(2-\frac{M_{0k}^2}{M_2^2}\right)
-10-\frac{M_{0k}^2}{M_2^2}
\Bigg]\nonumber\\
&&+\frac{E^2}{M_2^4}\frac{4}{9}g_{220k}^2\Bigg[
-2\cos^2\theta\left(7+4\frac{M_{0k}^2}{M_2^2}\right)
+26+8\frac{M_{0k}^2}{M_2^2}+\frac{M_{0k}^4}{M_2^4}
\Bigg]\nonumber
\end{eqnarray}
\end{minipage}
\begin{eqnarray}
&&-\frac{1}{M_2^2}\frac{1}{9}g^2_{220k}\Bigg[
4\cos^2\theta\left(20+17\frac{M_{0k}^2}{M_2^2}\right)
+16+20\frac{M_{0k}^2}{M_2^2}+12\frac{M_{0k}^4}{M_2^4}+2\frac{M_{0k}^6}{M_2^6}
\Bigg]
+\mathcal{O}\left(E^{-2}\right)
\end{eqnarray}
\end{minipage}

\subsection{\label{app:M20-10}Expanded $\mathcal{M}_{20-10}$ Diagrams}
In this section, we use the Feynman rules given in App.~\ref{sec:Feynman Rules} to explicitly calculate the $2\to2$ scattering amplitude with the helicities $\sigma_1=2, \sigma_2=0, \sigma_3=-1, \sigma_4=0$.  These are the diagrams that, along with the $\sigma_1=\sigma_2=\sigma_3=\sigma_4=0$ diagram, give the full set of linearly independent equations whose solution removes energy growth from all diagrams at tree-level.  We explicitly expand each diagram in energy and use the standard propagators ($F=1$, $C_{12}=\frac{1}{2}$, $C_3=-\frac{1}{3}$, $C_{47}=-\frac{1}{2}$, $C_{89}=\frac{1}{3}$ and $C_{10}=\frac{2}{3}$) for compactness and clarity in these expressions.  

Using the Feynman rules given in Eqs.~\eqref{eq:g_2222^1} and \eqref{eq:g_2222^2} gives the 4-point diagram\\
\begin{minipage}{7in}
\begin{minipage}{1.75in}
\includegraphics[scale=0.4]{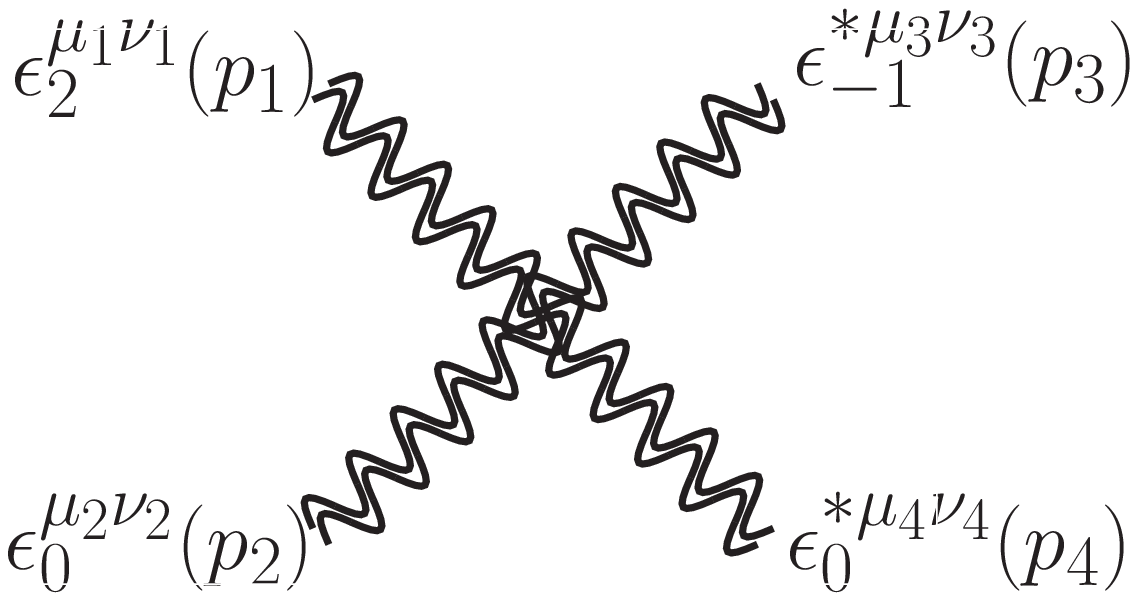}
\end{minipage}
\hfill
\begin{minipage}{5.25in}
\begin{eqnarray}
&&\mathcal{M}^4_{20\minus10} =
\frac{E^5}{M_2^5}\frac{8}{3}\left(\cos\theta-1\right)\sin\theta\Bigg[
\cos^2\theta\left(2g^{(1)}_{2222}+g^{(2)}_{2222}\right)
+\left(2g^{(1)}_{2222}-g^{(2)}_{2222}\right)
\Bigg]\nonumber
\end{eqnarray}
\end{minipage}
\begin{eqnarray}
&&+\frac{E^3}{M_2^3}\frac{2}{3}\left(\cos\theta-1\right)\sin\theta\Bigg[
4\cos^2\theta\left(2g^{(1)}_{2222}+g^{(2)}_{2222}\right)
+5\cos\theta\left(2g^{(1)}_{2222}+g^{(2)}_{2222}\right)
-\left(14g^{(1)}_{2222}-g^{(2)}_{2222}\right)
\Bigg]\nonumber\\
&&+\frac{E}{M_2}\frac{1}{3}\left(\cos\theta-1\right)\sin\theta\left(2g^{(1)}_{2222}+g^{(2)}_{2222}\right)\Bigg[
2\cos^2\theta-\cos\theta+3\Bigg]\nonumber
\end{eqnarray}
\end{minipage}

Using the Feynman rules given in Eqs.~\eqref{eq:g_222^1} and \eqref{eq:g_222^2} gives the S-channel diagram\\
\begin{minipage}{7in}
\begin{minipage}{2in}
\includegraphics[scale=0.4]{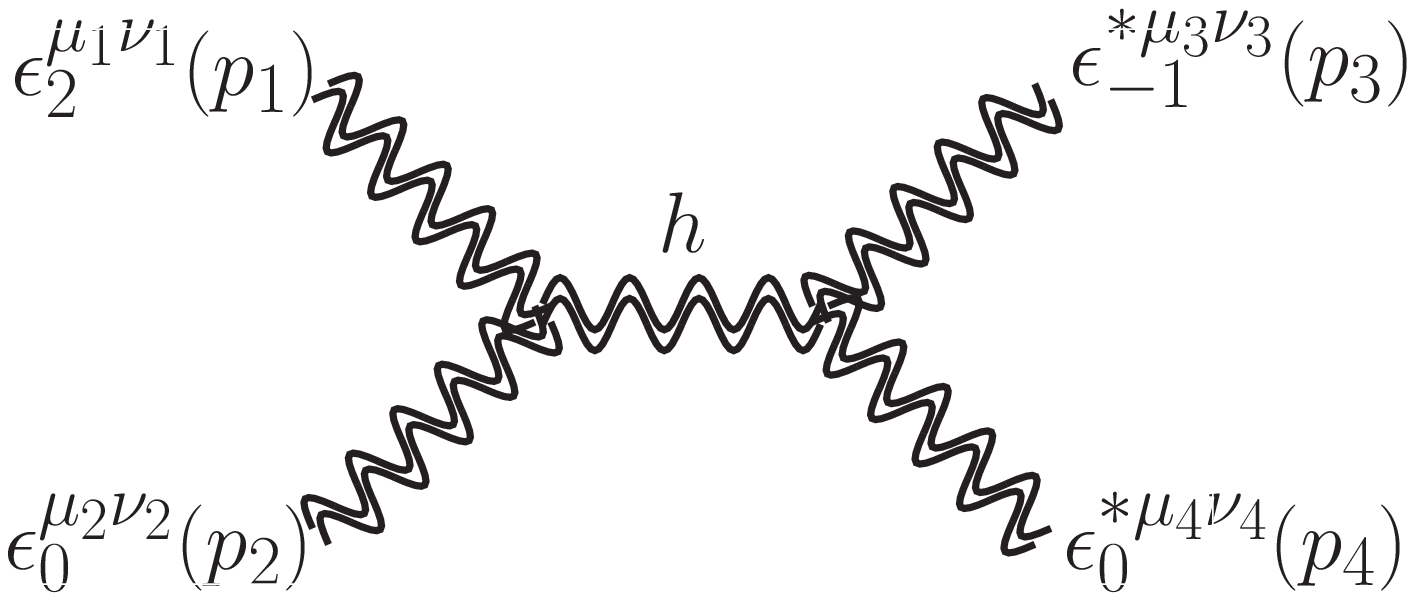}
\end{minipage}
\hfill
\begin{minipage}{5in}
\begin{eqnarray}
&&\mathcal{M}^{Sh}_{20\minus10} =
\frac{E}{M_2^3}\frac{3}{2}\left(\cos\theta-1\right)\sin\theta g^{(2)\ 2}_{222}
+\mathcal{O}\left(E^{-1}\right)
\end{eqnarray}
\end{minipage}
\end{minipage}
and the T+U-channel diagrams\\
\begin{minipage}{7in}
\begin{minipage}{2in}
\includegraphics[scale=0.4]{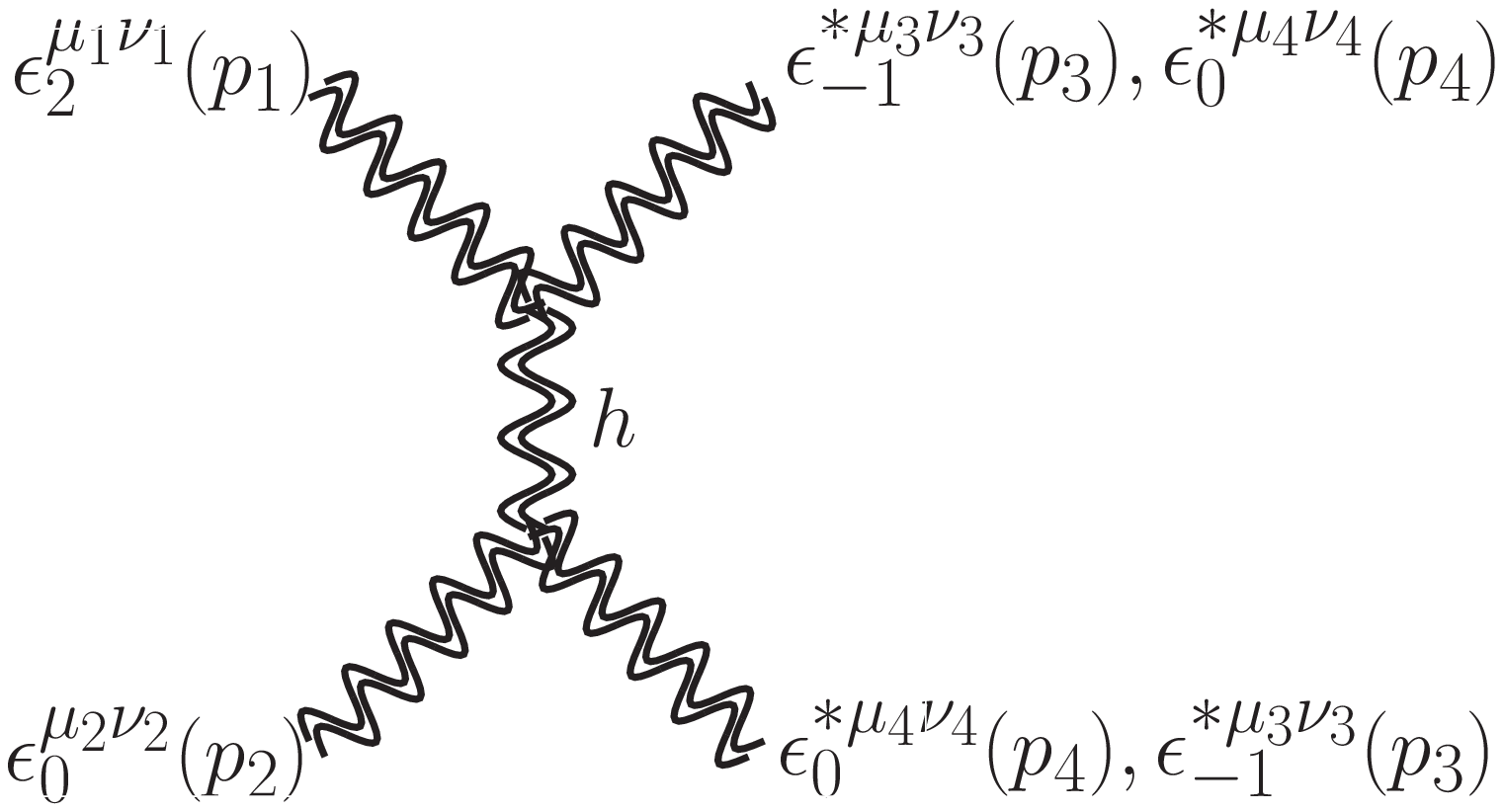}
\end{minipage}
\hfill
\begin{minipage}{4.75in}
\begin{eqnarray}
&&\mathcal{M}^{TUh}_{20\minus10} =
\frac{E^7}{M_2^9}\frac{4}{9}\left(\cos\theta-1\right)\sin\theta\left(2g^{(1)}_{222}+3g^{(2)}_{222}\right)\Bigg[
3\cos^2\theta\left(4g^{(1)}_{222}+5g^{(2)}_{222}\right)\nonumber\\
&&\hspace{1in}+6\cos\theta g^{(2)}_{222}
+\left(4g^{(1)}_{222}+3g^{(2)}_{222}\right)
\Bigg]\nonumber
\end{eqnarray}
\end{minipage}
\begin{eqnarray}
&&+\frac{E^5}{M_2^7}\frac{1}{9}\left(\cos\theta-1\right)\sin\theta\Bigg[
-3\cos^2\theta\left(2g^{(1)}_{222}+3g^{(2)}_{222}\right)\left(20g^{(1)}_{222}+19g^{(2)}_{222}\right)
+\cos\theta\left(80g^{(1)\ 2}_{222}+84g^{(1)}_{222}g^{(2)}_{222}-27g^{(2)\ 2}_{222}\right)\nonumber\\
&&\hspace{0.5in}-6\left(20g^{(1)\ 2}_{222}+33g^{(1)}_{222}g^{(2)}_{222}+18g^{(2)\ 2}_{222}\right)
\Bigg]\nonumber\\
&&-\frac{E^3}{M_2^5}\frac{1}{36}\left(\cos\theta-1\right)\sin\theta\Bigg[
6\cos^2\theta\left(2g^{(1)}_{222}+3g^{(2)}_{222}\right)\left(12g^{(1)}_{222}+5g^{(2)}_{222}\right)
+\cos\theta\left(608g^{(1)\ 2}_{222}+960g^{(1)}_{222}g^{(2)}_{222}+261g^{(2)\ 2}_{222}\right)\nonumber\\
&&\hspace{0.5in}-\left(496g^{(1)\ 2}_{222}+612g^{(1)}_{222}g^{(2)}_{222}+45g^{(2)\ 2}_{222}\right)
\Bigg]\nonumber\\
&&+\frac{E^1}{M_2^3}\frac{1}{24}\frac{\sin\theta}{\left(\cos\theta+1\right)}\Bigg[
-4\cos^4\theta\left(2g^{(1)}_{222}+3g^{(2)}_{222}\right)\left(4g^{(1)}_{222}-3g^{(2)}_{222}\right)
+3\cos^3\theta\left(32g^{(1)\ 2}_{222}+48g^{(1)}_{222}g^{(2)}_{222}+9g^{(2)\ 2}_{222}\right)\nonumber\\
&&\hspace{0.5in}-\cos^2\theta\left(32g^{(1)\ 2}_{222}+225g^{(2)\ 2}_{222}\right)
-3\cos\theta\left(32g^{(1)\ 2}_{222}+48g^{(1)}_{222}g^{(2)}_{222}+69g^{(2)\ 2}_{222}\right)\nonumber\\
&&\hspace{0.5in}+\left(64g^{(1)\ 2}_{222}+24g^{(1)}_{222}g^{(2)}_{222}-135g^{(2)\ 2}_{222}\right)
\Bigg]
+\mathcal{O}\left(E^{-1}\right)
\end{eqnarray}
\end{minipage}

Using the Feynman rules given in Eqs.~\eqref{eq:g_222i^1} and \eqref{eq:g_222i^2} gives the S-channel diagram\\
\begin{minipage}{7in}
\begin{minipage}{2in}
\includegraphics[scale=0.4]{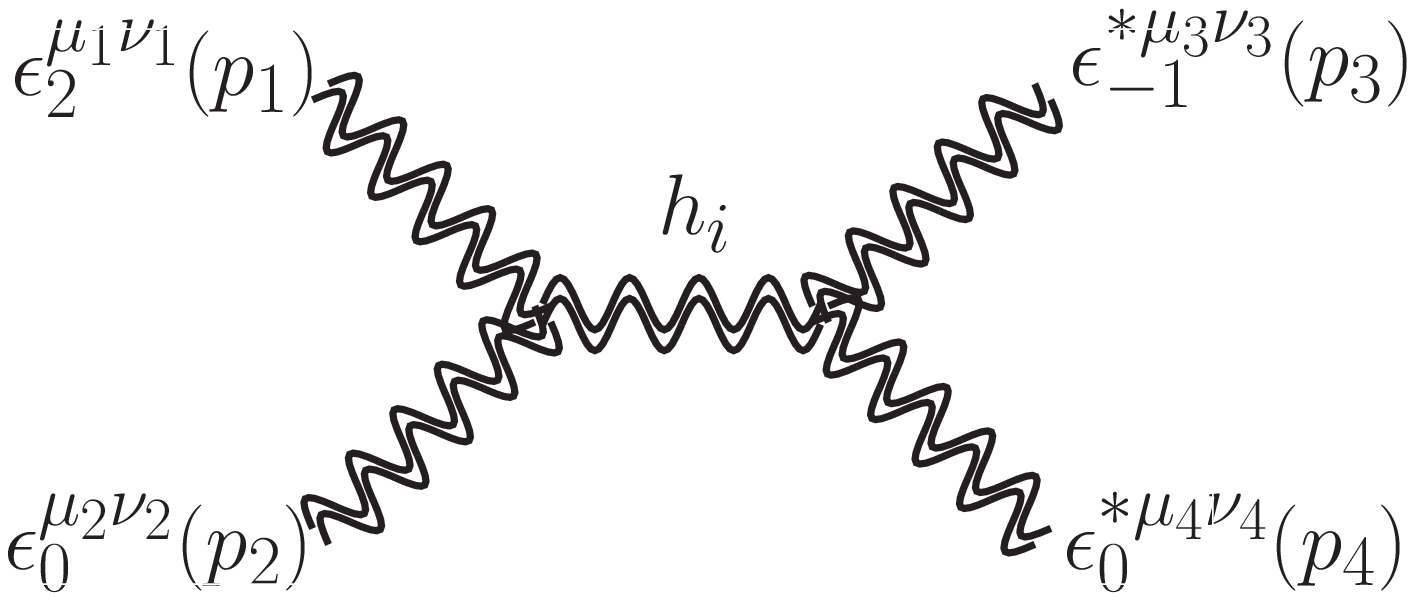}
\end{minipage}
\begin{minipage}{4in}
\begin{eqnarray}
&&\mathcal{M}^{Shi}_{20\minus10} =
\frac{E}{M_2^3}\frac{1}{6}\left(\cos\theta-1\right)\sin\theta g^{(2)\ 2}_{222i}
+\mathcal{O}\left(E^{-1}\right)
\end{eqnarray}
\end{minipage}
\end{minipage}
and the T+U-channel diagrams\\
\begin{minipage}{7in}
\begin{minipage}{2in}
\includegraphics[scale=0.4]{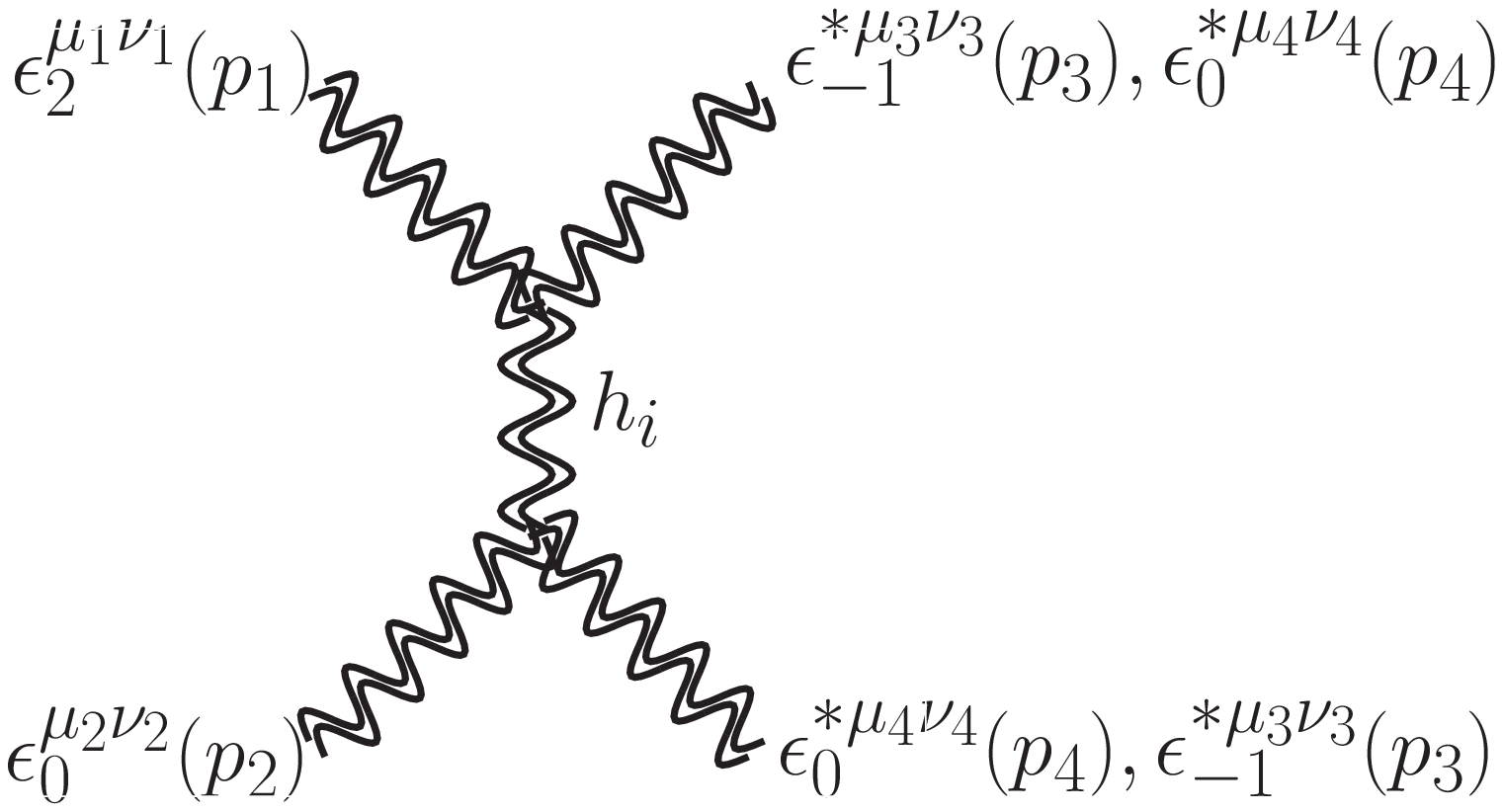}
\end{minipage}
\hfill
\begin{minipage}{4.5in}
\begin{eqnarray}
&&\mathcal{M}^{TUhi}_{20\minus10} =
\frac{E^7}{M_2^5M_{2i}^4}\frac{4}{9}\left(\cos\theta-1\right)\sin\theta\left(2g^{(1)}_{222i}+g^{(2)}_{222i}\right)\Bigg[\nonumber\\
&&\hspace{1in}\cos^2\theta\left(12g^{(1)}_{222i}+5g^{(2)}_{222i}\right)
+2\cos\theta g^{(2)}_{222i}+\left(4g^{(1)}_{222i}+g^{(2)}_{222i}\right)
\Bigg]
\nonumber
\end{eqnarray}
\end{minipage}
\begin{eqnarray}
&&+\frac{E^5}{M_2^5M_{2i}^2}\frac{1}{9}\left(\cos\theta-1\right)\sin\theta\Bigg[
-\cos^2\theta\left(2g^{(1)}_{222i}+g^{(2)}_{222i}\right)\Bigg(\frac{M_2^2}{M_{2i}^2}\left(44g^{(1)}_{222i}+19g^{(2)}_{222i}\right)
+16g^{(1)}_{222i}\Bigg)\nonumber\\
&&\hspace{0.5in}+\cos\theta\Bigg(2\frac{M_2^2}{M_{2i}^2}\left(40g^{(1)\ 2}_{222i}+14g^{(1)}_{222i}g^{(2)}_{222i}-g^{(2)\ 2}_{222i}\right)
-g^{(2)\ 2}_{222i}\Bigg)\nonumber\\
&&\hspace{0.5in}-\Bigg(\frac{M_2^2}{M_{2i}^2}\left(4g^{(1)}_{222i}+g^{(2)}_{222i}\right)\left(22g^{(1)}_{222i}+7g^{(2)}_{222i}\right)
+\left(32g^{(1)\ 2}_{222i}+16g^{(1)}_{222i}g^{(2)}_{222i}+5g^{(2)\ 2}_{222i}\right)\Bigg)
\Bigg]\nonumber\\
&&+\frac{E^3}{M_2^5}\frac{1}{36}\left(\cos\theta-1\right)\sin\theta\Bigg[
-2\cos^2\theta\left(2g^{(1)}_{222i}+g^{(2)}_{222i}\right)\Bigg(
\frac{M_2^4}{M_{2i}^4}\left(4g^{(1)}_{222i}+5g^{(2)}_{222i}\right)
+32\frac{M_2^2}{M_{2i}^2}g^{(1)}_{222i}\Bigg)\nonumber\\
&&\hspace{0.5in}-\cos\theta\Bigg(
2\frac{M_2^4}{M_{2i}^4}\left(224g^{(1)\ 2}_{222i}+120g^{(1)}_{222i}g^{(2)}_{222i}+15g^{(2)\ 2}_{222i}\right)
+\frac{M_2^2}{M_{2i}^2}\left(160g^{(1)\ 2}_{222i}+80g^{(1)}_{222i}g^{(2)}_{222i}-g^{(2)\ 2}_{222i}\right)
\Bigg)\nonumber\\
&&\hspace{0.5in}+\Bigg(
4\frac{M_2^4}{M_{2i}^4}\left(68g^{(1)\ 2}_{222i}+23g^{(1)}_{222i}g^{(2)}_{222i}+3g^{(2)\ 2}_{222i}\right)
+\frac{M_2^2}{M_{2i}^2}\left(224g^{(1)\ 2}_{222i}+112g^{(1)}_{222i}g^{(2)}_{222i}+23g^{(2)\ 2}_{222i}\right)
-30g^{(2)\ 2}_{222i}
\Bigg)
\Bigg]\nonumber\\
&&+\frac{E}{M_2^3}\frac{1}{72}\frac{\sin\theta}{\left(\cos\theta+1\right)}\Bigg[
-4\cos^4\theta\left(2g^{(1)}_{222i}+g^{(2)}_{222i}\right)\Bigg(
\frac{M_2^4}{M_{2i}^4}\left(4g^{(1)}_{222i}-3g^{(2)}_{222i}\right)
+8\frac{M_2^2}{M_{2i}^2}g^{(1)}_{222i}\Bigg)\nonumber\\
&&\hspace{0.5in}+\cos^3\theta\Bigg(
4\frac{M_2^4}{M_{2i}^4}\left(8g^{(1)}_{222i}+g^{(2)}_{222i}\right)\left(8g^{(1)}_{222i}+3g^{(2)}_{222i}\right)
+\frac{M_2^2}{M_{2i}^2}\left(32g^{(1)\ 2}_{222i}+16g^{(1)}_{222i}g^{(2)}_{222i}-3g^{(2)\ 2}_{222i}\right)
\Bigg)\nonumber\\
&&\hspace{0.5in}-\cos^2\theta\Bigg(
4\frac{M_2^4}{M_{2i}^4}\left(16g^{(1)\ 2}_{222i}-4g^{(1)}_{222i}g^{(2)}_{222i}+3g^{(2)\ 2}_{222i}\right)
+2\frac{M_2^2}{M_{2i}^2}\left(4g^{(1)}_{222i}+g^{(2)}_{222i}\right)^2
+g^{(2)\ 2}_{222i}\left(53+8\frac{M_{2i}^2}{M_2^2}\right)
\Bigg)\nonumber\\
&&\hspace{0.5in}-\cos\theta\Bigg(
4\frac{M_2^4}{M_{2i}^4}\left(8g^{(1)}_{222i}+g^{(2)}_{222i}\right)\left(8g^{(1)}_{222i}+3g^{(2)}_{222i}\right)
+\frac{M_2^2}{M_{2i}^2}\left(32g^{(1)\ 2}_{222i}+16g^{(1)}_{222i}g^{(2)}_{222i}+5g^{(2)\ 2}_{222i}\right)\nonumber\\
&&\hspace{1in}+2g^{(2)\ 2}_{222i}\left(25+\frac{M_{2i}^2}{M_2^2}\right)
\Bigg)\nonumber\\
&&\hspace{0.5in}+\Bigg(
24\frac{M_2^4}{M_{2i}^4}g^{(1)}_{222i}\left(4g^{(1)}_{222i}-g^{(2)}_{222i}\right)
+6\frac{M_2^2}{M_{2i}^2}\left(16g^{(1)\ 2}_{222i}+8g^{(1)}_{222i}g^{(2)}_{222i}-g^{(2)\ 2}_{222i}\right)
-g^{(2)\ 2}_{222i}\left(9+30\frac{M_{2i}^2}{M_2^2}\right)
\Bigg)
\Bigg]\nonumber\\
&&+\mathcal{O}\left(E^{-1}\right)
\end{eqnarray}
\end{minipage}

Using the Feynman rules given in Eqs.~\eqref{eq:g_221j^1} and \eqref{eq:g_221j^2} gives the S-channel diagram\\
\begin{minipage}{7in}
\begin{minipage}{2in}
\includegraphics[scale=0.4]{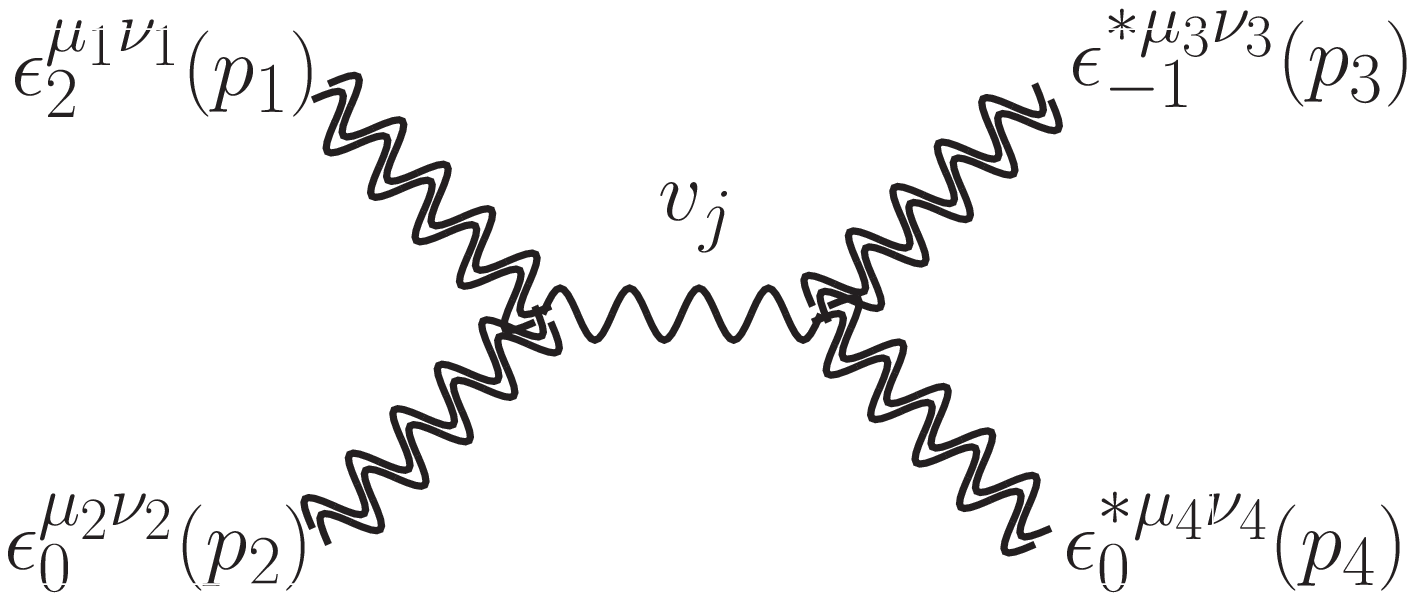}
\end{minipage}
\begin{minipage}{4in}
\begin{eqnarray}
&&\mathcal{M}^{Svj}_{20\minus10} =0
\end{eqnarray}
\end{minipage}
\end{minipage}
and the T+U-channel diagrams\\
\begin{minipage}{7in}
\begin{minipage}{2in}
\includegraphics[scale=0.4]{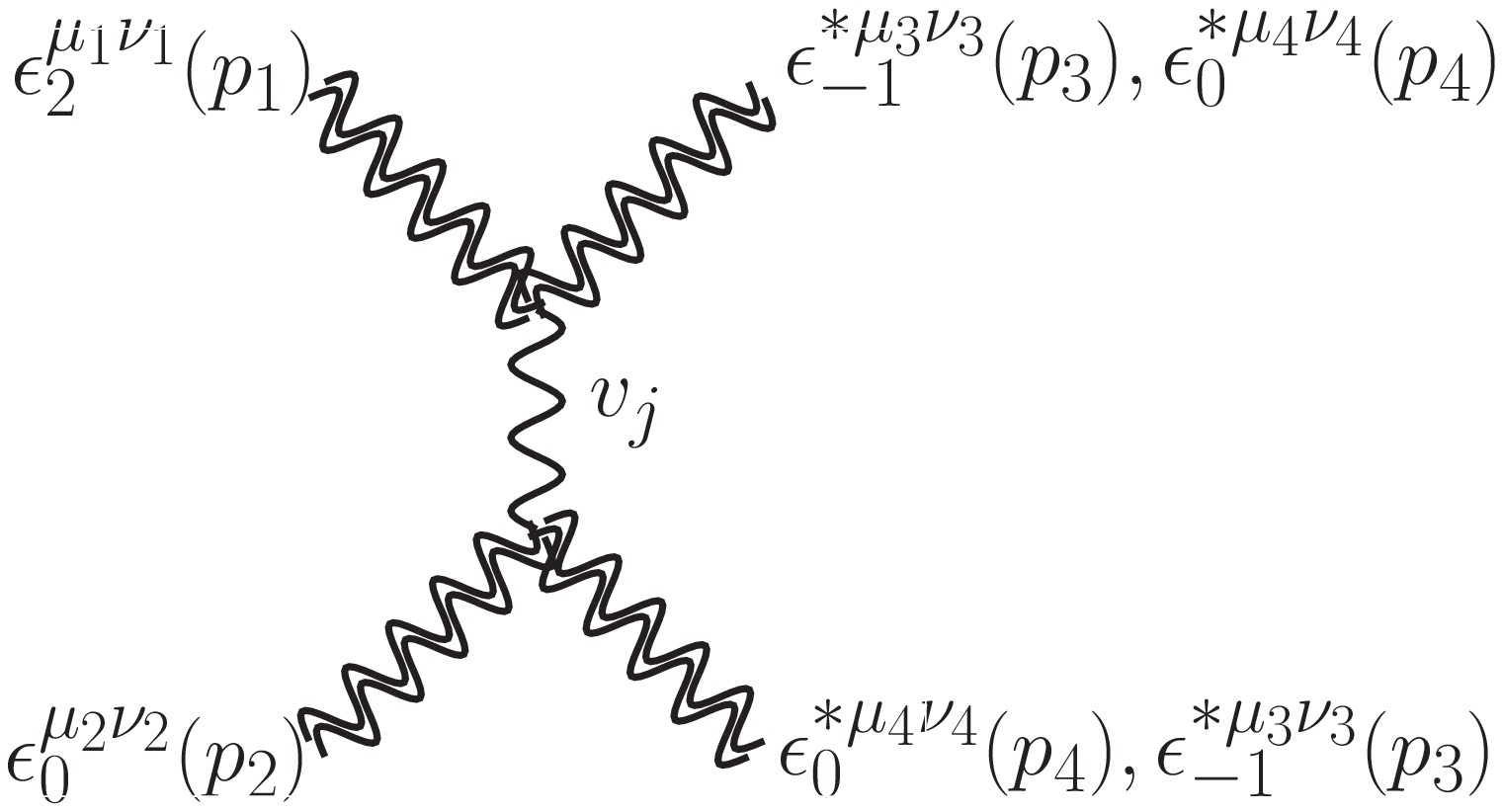}
\end{minipage}
\hfill
\begin{minipage}{4.75in}
\begin{eqnarray}
&&\mathcal{M}^{TUvj}_{20\minus10} =
-\frac{E^7}{M_2^5M_{1j}^2}\frac{2}{3}\left(\cos\theta-1\right)\sin\theta\left(g^{(1)}_{221j}+2g^{(2)}_{221j}\right)\Bigg[\nonumber\\
&&\hspace{1in}\cos^2\theta\left(5g^{(1)}_{221j}+12g^{(2)}_{221j}\right)
+2\cos\theta g^{(1)}_{221j}
+\left(g^{(1)}_{221j}+4g^{(2)}_{221j}\right)
\Bigg]
\nonumber
\end{eqnarray}
\end{minipage}
\begin{eqnarray}
&&+\frac{E^5}{M_2^5}\frac{1}{6}\left(\cos\theta-1\right)\sin\theta\Bigg[
\cos^2\theta\frac{M_2^2}{M_{1j}^2}\left(g^{(1)}_{221j}+2g^{(2)}_{221j}\right)\left(19g^{(1)}_{221j}+44g^{(2)}_{221j}\right)\nonumber\\
&&\hspace{0.5in}+\cos\theta\Bigg(2\frac{M_2^2}{M_{1j}^2}\left(g^{(1)\ 2}_{221j}-14g^{(1)}_{221j}g^{(2)}_{221j}-40g^{(2)\ 2}_{221j}\right)
+g^{(1)\ 2}_{221j}\Bigg)\nonumber\\
&&\hspace{0.5in}+\Bigg(\frac{M_2^2}{M_{1j}^2}\left(g^{(1)}_{221j}+4g^{(2)}_{221j}\right)\left(7g^{(1)}_{221j}+22g^{(2)}_{221j}\right)
+g^{(1)\ 2}_{221j}\Bigg)
\Bigg]\nonumber\\
&&+\frac{E^3}{M_2^3}\frac{1}{24}\left(\cos\theta-1\right)\sin\theta\Bigg[
2\cos^2\theta\frac{M_2^2}{M_{1j}^2}\left(g^{(1)}_{221j}+2g^{(2)}_{221j}\right)\left(5g^{(1)}_{221j}+4g^{(2)}_{221j}\right)\nonumber\\
&&\hspace{0.5in}+\cos\theta\Bigg(
2\frac{M_2^2}{M_{1j}^2}\left(15g^{(1)\ 2}_{221j}+120g^{(1)}_{221j}g^{(2)}_{221j}+224g^{(2)\ 2}_{221j}\right)
-g^{(1)\ 2}_{221j}\Bigg)\nonumber\\
&&\hspace{0.5in}-\Bigg(
4\frac{M_2^2}{M_{1j}^2}\left(3g^{(1)\ 2}_{221j}+23g^{(1)}_{221j}g^{(2)}_{221j}+68g^{(2)\ 2}_{221j}\right)
+g^{(1)\ 2}_{221j}\left(3+2\frac{M_{1j}^2}{M_2^2}\right)
\Bigg)
\Bigg]\nonumber\\
&&+\frac{E}{M_2}\frac{1}{48}\frac{\left(\cos\theta-1\right)\sin\theta}{\left(\cos\theta+1\right)}\Bigg[
-4\cos^3\theta\frac{M_2^2}{M_{1j}^2}\left(g^{(1)}_{221j}+2g^{(2)}_{221j}\right)\left(3g^{(1)}_{221j}-4g^{(2)}_{221j}\right)\nonumber\\
&&\hspace{0.5in}+\cos^2\theta\Bigg(-8\frac{M_2^2}{M_{1j}^2}\left(3g^{(1)\ 2}_{221j}+17g^{(1)}_{221j}g^{(2)}_{221j}+28g^{(2)\ 2}_{221j}\right)
+3g^{(1)\ 2}_{221j}\Bigg)\nonumber\\
&&\hspace{0.5in}-\cos\theta\Bigg(
4\frac{M_2^2}{M_{1j}^2}\left(3g^{(1)\ 2}_{221j}+38g^{(1)}_{221j}g^{(2)}_{221j}+40g^{(2)\ 2}_{221j}\right)
+3g^{(1)\ 2}_{221j}\left(-1+\frac{M_{1j}^2}{M_2^2}\right)
\Bigg)\nonumber\\
&&\hspace{0.5in}+\Bigg(24\frac{M_2^2}{M_{1j}^2}g^{(2)}_{221j}\left(-g^{(1)}_{221j}+4g^{(2)}_{221j}\right)
+g^{(1)\ 2}_{221j}\left(-\frac{M_{1j}^2}{M_2^2}+2\frac{M_{1j}^4}{M_2^4}\right)
\Bigg)
\Bigg]
+\mathcal{O}\left(E^{-1}\right)
\hspace{1.75in}
\end{eqnarray}
\end{minipage}

Using the Feynman rules given in Eq.~\eqref{eq:g_220k} gives the S-channel diagram\\
\begin{minipage}{7in}
\begin{minipage}{2in}
\includegraphics[scale=0.4]{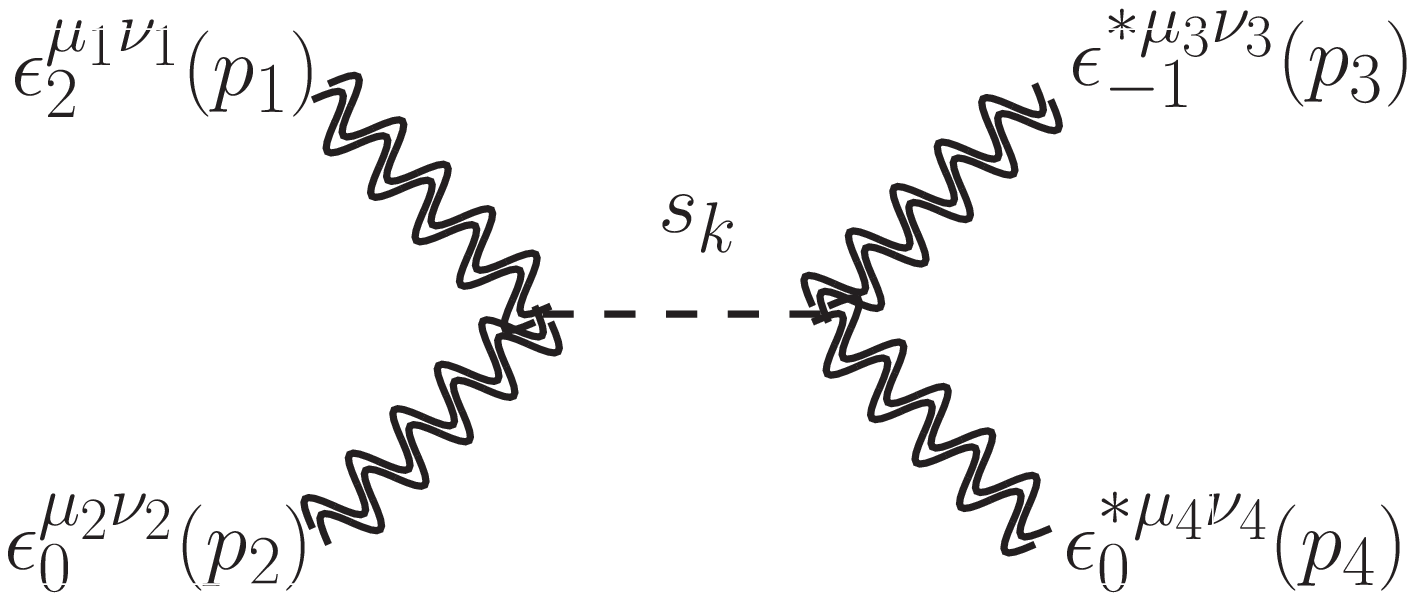}
\end{minipage}
\begin{minipage}{5in}
\begin{eqnarray}
&&\mathcal{M}^{Ssk}_{20\minus10} =0
\end{eqnarray}
\end{minipage}
\end{minipage}\\
and the T+U-channel diagrams\\
\begin{minipage}{7in}
\begin{minipage}{2in}
\includegraphics[scale=0.4]{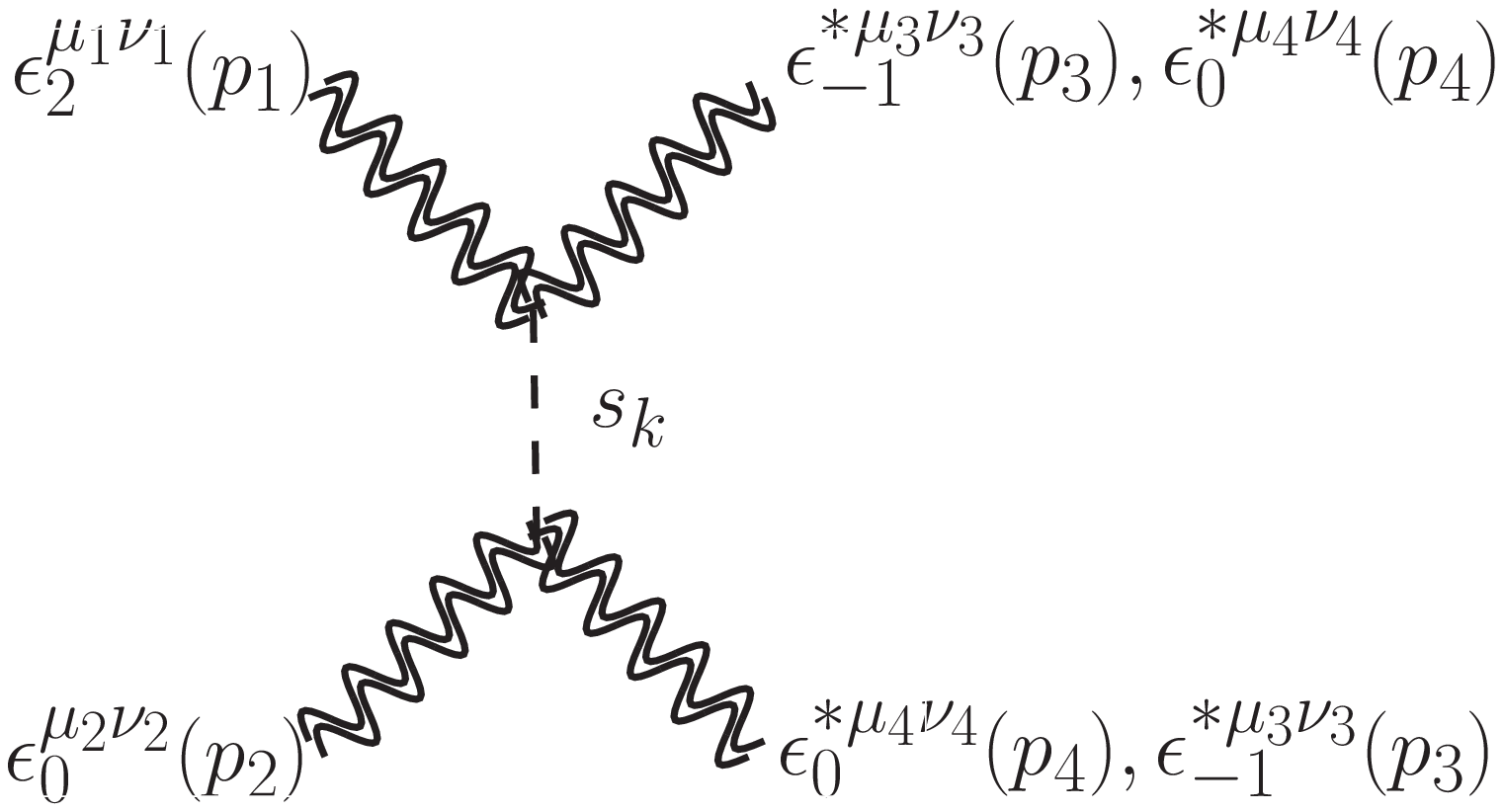}
\end{minipage}
\hfill
\begin{minipage}{4.75in}
\begin{eqnarray}
&&\mathcal{M}^{TUsk}_{20\minus10} =
\frac{E^{3}}{M_2^{5}}\frac{4}{3}g_{220k}^2\left(\cos\theta-1\right)\sin\theta\nonumber\\
&&-\frac{E^1}{M_2^{3}}g_{220k}^2\left(\cos\theta-1\right)\sin\theta\Bigg[
1+\frac{2}{3}\frac{M_{0k}^2}{M_2^2}
\Bigg]
+\mathcal{O}\left(E^{-1}\right)
\end{eqnarray}
\end{minipage}
\end{minipage}

\end{widetext}

\end{document}